\documentclass[aps,prd,10pt,notitlepage,nofootinbib,superscriptaddress,showkeys,showpacs,preprintnumbers]{revtex4-1}
\usepackage{amstext,amsmath,amssymb,amsfonts,bbm,euscript,color}
\usepackage{epsfig}
\usepackage{hyperref}
\usepackage{amsthm}
\usepackage{color}
\usepackage{multirow}
\usepackage{tikz}
\usepackage{bbold}
\usepackage{bm}
\usepackage{ulem}
\usepackage{comment}
\usepackage{graphicx}
\usepackage{subcaption}
\usepackage{diagbox}

\usetikzlibrary{arrows,backgrounds}

\usepackage{verbatim}

\usepackage{graphicx}

\usepackage{latexsym}

\newcommand{\bea}{\begin{eqnarray}}	
\newcommand{\eea}{\end{eqnarray}}
\newcommand{\be}{\begin{equation}}	
\newcommand{\ee}{\end{equation}}
\newcommand{\beq}{\begin{equation}}	
\newcommand{\eeq}{\end{equation}}

\newcommand{\Z}{{\mathbb Z}}
\newcommand{\C}{{\mathbb C}}

\newcommand{\br}{\bar R}
\newcommand{\bg}{\bar g}
\newcommand{\bnabla}{\bar\nabla}

\def \CDbar{\bar{\nabla}}

\newcommand{\al}[1]{\begin{align}#1\end{align}}

\newcommand{\Tr}{{\rm Tr}}

\newcommand{\dd}{{\textrm{d}}}

\def\R{\relax\ifmmode {\mathbb R}  \else${\mathbb R}$\fi}
\def\C{\relax\ifmmode {\mathbb C}  \else${\mathbb C}$\fi}
\def\Z{\relax\ifmmode {\mathbb Z}  \else${\mathbb Z}$\fi}
\def\N{\relax\ifmmode {\mathbb N}  \else${\mathbb N}$\fi}
\def\I{\relax\ifmmode {\mathbb I}  \else${\mathbb I}$\fi}

\begin{document}

\preprint{KU-TP 073}

\title{Asymptotic safety and field parametrization dependence in the \texorpdfstring{$f(R)$}{TEXT} truncation}


\author{Gustavo P. de Brito}\email{gpbrito@cbpf.br}

\affiliation{CBPF $-$ Centro Brasileiro de Pesquisas F\'isicas,
Rua Dr. Xavier Sigaud 150, 22290-180,
Rio de Janeiro, RJ, Brazil}

\author{Nobuyoshi~Ohta}\email{ohtan@phys.kindai.ac.jp}

\affiliation{Department of Physics, Kindai University,
Higashi-Osaka, Osaka 577-8502, Japan}

\affiliation{Maskawa Institute for Science and Culture,
Kyoto Sangyo University, Kyoto 603-8555, Japan}

\author{Antonio~D.~Pereira}\email{a.pereira@thphys.uni-heidelberg.de}

\affiliation{Institute for Theoretical Physics, University of Heidelberg,\\ 
Philosophenweg 12, 69120 Heidelberg, Germany}

\author{Anderson~A.~Tomaz}\email{tomaz@cbpf.br}

\affiliation{CBPF $-$ Centro Brasileiro de Pesquisas F\'isicas,
Rua Dr. Xavier Sigaud 150, 22290-180,
Rio de Janeiro, RJ, Brazil}

\author{Masatoshi~Yamada}\email{m.yamada@thphys.uni-heidelberg.de}

\affiliation{Institute for Theoretical Physics, University of Heidelberg,\\ 
Philosophenweg 12, 69120 Heidelberg, Germany}


\begin{abstract}
We study the dependence on field parametrization of the functional renormalization group equation in the $f(R)$ truncation for the effective average action. We perform a systematic analysis of the dependence of fixed points and critical exponents in polynomial truncations. We find that, beyond the Einstein-Hilbert truncation, results are qualitatively different depending on the choice of parametrization. In particular, we observe that there are two different classes of fixed points, one with three relevant directions and the other with two. The computations are performed in the background approximation. We compare our results with the available literature and analyze how different schemes in the regularizations can affect the fixed point structure. 

 

\end{abstract}

\maketitle


\section{Introduction}
The well-known fact that quantum gravity based on the Einstein-Hilbert theory is not perturbatively renormalizable \cite{tHooft:1974toh,Christensen:1979iy,Goroff:1985th} has motivated the construction of different approaches to quantum gravity beyond the standard quantum field theory realm. Within the local quantum field theory toolbox, a famous attempt to circumvent the perturbative nonrenormalizability of general relativity is the introduction of higher-derivative terms \cite{Stelle:1976gc,Julve:1978xn,Fradkin:1981hx,Fradkin:1981iu,Tomboulis:1983sw,Avramidi:1985ki}. As is well known, such a theory is perturbatively renormalizable. On the other hand, the existence of ghosts at the perturbative level is the main obstacle to make this theory a suitable candidate for quantum gravity. Recently, this issue has been investigated from different point of views, see \cite{deBerredoPeixoto:2003pj,deBerredoPeixoto:2004if,Mannheim:2006rd,Codello:2006in,Ohta:2013uca,Ohta:2015zwa,Alvarez-Gaume:2015rwa,Modesto:2015ozb,Holdom:2015kbf,Donoghue:2017fvm}.
The possibility for a consistent quantum theory of quadratic gravity can be realized within the quantum field
theory realm in a nonperturbative fashion. That is, asymptotic safety program for the quantization of the gravitational interaction, which was introduced by Weinberg in \cite{Hawking:1979ig}, is a possible candidate of the formulation for quantum gravity as a continuum quantum field theory, see review papers~\cite{Niedermaier:2006wt,Niedermaier:2006ns,Percacci:2007sz,Reuter:2012id,Codello:2008vh} and the recent \cite{Eichhorn:2017egq}.
It is crucial for the asymptotic safety scenario that the theory has a nontrivial fixed point in the renormalization group (RG) flow, a property which ensures the ``nonperturbative'' renormalizability of the theory.
In order for this approach to be predictive, the critical surface spanned by relevant operators of the RG flow, i.e. the surface made of the set of points which are attracted to the nontrivial fixed point, has to be finite dimensional.
The coupling constants of the operators corresponding to the relevant directions become free parameters to be fixed by experiments.
In this way, asymptotically safe quantum gravity could lead to an ultraviolet (UV) complete theory.
To look for such a nontrivial fixed point, nonperturbative techniques have to be employed.

In practice, the functional renormalization group (FRG) is a powerful method to study systems without relying on any perturbative expansion parameter such as coupling constants and the spacetime dimensions, see review papers~\cite{Morris:1998da,Berges:2000ew,Aoki:2000wm,Bagnuls:2000ae,Polonyi:2001se,Pawlowski:2005xe,Gies:2006wv,Delamotte:2007pf,Rosten:2010vm,Braun:2011pp}.
A lot of analyses have been devoted to investigating the asymptotic safety scenario for quantum gravity using the FRG after its pioneering work~\cite{Reuter:1996cp}.
Indeed, it has been shown that the pure-gravity system could have a nontrivial fixed point at which the dimension of the critical surface becomes finite~\cite{Souma:1999at,Reuter:2001ag,Litim:2003vp,Codello:2006in,Benedetti:2009rx,Benedetti:2009gn,Manrique:2010am,Manrique:2011jc,Christiansen:2012rx,Falls:2013bv,Benedetti:2013jk,Codello:2013fpa,Falls:2014tra,Christiansen:2014raa,Christiansen:2015rva,Gies:2015tca,Gies:2016con,Biemans:2016rvp,Christiansen:2016sjn,Denz:2016qks,Knorr:2017fus,Knorr:2017mhu,Christiansen:2017bsy,Falls:2017lst}.
One of the strengths of asymptotically safe quantum gravity is the predictability for low-energy physics.
The finiteness of the critical surface dimensionality at the fixed point in the system where gravity is coupled to matter could strongly constrain the low-energy dynamics of matter~\cite{Percacci:2002ie,Percacci:2003jz,Narain:2009fy,Eichhorn:2011pc,Eichhorn:2012va,Dona:2013qba,Dona:2014pla,Labus:2015ska,Oda:2015sma,Meibohm:2015twa,Dona:2015tnf,Meibohm:2016mkp,Eichhorn:2016esv,Eichhorn:2016vvy,Biemans:2017zca,Hamada:2017rvn,Christiansen:2017qca,Eichhorn:2017eht,Eichhorn:2017egq,Eichhorn:2017sok,Christiansen:2017cxa,Eichhorn:2017als,Eichhorn:2018akn}.
Indeed, it has been successful in predicting quantities such as the Higgs and top-quark masses and the charges in low-energy regimes~\cite{Shaposhnikov:2009pv,Harst:2011zx,Christiansen:2017gtg,Eichhorn:2017ylw,Eichhorn:2017lry,Eichhorn:2017muy,Eichhorn:2018whv} and to give hints towards a solution to problems such as the gauge hierarchy problem~\cite{Wetterich:2016uxm}, the U(1) Landau-pole problem~\cite{Christiansen:2017gtg} and the cosmological constant problem~\cite{Wetterich:2017ixo}.
Cosmological consequences from asymptotically safe gravity for the early universe are also discussed in \cite{Bonanno:2015fga,Bonanno:2017pkg,Bonanno:2018gck}.
These facts not only encourage the asymptotic safety scenario for quantum gravity but also could give a test of it from the low-energy physics observations.

As already mentioned, a crucial point in the asymptotically safe scenario is the finiteness of the number of relevant
operators with positive critical exponents around the fixed point, from which the RG flow goes away.
To study this, we have to extend our theory space as much as possible and identify this number.
However the theory space is in general infinite dimensional, and this is practically impossible.
The usual strategy is to make some truncations, and extend the space slightly and check that the extension
does not much affect the result.
In pure gravity system, the largest truncation to date is the $f(R)$-type one, namely, the theory space spanned
by a function of the Ricci scalar $R$.
If asymptotic safety is realized, we should see that the fixed point values of the relevant couplings
converge to certain numbers and the number of relevant directions does not change upon enlargements of the truncations of theory space.
See \cite{Codello:2007bd,Machado:2007ea,Codello:2008vh,Benedetti:2012dx,Benedetti:2013jk,Falls:2014tra,Dietz:2012ic,Dietz:2013sba,Demmel:2014sga,Ohta:2015efa,Ohta:2015fcu,Falls:2016msz,Christiansen:2017bsy,Alkofer:2018fxj} for investigations of the asymptotic safety scenario in the $f(R)$ truncation and \cite{Eichhorn:2015bna} in the unimodular case.

Recent works on this type of truncation in four-dimensional spacetime indeed show the good convergence of the values of the critical exponents~\cite{Falls:2013bv,Falls:2014tra}.
In particular, it is found that the number of the relevant directions is three.
That is, asymptotically safe quantum gravity could describe the low-energy dynamics with three free parameters.
This is also observed in the $R^2+C^2$-type truncation~\cite{Codello:2006in,Benedetti:2009rx,Benedetti:2009gn,Falls:2017lst}, where $C^2$ is the squared Weyl tensor.

Apart from the fact that the existence of the fixed point has to be tested for richer truncations, there is an issue that has been under investigation in recent years: The fixed point is computed by demanding that all beta functions of the theory vanish simultaneously. Beta functions are off-shell quantities and as such there could be spurious parameter-dependences such as gauge choice and/or field parametrization. 
The following question then naturally comes to mind: what is a suitable choice of these parameters?
A list of works on this topic and related aspects is \cite{Kalmykov:1995fd,Kalmykov:1998cv,Falls:2015qga,Ohta:2016npm,Ohta:2016jvw,Gies:2015tca,Falls:2017cze,Goncalves:2017jxq}.
To calculate the beta functions, we usually use the background field method where the metric $g_{\mu\nu}$ is decomposed into a background part $\bar g_{\mu\nu}$ and a fluctuation $\delta g_{\mu\nu}$.
Many earlier works including \cite{Falls:2013bv,Falls:2014tra} have employed a linear split, namely, $g_{\mu\nu}= \bar g_{\mu\nu}+\delta g_{\mu\nu}=\bar g_{\mu\nu}+h_{\mu\nu}$. As reported, they obtain a fixed point with very good convergent properties and three relevant directions.
On the other hand, one can also split the metric in the exponential form, $g_{\mu\nu}={\bar g}_{\mu\alpha}(e^{h})^\alpha_{~\nu}$.
Studies with the exponential parametrization in $f(R)$-type truncations were performed in \cite{Ohta:2015efa,Ohta:2015fcu,Alkofer:2018fxj} and have found only two relevant directions at the fixed point.
Even though the number of relevant directions is finite, it is an urgent problem to clarify why and how the number changes depending on the way of the metric parametrization since its information is important for the asymptotically safe gravity scenario. In particular, one can ask if different schemes in the field parametrization can affect the number of relevant directions. Here we should note that the previous results were obtained in the so-called background field approximation. So, one should also investigate whether such discrepancies are still present in more sophisticated truncations schemes.

In this work, we investigate how different choices of field parametrization affect the fixed point analysis in $f(R)$ truncations for the effective average action. In particular, we employ polynomial truncations up to sixth degree on $R$ for a one-parameter family of field parametrization as introduced in \cite{Ohta:2016npm,Ohta:2016jvw} and discuss the dependence of the number of relevant directions for different parametrizations. Also, we perform our analysis for two different values of the gauge parameter $\beta$ - introduced in Eq.~\eqref{gf2} - namely $\beta=0$ and $\beta\to-\infty)$. We also comment on different choices of basis for the computations and different prescriptions to deal with spurious modes coming from field decompositions. One of the main conclusions is that we do find, upon changes in the parameter which interpolates between different field parametrizations, that the number of relevant directions change. In particular, we are able to find fixed points with two and three relevant directions under certain conditions, in agreement with the existing literature.

The structure of the paper is the following: In the next section, we write a brief review of the functional renormalization group equation (FRGE) in order to fix our conventions. After that, we introduce the truncation for the effective average action employed in this work and discuss the derivation of the Hessians. Subsequently, we briefly discuss the existence of a duality discovered and discussed in \cite{Ohta:2016npm,Ohta:2016jvw,Ohta:2018sze} for the theory considered in this paper. Then, in Sect.~V we set up the flow equation and in Sect.~VI we collect our results for fixed points and critical exponents. Finally, we present some discussions and perspectives in Sect.~\ref{Discussion} and draw our conclusions in Sect.~\ref{Conclusions}. Technicalities relevant for a self-consistent presentation are presented in the appendices.

\section{Brief overview of the flow equation}

The search for suitable nontrivial fixed points for the asymptotic safety scenario is mostly carried out through the FRGE, see \cite{Wetterich:1992yh,Morris:1993qb}. This equation encodes the scale dependence of the so-called \textit{effective average action} $\Gamma_k$, obtained by the integration of modes with momentum greater than $k$. For $k=0$, the effective average action corresponds to the effective action (1PI generating functional) $\Gamma$, while for $k\to\Lambda$, with $\Lambda$ a UV cutoff, to the bare action $S_{\Lambda}$. In this perspective, $k$ plays the role of an infrared cutoff. 
The Wetterich equation (or \textit{flow} equation) is formally written as
\begin{equation}
\partial_t \Gamma_k = \frac{1}{2}\mathrm{STr}\left[\frac{\partial_t \textbf{R}_k}{\Gamma^{(2)}_k+\textbf{R}_k}\right]\,,
\label{fe01}
\end{equation}
where $\partial_t \equiv k\partial_k$, $\Gamma^{(2)}_k$ is a notation for the Hessian\footnote{We use $\Phi$ as a shorthand notation for all the fields of the theory.} $\Gamma^{(2)}_k = \delta^{2}\Gamma_k/\delta \Phi \delta \Phi$, $\mathrm{STr}$ denotes the supertrace which takes into account appropriate numerical factors depending on the nature of the superfield $\Phi$ and $\textbf{R}_k$ is a cutoff function responsible for the suppression of modes with momentum smaller than $k$. The effective average action contains all field operators compatible with the symmetries of the underlying theory and can be generically expressed as
\begin{equation}
\Gamma_k =  \sum_i g_i (k)\EuScript{O}^i (\Phi)\,,
\label{fe02}
\end{equation}
where $\EuScript{O}^i (\Phi)$ is an integrated field operator and $g_i (k)$ the corresponding coupling constant for such an operator. Taking the scale derivative $\partial_t$ of \eqref{fe02} leads to
\begin{equation}
\partial_t \Gamma_k =\sum_i \partial_t g_i (k) \EuScript{O}^i (\Phi)
\equiv \sum_i \beta_i (k) \EuScript{O}^i (\Phi)\,,
\label{fe03}
\end{equation}
with $\beta_i$ being the beta function of the coupling $g_i$.
One can obtain the explicit forms of the beta functions from \eqref{fe01}.

After the computation of the beta functions, one can immediately look for fixed points by demanding that all beta functions vanish simultaneously, namely, $\beta_i=0$ for all $i$.
Since exactly solving Eq.~\eqref{fe01} for the complete theory space is far from our capabilities, some approximations must be employed. A useful scheme is the implementation of truncations, namely, a particular basis which spans part of the theory space $\left\{\EuScript{O}\right\}$ is chosen. Note that even though we make approximations, they do not correspond to a perturbative expansion and nonperturbative effects turn out to be accessible. 

Once an interacting fixed point is found, one still has to check whether it leads to a predictive theory or not. 
The fixed point will lead to a predictive theory if the critical surface is finite dimensional. 
This can be studied by the linearized flow around the fixed point, let us say $g^{\ast} = \left\{g^{\ast}_1,g^{\ast}_2,\ldots,g^{\ast}_n\right\}$, with $n$ being the dimensionality of the theory space.
The linearized flow equation is obtained by
\begin{equation}
\partial_t (g_l - g^\ast_l) = -\sum_m^n B_{lm}(g_m - g^\ast_m)\,\qquad \text{with}\qquad B_{lm} = -\frac{\partial \beta_l}{\partial g_m}\Bigg|_{g=g^\ast}\,.
\label{fe04}
\end{equation}
The matrix $B_{lm}$ is known as stability matrix and its eigenvalues are called critical exponents.
The dimensionality of the critical surface is the number of positive (relevant) critical exponents. If it is finite, we can define a UV complete theory associated with such a fixed point.
More specifically, the solution of \eqref{fe04} is given by
\al{
g_i=g_i^*+\sum_{j=1}^{n}V_{ij}C_j\left( \frac{k}{\Lambda}\right)^{-\theta_j},
}
where $V_{ij}$ is the matrix diagonalizing the stability matrix, $C_j$ are undetermined constants for the integration of the scale and the critical exponents are denoted by $\theta_j$.
In the context of asymptotic safety, the coupling constants with $\theta_j>0$ are relevant and then the corresponding $C_j$ become free parameters to be fixed.
Note that in the perturbative approach, these operators just correspond to the renormalizable interactions with four and less dimensions.

\section{\texorpdfstring{$f(R)$}{TEXT}-truncation: Preliminaries}

\subsection{Gravitational sector}

The gravitational action we consider in this work has the form
\begin{equation}
\Gamma_k [g] = \int \dd^dx\sqrt{g}\,f_k (R)\,,
\label{fr1}
\end{equation}
with $k$ encoding the scale dependence of the action and $f$ is a generic function of the Ricci scalar $R$. For simplicity, we assume that $f$ admits a Taylor expansion on $R$. 
Employing \eqref{ap11}, one obtains
\al{
\Gamma_k [h_{\mu\nu};\bg_{\mu\nu}] &= \int \dd^d x\, \sqrt{\bg} \, \bigg[  f_k (\br) + \frac{1}{2} \delta g^{(1)}  f_k (\br) + R^{(1)}  f^\prime_k (\br)  +  \frac{1}{2} {R^{(1)}}^2 f^{\prime\prime}_k (\br) +  \frac{1}{2}\Big(\delta g^{(1)}  R^{(1)} + 2 R^{(2)} \Big)   f^\prime_k (\br) \nonumber \\
&\quad
+ \frac{1}{8} \Big( \left( \delta g^{(1)}\right)^2 - 2  \delta g^{(1)}_{\mu\nu}  \delta g^{(1){\mu\nu}} +  \delta g^{(2)}  \Big) f_k (\br)  + \mathcal{O}(h^3) \bigg] .
\label{fr2}
}
The explicit expressions for $R^{(1)}$ and $R^{(2)}$ are reported in Appendix~\ref{ap1}. In the present work, we restrict the background to be a maximally symmetric space:
\begin{equation}
\br_{\mu\nu}=\frac{\br}{d}\bg_{\mu\nu}\,,\qquad \br_{\mu\nu\alpha\beta}=\frac{\br}{d(d-1)}(\bg_{\mu\alpha}\bg_{\nu\beta}-\bg_{\mu\beta}\bg_{\nu\alpha})\,,
\label{fr3}
\end{equation}
with $\br$ a constant. Also, we retain the terms of \eqref{fr2} up to second order in the fluctuation $h_{\mu\nu}$ and neglect total derivatives. Then the second order expression in $h$ for $\Gamma_k$ is given by
\al{
\Gamma_{k,2}[h_{\mu\nu};\bg_{\mu\nu}] &= \int \dd^d x\, \sqrt{\bg} \, \bigg[\frac{1}{8} \Big( \left(\delta g^{(1)}\right)^2 - 2 \delta g^{(1)}_{\mu\nu} \delta g^{(1)\mu\nu} + 4\delta g^{(2)}  \Big) f_k (\br)+\left(\frac{1}{4} \delta g^{(1)\mu\nu} \bnabla^2 \delta g^{(1)}_{\mu\nu}   \right.\nonumber \\ 
&\quad
-\left.\frac{1}{4} \delta g^{(1)}\bnabla^2 \delta g^{(1)}  + \frac{1}{2} \delta g^{(1)}  \bnabla_\mu \bnabla_\nu \delta g^{(1)\mu\nu} + \frac{1}{2} \bnabla_\mu \delta g^{(1)\mu\nu} \bnabla_\alpha \delta g^{(1)}_{\nu}{}^{\alpha}+\frac{1}{2}\delta g^{(1)\mu\nu}\delta g^{(1)\alpha\beta}\br_{\mu\alpha\nu\beta} \right.\nonumber\\ 
&\quad
-\left. \frac{1}{2} \delta g^{(1)} \delta g^{(1)\mu\nu}\br_{\mu\nu}+\frac{1}{2}\delta g^{(1)\mu\nu}\delta g^{(1)}_{\mu}{}^{\alpha}\br_{\nu\alpha} - \delta g^{(2)\mu\nu}\br_{\mu\nu}\right) f^{\prime}_k (\br) +  \frac{1}{2} \Big( \bnabla_\mu \bnabla_\nu \delta g^{(1)\mu\nu} \bnabla_\alpha \bnabla_\beta \delta g^{(1)\alpha\beta} \nonumber\\
&\quad
+ \delta g^{(1)}  \bnabla^2 \delta g^{(1)} - 2\, \bnabla_\mu \bnabla_\nu \delta g^{(1)\mu\nu}  \bnabla^2 \delta g^{(1)} - 2\br_{\alpha\beta}\delta g^{(1)\alpha\beta}\bnabla_\mu \bnabla_\nu \delta g^{(1)\mu\nu}   + 2\br_{\mu\nu} \delta g^{(1)\mu\nu}  \bnabla^2 \delta g^{(1)}\nonumber\\
&\quad
+\br_{\mu\nu}\br_{\alpha\beta} \delta g^{(1)\mu\nu}\delta g^{(1)\alpha\beta}   \Big) f^{\prime\prime}_k (\br)   \bigg]\,,
\label{fr4}
}
where the subscript $2$ means that we keep only quadratic terms on the quantum fluctuation $h_{\mu\nu}$. Next, we employ the York decomposition for the fluctuation $h_{\mu\nu}$:
\begin{equation}
h_{\mu\nu} = h^{\mathrm{TT}}_{\mu\nu}+\bnabla_{\mu}\xi_{\nu}+\bnabla_{\nu}\xi_{\mu}+\bnabla_\mu \bnabla_\nu \sigma - \frac{1}{d}\bnabla^2 \sigma + \frac{1}{d}\bg_{\mu\nu}h\,,
\label{fr5}
\end{equation}
with
\begin{equation}
\bnabla^\mu h^{\mathrm{TT}}_{\mu\nu} = 0\,,\qquad \bg^{\mu\nu}h^{\mathrm{TT}}_{\mu\nu} = 0\,,\qquad \bnabla^\mu \xi_\mu = 0\,, \qquad h = \bg^{\mu\nu}h_{\mu\nu}\,.
\label{fr6}
\end{equation}
As is well known~\cite{Codello:2008vh,Lauscher:2001ya}, the York decomposition entails the introduction of Jacobians. 
We will introduce the effects of these Jacobians later on.
Another feature which has to be taken into account is the existence of solutions to the equations
\begin{equation}
\bnabla_{\mu}\xi_{\nu}+\bnabla_{\nu}\xi_{\mu}=0\qquad\mathrm{and}\qquad \bnabla_\mu \bnabla_\nu \sigma - \frac{1}{d}\bar{g}_{\mu\nu}\bnabla^2 \sigma=0\,.
\label{fr7}
\end{equation}
Configurations $(\xi,\sigma)$ that satisfy Eq.~\eqref{fr7} do not contribute to $h_{\mu\nu}$ and should be removed from the path integral. This procedure is explained in details in \cite{Codello:2008vh,Lauscher:2001ya,Machado:2007ea} and will be used in this paper.

\subsection{Gauge-fixing and ghost sectors}

For the present computation we employ the same gauge-fixing term as in \cite{Ohta:2016npm}. We add to \eqref{fr4} the term
\begin{equation}
S_{\mathrm{gf}} = \frac{Z_N}{2\alpha}\int \dd^dx\,\sqrt{\bg}\, F_{\mu}\bg^{\mu\nu}F_{\nu}\,,
\label{gf1}
\end{equation}
with
\begin{equation}
F_{\mu} = \bnabla_\alpha h^{\alpha}{}_{\mu}-\frac{1+\bar{\beta}}{d}\bnabla_\mu h\,,
\label{gf2}
\end{equation}
$\alpha$ and $\bar{\beta}$ are gauge parameters and $Z_N$ is chosen such that $\alpha$ is dimensionless. We choose $Z_N = 1/(16\pi G_{\mathrm{bare}})$ with $G_{\mathrm{bare}}$ being the bare or classical Newton's constant, see \cite{Machado:2007ea}. As in \cite{Ohta:2016npm}, we rescale $\bar{\beta}$ such that $\bar{\beta} = \beta (1+dm)$. Eq.~\eqref{gf1} can be expressed as
\begin{equation}
S_{\mathrm{gf}} = -\frac{Z_N}{2\alpha}\int \dd^dx\,\sqrt{\bg}\,\left[h_{\mu\nu}\bnabla^{\nu}\bnabla^{\alpha}h_{\alpha}{}^{\mu}-2\frac{1+\beta (1+dm)}{d}h\bnabla^{\mu}\bnabla^{\nu}h_{\mu\nu}+\left(\frac{1+\beta (1+dm)}{d}\right)^2 h\bnabla^2 h\right]\,.
\label{gf3}
\end{equation}
Applying the York decomposition \eqref{fr5} leads to
\begin{eqnarray}
S_{\mathrm{gf}} &=& \frac{Z_N}{2\alpha}\int \dd^dx\,\sqrt{\bg}\,\left[\xi_\mu\left(-\bnabla^2-\frac{\br}{d}\right)^2\xi^\mu-\left(\frac{d-1}{d}\right)^2\sigma\left(-\bnabla^2-\frac{\br}{d-1}\right)^2\bnabla^2\sigma-\left(\frac{1+dm}{d}\right)^2\beta^2 h\bnabla^2 h\right.\nonumber\\
&+&\left.2\beta\frac{(d-1)(1+dm)}{d^2} \sigma\left(\bnabla^2+\frac{\br}{d-1}\right)\bnabla^2 h\right].
\label{gf4}
\end{eqnarray}
Collecting the contributions coming from \eqref{fr4} and \eqref{gf4} results in
\begin{equation}
\Gamma^{\mathrm{grav}}_k = \frac{1}{2}\int \dd^dx\,\sqrt{\bg}\,\bigg[h^{\mathrm{TT}}_{\mu\nu}\Gamma^{\mu\nu\alpha\beta}_{\mathrm{TT}} h^{\mathrm{TT}}_{\alpha\beta}+\xi_\mu\Gamma^{\mu\nu}_{\xi\xi} \xi_\nu+\sigma \Gamma_{\sigma\sigma}\sigma+h\Gamma_{hh}h+\sigma\Gamma_{\sigma h}h+h\Gamma_{h\sigma}\sigma\bigg]\,,
\label{fr8}
\end{equation}
where Jacobians and spurious configurations that satisfy Eq.~\eqref{fr7} still have to be taken into account. We collect the explicit complete expressions for the Hessians in Appendix~\ref{hessians&Cutoffs}.

The introduction of the gauge-fixing term \eqref{gf1} demands the introduction of the Faddeev-Popov contribution which is 
\begin{equation}
S_{\mathrm{gh}} =  \int \dd^dx\,\sqrt{\bg}\,\bar{C}^{\mu}\left[\delta^{\nu}_\mu \bnabla^2 + \left(1-2\frac{1+\beta}{d}\right)\bnabla_\mu\bnabla^\nu+\frac{\br}{d}\delta^\nu_\mu\right]C_\nu,
\label{gf5}
\end{equation}
with $(\bar{C},C)$ being the Faddeev-Popov ghosts. We employ the York decomposition for the ghost fields, i.e.
\begin{equation}
\bar{C}^{\mu} = \bar{C}^{\mathrm{T}\mu}+\bnabla^{\mu}\bar{C}\,,\qquad C_{\mu} = C^{\mathrm{T}}_{\mu}+\bnabla_{\mu}C,
\label{gf6}
\end{equation}
where $\bnabla_\mu \bar{C}^{\mathrm{T}\mu} = \bnabla^\mu C^{\mathrm{T}}_{\mu} = 0$. 
As before, the decomposition \eqref{gf6} generates Jacobians which will be introduced later on. The decomposed ghost action is given by
\begin{equation}
S_{\mathrm{gh}} = \int \dd^dx\,\sqrt{\bg}\,\left[\bar{C}^{\mathrm{T}\mu}\left(\bnabla^2+\frac{\br}{d}\right)C^{\mathrm{T}}_{\mu}-2\frac{d-1-\beta}{d}\bar{C}\left(\bnabla^2+\frac{\br}{d-1-\beta}\right)\bnabla^2 C\right].
\label{gf7}
\end{equation}
We emphasize that in Eq.~\eqref{gf7} the gauge parameter $\beta$ appears without the factor $(1+dm)$ unlike in the gauge-fixing action \eqref{gf3}. In this work, we focus on the Landau gauge condition which corresponds to setting  $\alpha = 0$. In the next subsection we introduce the contributions coming from the Jacobians of the York decompositions \eqref{fr5} and \eqref{gf6}. 

\subsection{Auxiliary sector}

As mentioned in the previous subsections, the York decomposition brings in nontrivial Jacobians that should be taken into account in the computations. Also, spurious modes that satisfy \eqref{fr7} have to be discarded in the evaluation of the path integral since they do not contribute to $h_{\mu\nu}$. The later procedure is well discussed in the literature~\cite{Codello:2008vh,Lauscher:2001ya,Machado:2007ea}. The Jacobians can be easily derived~\cite{Codello:2008vh,Lauscher:2001ya,Machado:2007ea}, and lead to the following contributions:
\begin{equation}
J_{\mathrm{grav}}=\left[\mathrm{det}^\prime_{\mathrm{(1)}}\left(-\bnabla^2-\frac{\br}{d}\right)\right]^{1/2}\left\{\mathrm{det}^{\prime\prime}_{\mathrm{(0)}}\left[-\bnabla^{2}\left(-\bnabla^2-\frac{\br}{d-1}\right)\right]\right\}^{1/2}\,,\qquad J_{\mathrm{gh}}=\left[\mathrm{det}^\prime_{\mathrm{(0)}}(-\bnabla^2)\right]^{-1},
\label{aux1}
\end{equation}
where $J_{\mathrm{grav}}$ is the resulting Jacobian coming from the York decomposition of $h_{\mu\nu}$ while $J_{\mathrm{gh}}$ is that from the decomposition of the Faddeev-Popov ghosts. The primes denote the appropriate elimination of spurious modes. These determinants can be expressed as functional integrals:
\al{
J_{\mathrm{grav}}&=\int \EuScript{D}\zeta\EuScript{D}\bar{\zeta}\EuScript{D}\psi\EuScript{D}\bar{\psi}\EuScript{D}\chi\EuScript{D}\theta\,\mathrm{exp}\left\{-\int \dd^dx\,\sqrt{\bg}\,\left[\frac{1}{2}\chi_\mu\left(-\bnabla^2-\frac{\br}{d}\right)^\prime\chi^\mu+\frac{1}{2}\theta\left[-\bnabla^{2}\left(-\bnabla^2-\frac{\br}{d-1}\right)\right]^{\prime\prime}\theta\right.\right.\nonumber\\
&\quad
-\left.\left.\bar{\zeta}_{\mu}\left(-\bnabla^2-\frac{\br}{d}\right)^{\prime}\zeta^{\mu} -\bar{\psi}\left[-\bnabla^{2}\left(-\bnabla^2-\frac{\br}{d-1}\right)\right]^{\prime\prime}\psi\right]\right\},
\label{aux2}
} 
with $\chi_\mu$ a real bosonic vector field, $\theta$ a real bosonic scalar field, $(\bar{\zeta}_\mu,\zeta^\nu)$ vector anticommuting ghosts and $(\bar{\psi},\psi)$ scalar anticommuting ghosts. For the Jacobian coming from the ghost sector $J_{\mathrm{gh}}$, one writes
\begin{equation}
J_{\mathrm{gh}} = \int \EuScript{D}\bar{\phi}\EuScript{D}\phi\,\mathrm{exp}\left[-\int\dd^dx\,\sqrt{\bg}\,\bar{\phi}\,(-\bnabla^2)^{\prime}\phi\right],
\label{aux3}
\end{equation}
where $(\bar{\phi},\phi)$ are complex scalar fields. Hence, the contribution coming from the Jacobians of the York decomposition can be taken into account by the introduction of a set of auxiliary fields $\Phi = \left\{\chi_\mu,\theta,\bar{\zeta}_\mu,\zeta^\mu,\bar{\psi},\psi,\bar{\phi},\phi\right\}$. The complete set of fields including the gravitational, Faddeev-Popov ghosts and auxiliary sector define the following functional measure
\begin{equation}
\EuScript{D}\mu = \EuScript{D}{h^{\mathrm{TT}}}\EuScript{D}{\xi}\EuScript{D}{\sigma}\EuScript{D}{h}\EuScript{D}{\bar{C}^{\mathrm{T}}}\EuScript{D}{\bar{C}}\EuScript{D}{C^{\mathrm{T}}}\EuScript{D}{C}\EuScript{D}\zeta\EuScript{D}\bar{\zeta}\EuScript{D}\psi\EuScript{D}\bar{\psi}\EuScript{D}\chi\EuScript{D}\theta\EuScript{D}\bar{\phi}\EuScript{D}\phi\,.
\label{aux4}
\end{equation}
We are now ready to discuss FRGE and our results.

\section{Duality}

In references \cite{Ohta:2016npm,Ohta:2016jvw}, gauge and parametrization dependence of one-loop divergences were studied for Einstein-Hilbert and higher derivative theories. An interesting feature that was observed is the existence of a ``duality", namely, the invariance of the results under a discrete transformation of the parameters that characterizes different parametrizations of the fluctuation viz. $(\omega,m)$. Specifically, this duality transformation is defined by
 \begin{equation}
(\omega,m)\,\,\,\longrightarrow\,\,\, \left(1-\omega,-m-\frac{2}{d}\right).
\label{dual1}
\end{equation} 
In the present case, it is possible to prove that the duality also holds, the reason being it is an invariance of the Hessian. This is easily seen by noticing that the ghost and auxiliary sectors are independent of $(\omega,m)$ and therefore are obviously invariant. The pure gravitational sector and the gauge-fixing terms are nearly diagonal apart from the $\sigma - h$ mixing. Note that the combination $(2\omega-1)(1+dm)$ is duality invariant. Hence, the traceless-transverse $h^{\mathrm{TT}}$ and transverse $\xi$ sectors are clearly duality invariant. The diagonal scalar sector is invariant while the off-diagonal flips a sign. However, what matters for our computations is the determinant of the mixed scalar matrix which is duality invariant since the flipped sign cancels out. Therefore, our results for beta functions, fixed points and critical exponents should be explicitly invariant under \eqref{dual1}. 

\section{Setting the flow equation for the \texorpdfstring{$f(R)$}{TEXT} truncation}

\subsection{Choice of cutoff function}
In this paper we use the Type I cutoff functions, see \cite{Codello:2008vh}. 
That is, the regulator functions $R_k$ takes the following form:
\begin{equation}
-\bnabla^2\quad\longrightarrow\quad P_k (-\bnabla^2) = -\bnabla^2 + R_k (-\bnabla^2),
\label{fe1}
\end{equation}
with $P_k (-\bnabla^2)$ being the regularized Laplacian operator. This is easily achieved by defining the regulator $\textbf{R}_k (-\bnabla^2)$ as
\begin{equation}
\textbf{R}^{\phi_i \phi_j}_k (-\bnabla^2) = \Gamma^{\phi_i \phi_j}_k (P_k (-\bnabla^2))- \Gamma^{\phi_i \phi_j}_k(-\bnabla^2),
\label{fe2}
\end{equation}
with $\Gamma^{\phi_i \phi_j}_k$ denoting the second derivative of $\Gamma_k$ with respect to the fields $\phi_i$ and $\phi_j$. Finally, we choose for the profile function of the regulator, the optimized or Litim's cutoff \cite{Litim:2001up}, given by
\begin{equation}
R_k (z) = (k^2 - z)\theta (k^2 - z).
\label{fe3}
\end{equation}
An interesting and important task is the investigation of the stability of the results reported in this work under modifications of the cutoff type (see the recent paper \cite{Alkofer:2018fxj} for this kind of analysis in the context of gravity-matter systems within the $f(R)$ truncation in the exponential parametrization) and profile function. 
We leave this detailed analysis for future study. Here instead we simply point out that results computed in the linear parametrization of the quantum fluctuation point to a fixed point with three relevant directions, see \cite{Codello:2007bd,Codello:2008vh,Falls:2014tra} while in the exponential parametrization a fixed point with two relevant directions is reported in the literature, see \cite{Ohta:2015fcu,Alkofer:2018fxj}. We verify that such discrepancy is due to different choices of regularization schemes.  

\subsection{Flow equation}
The flow equation \eqref{fe01} is written explicitly in terms of all the fields coming from the York decomposition of $h_{\mu\nu}$, Faddeev-Popov ghosts and auxiliary fields as 
\begin{align}
&\partial_t \Gamma_k[0;\bg] = \frac{1}{2} \Tr_{(2)} \Bigg[ \frac{\partial_t \textbf{R}_k^{\mathrm{TT}} }{ \Gamma^{\mathrm{TT}}_k + \textbf{R}_k^{\mathrm{TT}} } \Bigg] + \frac{1}{2} \Tr^\prime_{(1)} \Bigg[ \frac{\partial_t \textbf{R}_{k}^{\xi\xi} }{ \Gamma^{\xi\xi}_k + \textbf{R}_{k}^{\xi\xi} } \Bigg]  +  \frac{1}{2} \sum_{j=1}^2 \frac{\partial_t \textbf{R}^{hh}_k(\lambda_j)}{\Gamma^{hh}_k (\lambda_j) + \textbf{R}^{hh}_k(\lambda_j)} \nonumber\\
&+ \frac{1}{2} \Tr_{(0)}^{\prime\prime} \Bigg[ \begin{pmatrix}
\Gamma^{\sigma \sigma}_k + \textbf{R}^{\sigma\sigma}_k & \Gamma^{\sigma h}_k + \textbf{R}^{\sigma h}_k\\ 
\Gamma^{h \sigma}_k + \textbf{R}^{h \sigma}_k & \Gamma^{hh}_k + \textbf{R}^{hh}_k
\end{pmatrix}^{-1}  \begin{pmatrix}
\partial_t \textbf{R}^{\sigma \sigma}_k & \partial_t \textbf{R}^{\sigma h}_k\\ 
\partial_t \textbf{R}^{h \sigma}_k & \partial_t \textbf{R}^{hh}_k
\end{pmatrix}\Bigg]  - \Tr_{(1)} \Bigg[ \frac{ \partial_t \textbf{R}_k^{C^\mathrm{T}\!C^\mathrm{T}} }{ \Gamma^{C^\mathrm{T}\!C^\mathrm{T}}_k + \textbf{R}_k^{C^\mathrm{T} \!C^\mathrm{T}} }  \Bigg]   \nonumber\\
&-\! \Tr_{(0)}^\prime \Bigg[ \frac{ \partial_t \textbf{R}_k^{C\,C} }{ \Gamma^{C\,C}_k + \textbf{R}_k^{C \,C} }  \Bigg] 
+ \frac{1}{2} \Tr_{(1)}^\prime \Bigg[ \frac{\partial_t \textbf{R}_{k}^{\chi\chi} }{ \Gamma^{\chi\chi}_k + \textbf{R}_{k}^{\chi\chi} } \Bigg] \!+\! \frac{1}{2} \Tr_{(0)}^{\prime\prime} \Bigg[ \frac{\partial_t \textbf{R}_{k}^{\theta\theta} }{ \Gamma^{\theta\theta}_k + \textbf{R}_{k}^{\theta\theta} } \Bigg] \!+\!  \Tr_{(0)}^\prime \Bigg[ \frac{\partial_t \textbf{R}_{k}^{\bar{\phi}\!\phi} }{ \Gamma^{\bar{\phi}\,\phi}_k + \textbf{R}_{k}^{\bar{\phi}\,\phi} } \Bigg] \nonumber\\
&-\Tr_{(1)}^\prime \Bigg[ \frac{\partial_t \textbf{R}_{k}^{\zeta\zeta} }{ \Gamma^{\zeta\zeta}_k + \textbf{R}_{k}^{\zeta\zeta} } \Bigg]  - \Tr_{(0)}^{\prime\prime} \Bigg[ \frac{\partial_t \textbf{R}_{k}^{\psi\psi} }{ \Gamma^{\psi\psi}_k + \textbf{R}_{k}^{\psi\psi} } \Bigg]\,,
\label{fe4}
\end{align}
where we explicitly point out that we are working within the background approximation by setting $h_{\mu\nu}=0$. In comparison with expression \eqref{fe01} we see that suitable factors of $1/2$ and $-1$ must be taken into account depending on the nature of the field. Finally, the Hessians for the $(\sigma,h)$ sector contribute as
\begin{gather}
 \Tr_{(0)}^{\prime\prime} \Bigg[ \begin{pmatrix}
\Gamma^{\sigma \sigma}_k + \textbf{R}^{\sigma\sigma}_k & \Gamma^{\sigma h}_k + \textbf{R}^{\sigma h}_k\\ 
\Gamma^{h \sigma}_k + \textbf{R}^{h \sigma}_k & \Gamma^{hh}_k + \textbf{R}^{hh}_k
\end{pmatrix}^{-1}  \begin{pmatrix}
\partial_t \textbf{R}^{\sigma \sigma}_k & \partial_t \textbf{R}^{\sigma h}_k\\ 
\partial_t \textbf{R}^{h \sigma}_k & \partial_t \textbf{R}^{hh}_k
\end{pmatrix}\Bigg] =  \Tr_{(0)}^{\prime\prime}\left[W^{\sigma h}_k(-\bnabla^2)\right]\,,
\label{fe5}
\end{gather}
with
\begin{equation}
W^{\sigma h}_k(-\bnabla^2) = \frac{\Gamma^{hh}_k \partial_t\mathbf{R}^{\sigma\sigma}_k+\Gamma^{\sigma\sigma}_k \partial_t\textbf{R}^{hh}_k - \Gamma^{\sigma h}_k\partial_t \mathbf{R}^{h\sigma}_k-\Gamma^{h\sigma}_k \partial_t \mathbf{R}^{\sigma h}_k}{\Gamma^{\sigma\sigma}_k \Gamma^{hh}_k-\Gamma^{\sigma h}_k \Gamma^{h \sigma}_k}\,.
\label{fe6}
\end{equation}

Using the heat kernel coefficients given in Appendix~\ref{heatkernel} together with the above results, we can write down the FRGE. Since its final form is lengthy we do not write it explicitly.

\section{Fixed points and critical exponents}

In order to obtain concrete results, we consider polynomial truncations up to $N$-th order in $R$, namely
\begin{equation}
f_k (R) = \sum^{N}_{i=0}g_{i}(k) R^i,
\label{fp1}
\end{equation}
where $g_{i}(k)$ denotes the scale dependent couplings. In particular, $g_0 = \Lambda_k/(8\pi G_k)$ and $g_1 = -1/(16\pi G_k)$ with $G_k$ and $\Lambda_k$ the dimensionful Newton and cosmological constant, respectively. We analyze up to $N=6$
in $d=4$ and separate the analysis for $\beta = 0$ and $\beta=-\infty$ but always choose the Landau gauge $\alpha=0$. As is well known, this choice corresponds to a sharp imposition of the gauge condition and also corresponds to fixed point of the renormalization group flow~\cite{Ellwanger:1995qf}. In the next subsections we collect our results for the different values of $\beta$.

The right hand side of the flow equation \eqref{fe4} is
\begin{equation}
\partial_t \Gamma_k [0;\bg] = \partial_t \int \dd^dx~\sqrt{\bg}\,f_k (\br).
\label{fp2}
\end{equation}
The couplings $g_i (k)$ are dimensionful in general and can be expressed in terms of dimensionless ones $\tilde{g}_{i}(k)$ as
\begin{equation}
g_i (k) = k^{d-2i}\tilde{g}_i (k)\qquad\Rightarrow\qquad\partial_t g_i (k) = k^{d-2i}(d-2i+\partial_t\tilde{g}_i (k)).
\label{fp3}
\end{equation}
Using the ansatz \eqref{fp1}, Eq.~\eqref{fp2} is expressed as
\begin{equation}
\partial_t \Gamma_k [0;\bg] = \sum^{N}_{i=0}k^{d-2i}(d-2i+\partial_t\tilde{g}_i (k))\int \dd^dx\,\sqrt{\bg}\,\br^i \equiv \sum^{N}_{i=0}k^{d-2i}(d-2i+\tilde{\beta}_i)\int \dd^dx\,\sqrt{\bg}\,\br^i,
\label{fp4}
\end{equation}
where $\tilde{\beta}_i$ stands for the beta function of the coupling $\tilde{g}_i$. The flow equation \eqref{fe4} enables the extraction of beta functions by the computation of the trace on its right hand side and a suitable projection rule.

\subsection{\texorpdfstring{$\beta=0$}{TEXT}} \label{beta0sub}
This particular choice of $\beta$ is motivated by previous works on the $f(R)$ truncation. In \cite{Codello:2008vh,Machado:2007ea,Falls:2014tra}, this truncation was analyzed for polynomials of $\br$ using this choice (on top of the Landau gauge condition). These works have employed the linear split of the metric, which in our notation corresponds to $(\omega,m)=(0,0)$. In the following, we report our results for fixed points for different choices of $(\omega,m)$ in this gauge and make some comments about stability of the results.

Before that, however, let us mention a subtlety in this gauge. From the full expressions for the Hessians, see Appendix~\ref{hessians&Cutoffs}, one easily notices that the parameter $\beta$ enters the mixed contributions $(\sigma,h)$, the pure trace part $(h,h)$ and the longitudinal ghost contributions. When one sets $\beta=0$, the only gauge-parameter dependent contribution in the scalar sector comes from the $(\sigma,\sigma)$ sector with $1/\alpha$ dependence. By taking into account that in the Landau gauge limit terms with $1/\alpha$ dominate, we have a disentanglement of the (gravitational) scalar sector (see Eq.~\eqref{fe6}):
\begin{equation}
\lim_{\alpha \to 0}\mathrm{Tr}^{\prime\prime}_{(0)}\left[W^{\sigma h}_k(-\bnabla^2)\right]\Big|_{\beta=0} = \mathrm{Tr}^{\prime\prime}_{(0)}\left(\frac{\partial_t\mathbf{R}^{\sigma\sigma\prime}_k}{\Gamma^{\sigma\sigma\prime}_k}\right) + \mathrm{Tr}^{\prime\prime}_{(0)}\left(\frac{\partial_t\mathbf{R}^{hh}_k}{\Gamma^{hh}_k}\right),
\label{betazero1}
\end{equation} 
where $\Gamma^{\sigma\sigma\prime}_k$ is the last term in (\ref{hess3}) and $\mathbf{R}^{\sigma\sigma\prime}_k$ is the regulator obtained from it by (\ref{fe2}). Also, for the $(\xi,\xi)$ sector, the Landau gauge limit makes the contribution\footnote{For $\omega=1/2$ or $m=-1/d$ the $(\xi,\xi)$ contribution to the pure gravitational sector vanishes.} coming from the pure gravitational action suppressed with respect to the gauge-fixing contribution. Effectively, the $(\xi,\xi)$ term leads to
\begin{equation}
\lim_{\alpha\to 0}\, \frac{1}{2}\,\Tr^\prime_{(1)} \Bigg[ \frac{\partial_t \textbf{R}_{k}^{\xi\xi} }{ \Gamma^{\xi\xi}_k + \textbf{R}_{k}^{\xi\xi} } \Bigg] = \Tr^\prime_{(1)}\Bigg[\frac{\partial_t P_k}{P_k+\frac{\br}{d}}\Bigg],
\label{betazero2}
\end{equation}
see, for instance, \cite{Machado:2007ea}. On the other hand, the contribution to the flow equation from the transverse ghost sector is expressed as
\begin{equation}
-\lim_{\alpha\to 0}\,\Tr_{(1)} \Bigg[ \frac{ \partial_t \textbf{R}_k^{C^\mathrm{T}\!C^\mathrm{T}} }{ \Gamma^{C^\mathrm{T}\!C^\mathrm{T}}_k + \textbf{R}_k^{C^\mathrm{T} \!C^\mathrm{T}} }  \Bigg] = - \Tr_{(1)} \Bigg[\frac{\partial_t P_k}{P_k + \frac{\br}{d}}\Bigg].
\label{betazero3}
\end{equation}
One sees that the contributions \eqref{betazero2} and \eqref{betazero3} nearly cancel each other apart from the fact that the former trace is primed and the latter is not. A common strategy is to ``prime" the trace over the transverse ghost sector to have an exact cancellation of these sectors. The contribution from the spin-1 sector in the gauge-fixing is  exactly canceled by that from the transverse ghost. This was implemented in \cite{Codello:2008vh,Machado:2007ea,Codello:2007bd}. We present results without employing this procedure. However, as a consistency check, when this prescription is used and the linear parametrization is fixed, we reproduce the results obtained in e.g. \cite{Machado:2007ea}.

\subsubsection{\texorpdfstring{$\omega$}{TEXT}-dependence}

In the following, we exhibit our results. First, for the truncation \eqref{fp1} with a finite $N$, we plot the values of the fixed points against the parameter $\omega$. For simplicity, we take $m=0$. 

\begin{figure}[b!]
  \centering
  \begin{subfigure}[b]{0.49\linewidth}
    \includegraphics[width=\linewidth]{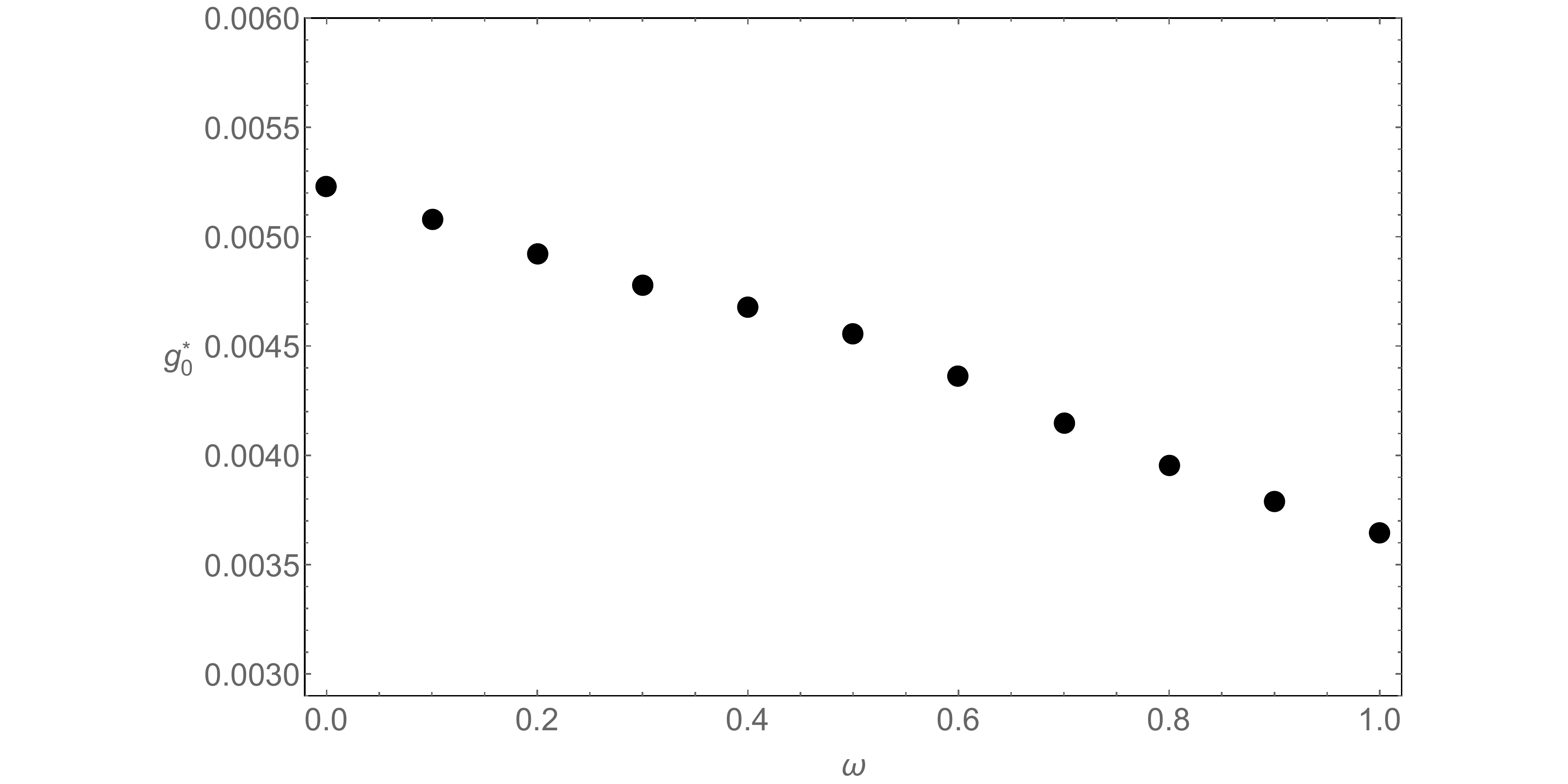}
  \end{subfigure}
  \begin{subfigure}[b]{0.49\linewidth}
    \includegraphics[width=\linewidth]{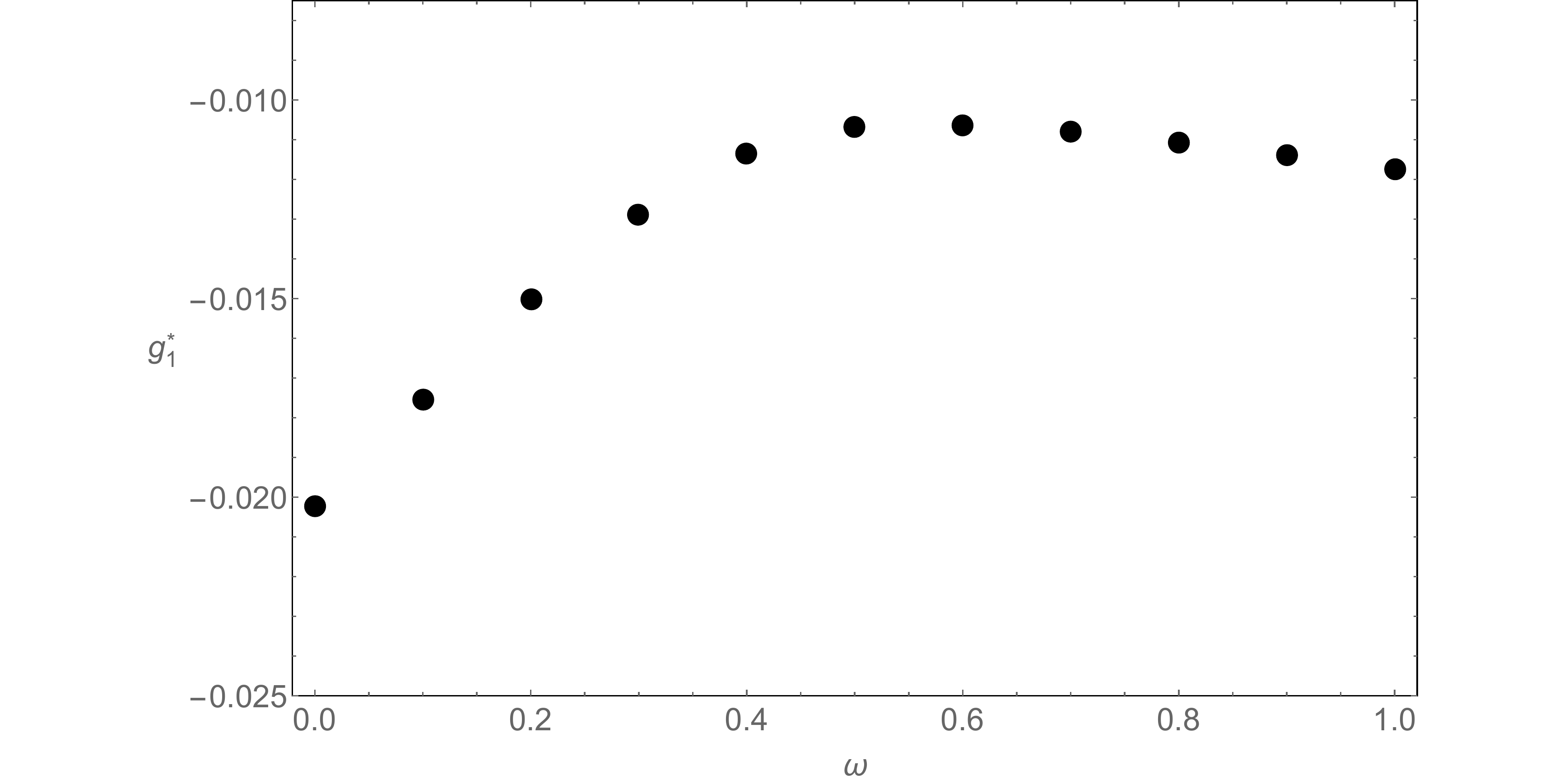}
  \end{subfigure}
  \caption{Fixed point values for the couplings $g_0$ and $g_1$ in the Einstein-Hilbert truncation.}
  \label{EHbeta0}
\end{figure}

\begin{figure}[t!]
  \centering
  \begin{subfigure}[b]{0.49\linewidth}
    \includegraphics[width=\linewidth]{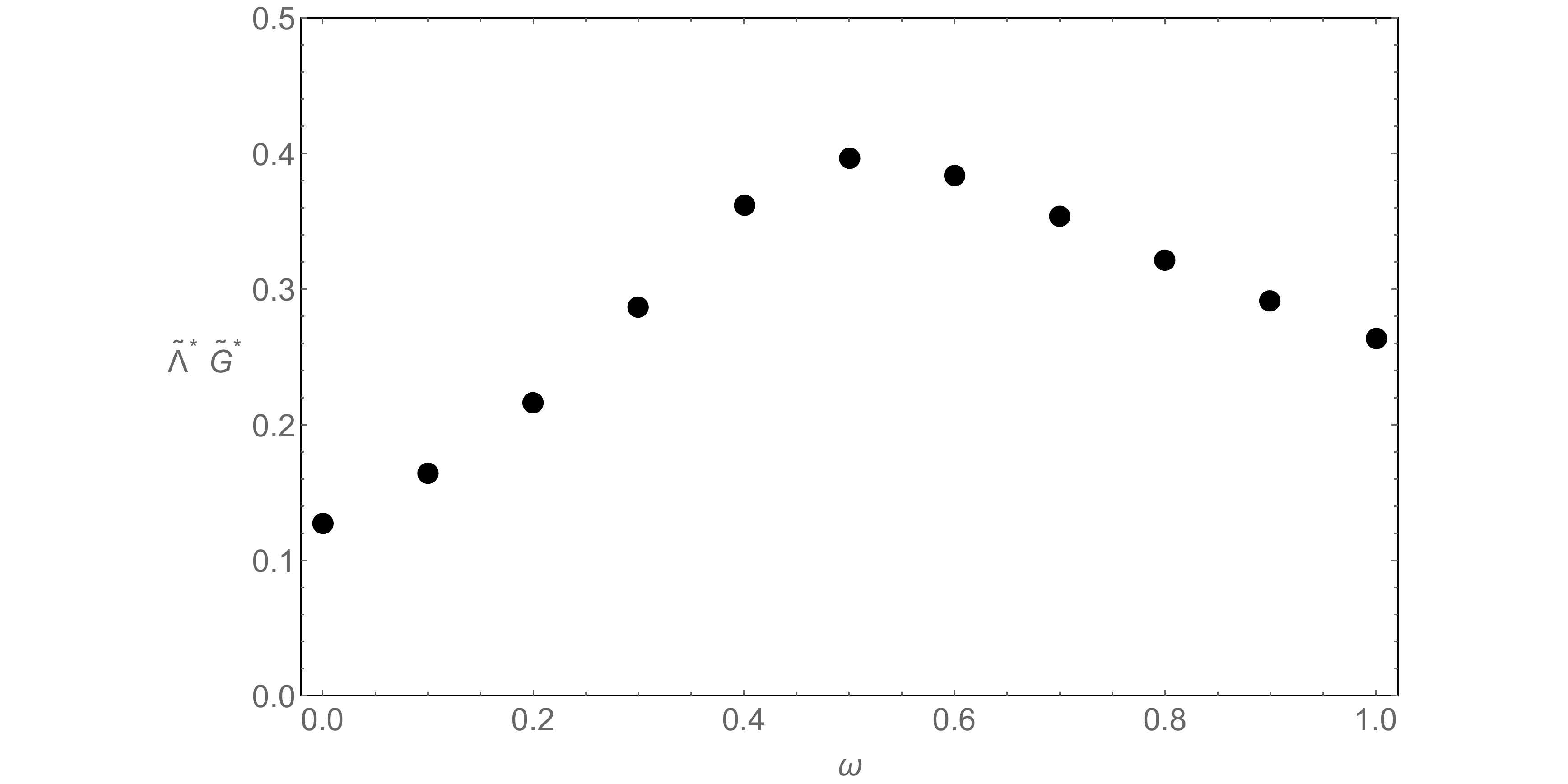}
    \caption{Product $\tilde{\Lambda}^\ast \tilde{G}^\ast$.}
  \end{subfigure}
  \begin{subfigure}[b]{0.49\linewidth}
    \includegraphics[width=\linewidth]{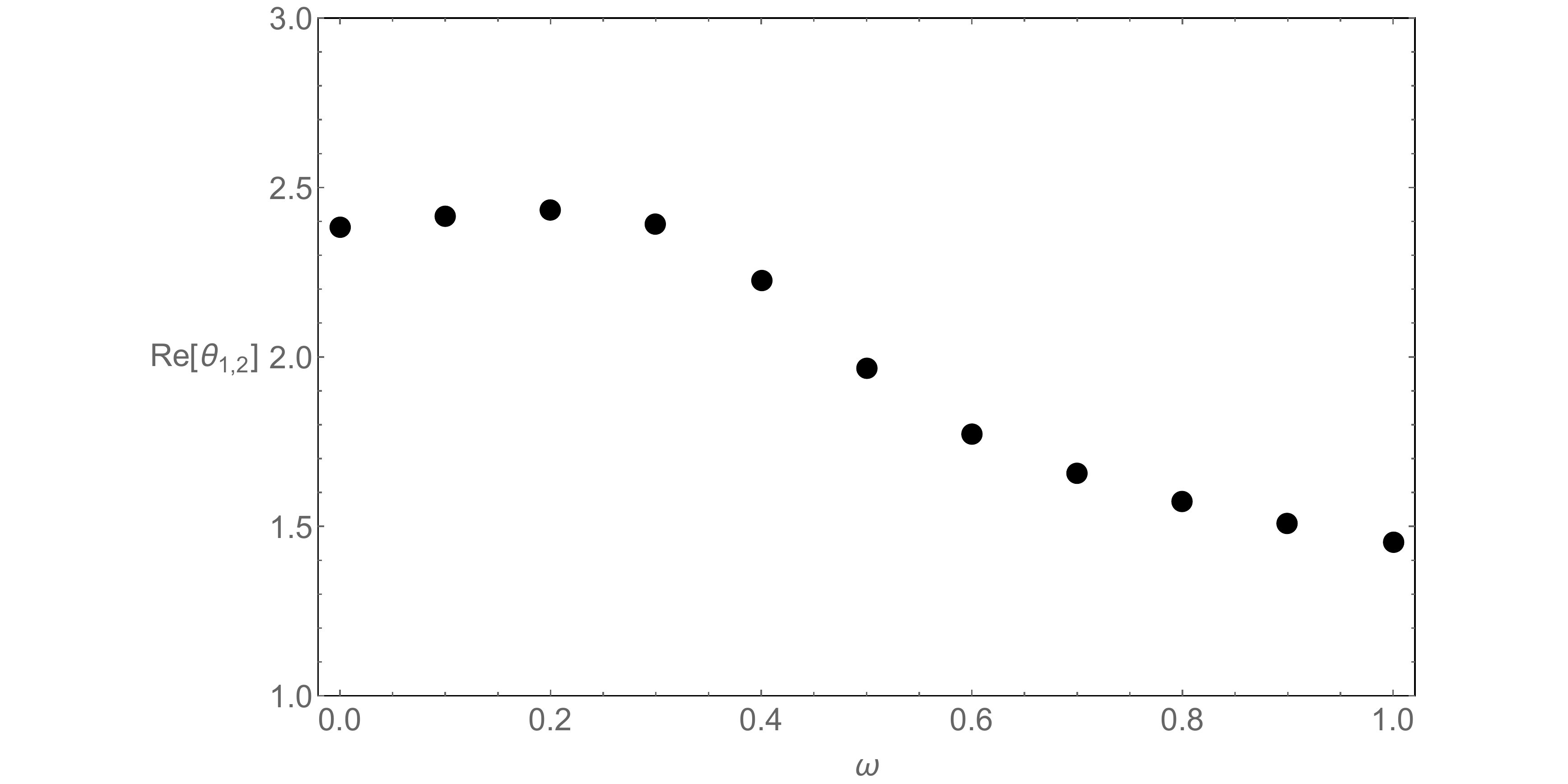}
    \caption{Real part of the critical exponents.}
  \end{subfigure}
  \caption{Einstein-Hilbert truncation in the $\beta=0$ gauge.}
  \label{lambdagEHbeta0}
\end{figure}

In the Einstein-Hilbert truncation i.e. $N=1$, we show our results in Fig~\ref{EHbeta0}.
One can see that the numerical value of the fixed points $(g^\ast_0,g^\ast_1)$ changes for different choices of $\omega$. However, the numerical variation is relatively small and the qualitative behavior is the same: For all choices of $\omega$ in the interval $[0,1]$, the fixed point value for $g^\ast_0$ is positive, while $g^\ast_1$ is negative. Moreover, the values of $g^\ast_1$ are very stable for $1/2\le \omega \le 1$ and we find a local maximum around $\omega=1/2$. We also plot the product $\tilde{\Lambda}^\ast \tilde{G}^\ast$ against the parameter $\omega$ in Fig.~\ref{lambdagEHbeta0}. Again, one sees that the product reaches a local maximum around the exponential parametrization $\omega=1/2$. One can conclude that within this truncation, the exponential parametrization gives the most stable results in the sense that small perturbations of this choice give very similar results. For all values of $\omega\in [0,1]$, one obtains two relevant directions for the UV fixed point in this truncation as shown in Fig.\ref{lambdagEHbeta0}.

\begin{figure}[h!]
  \centering
  \begin{subfigure}[b]{0.49\linewidth}
    \includegraphics[width=\linewidth]{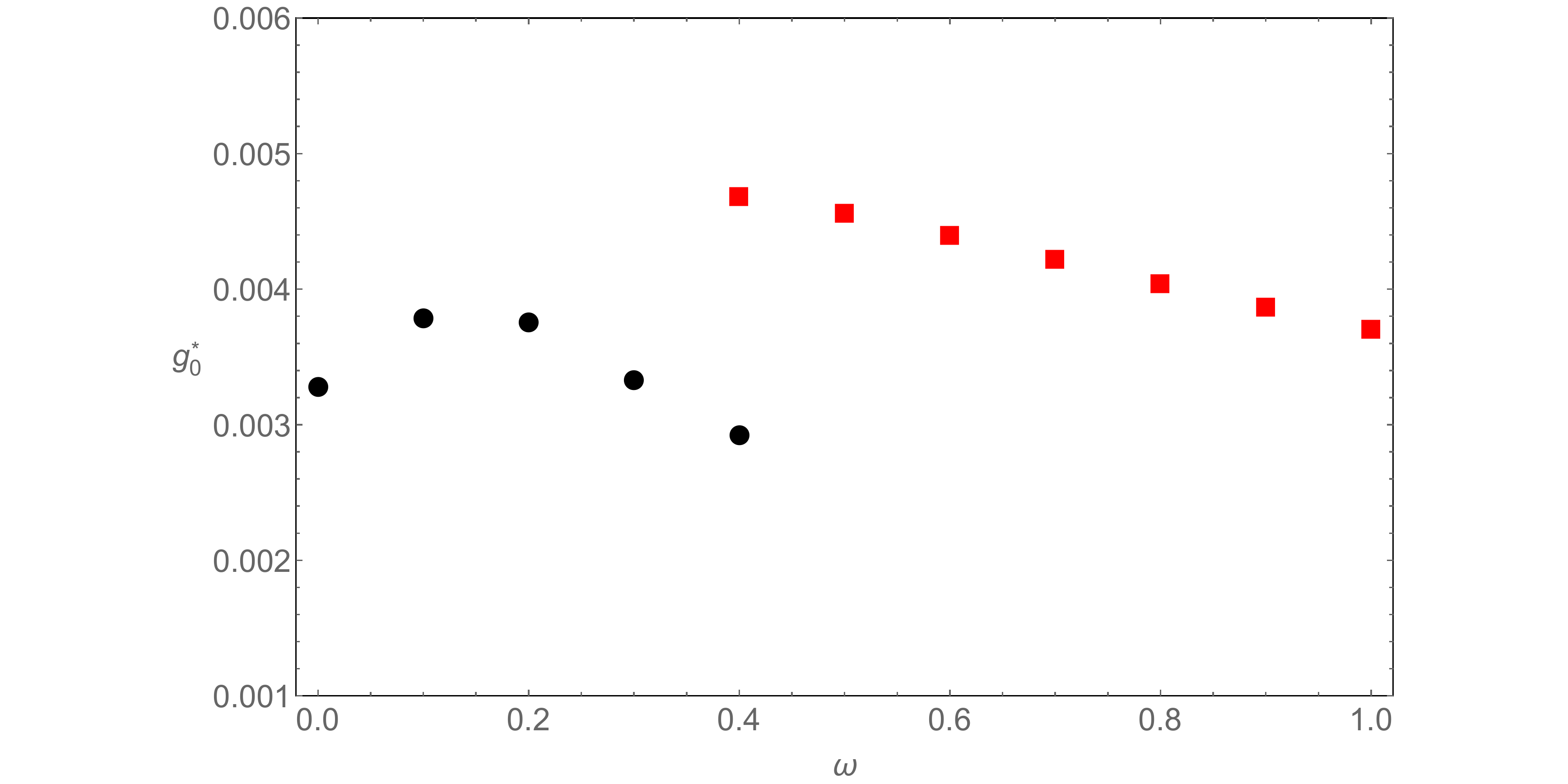}
  \end{subfigure}
  \begin{subfigure}[b]{0.49\linewidth}
    \includegraphics[width=\linewidth]{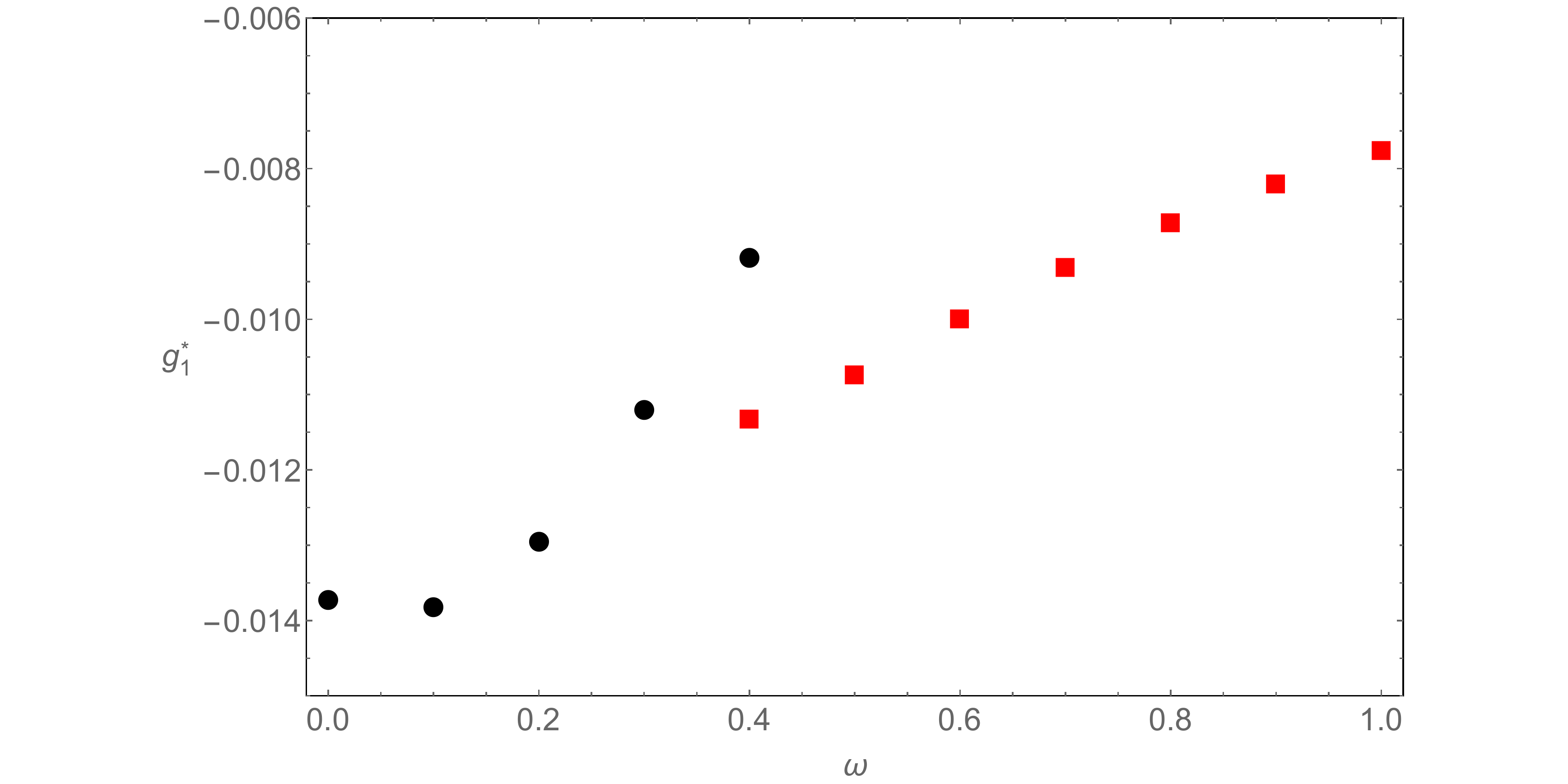}
  \end{subfigure}
    \begin{subfigure}[b]{0.49\linewidth}
    \includegraphics[width=\linewidth]{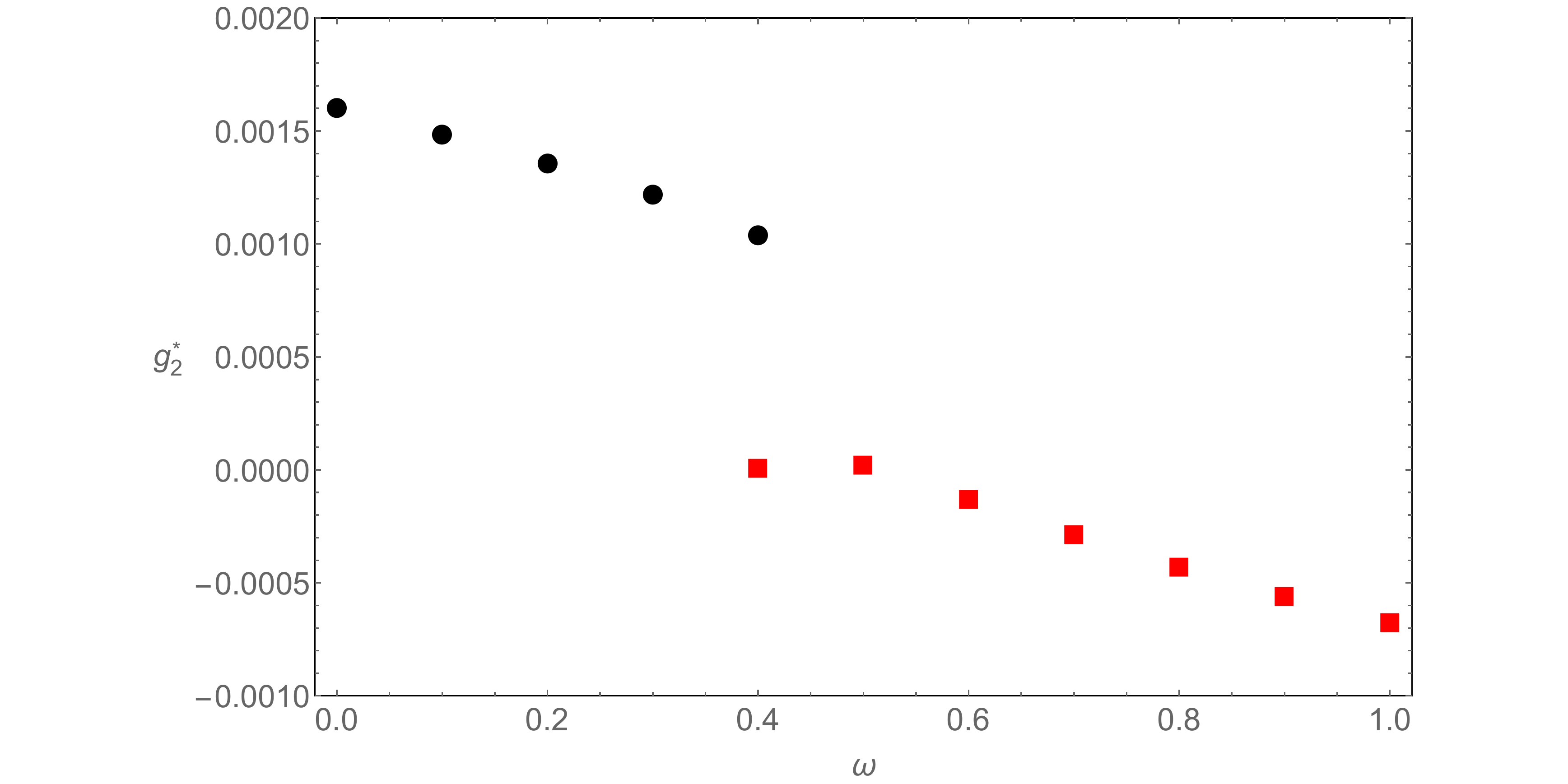}
  \end{subfigure}
  \caption{Fixed point values for the couplings $g_0$, $g_1$ and $g_2$ in the $R^2$ truncation in the $\beta=0$ gauge.}
  \label{R2beta0}
\end{figure}

\begin{figure}[t!]
  \centering
  \begin{subfigure}[b]{0.49\linewidth}
    \includegraphics[width=\linewidth]{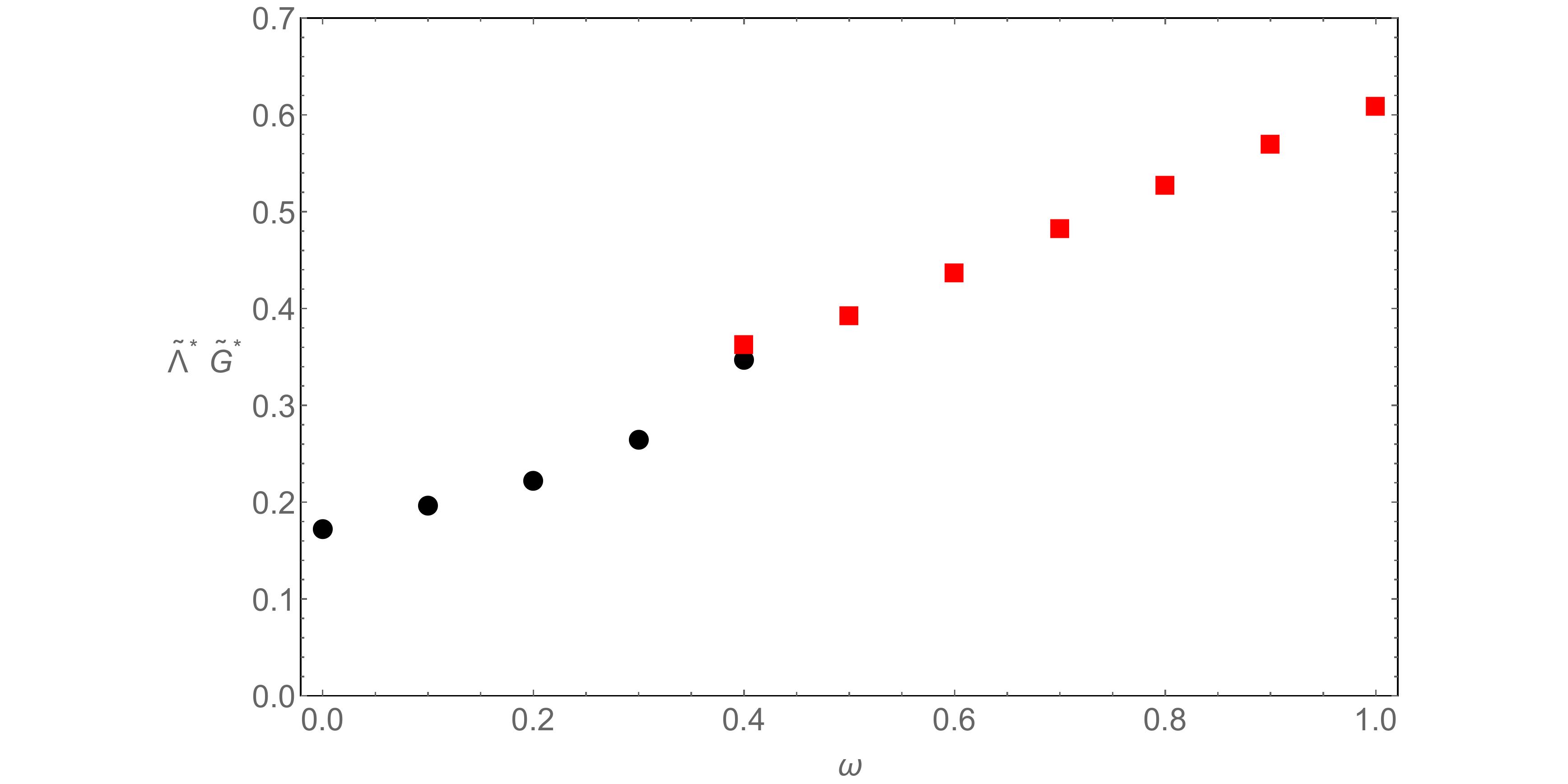}
    \caption{Product $\tilde{\Lambda}^\ast \tilde{G}^\ast$.}
  \end{subfigure}
  \begin{subfigure}[b]{0.49\linewidth}
    \includegraphics[width=\linewidth]{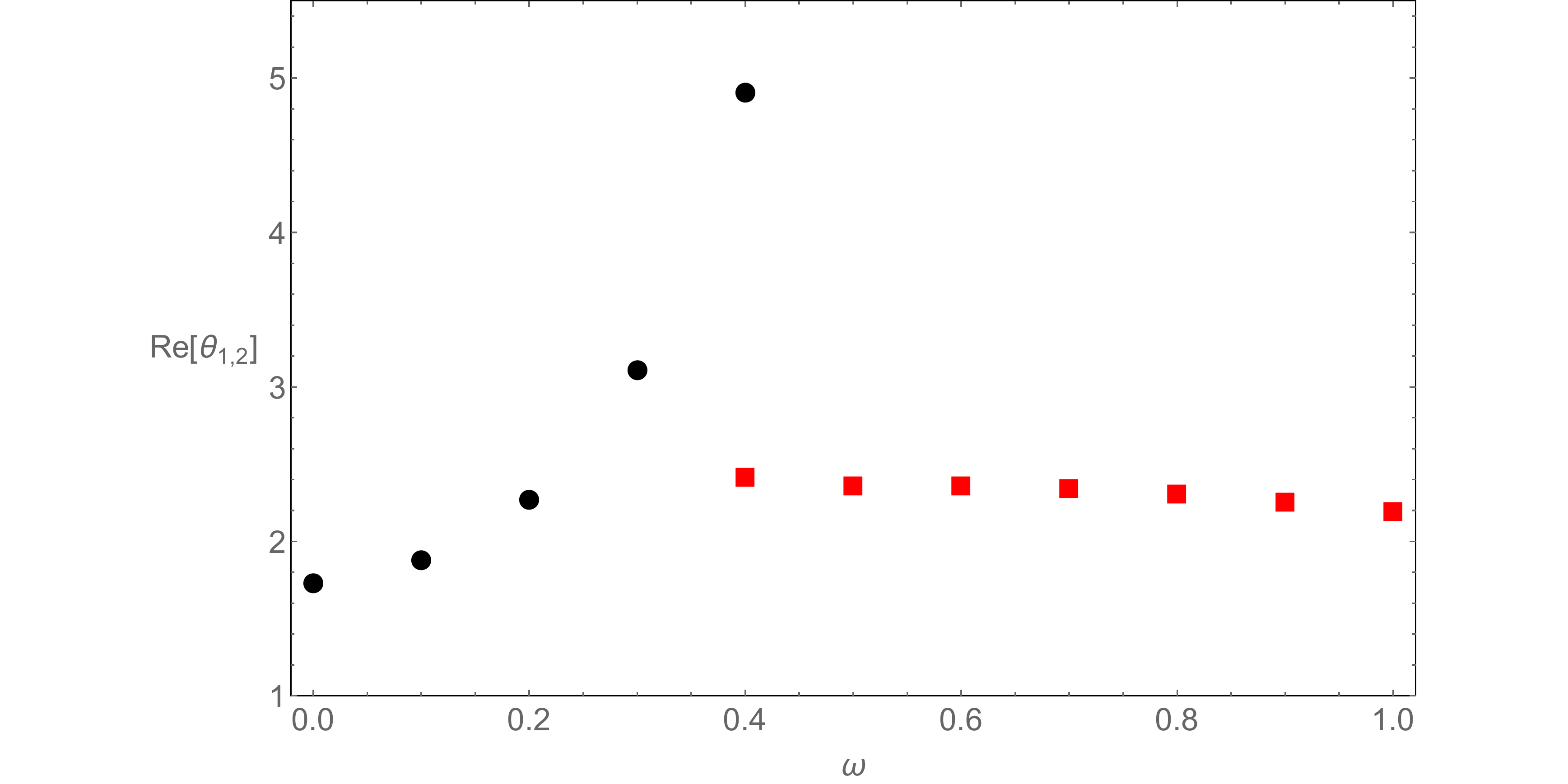}
    \caption{Real part of the critical exponents $\theta_{1,2}$.}
  \end{subfigure}
\begin{subfigure}[b]{0.49\linewidth}
    \includegraphics[width=\linewidth]{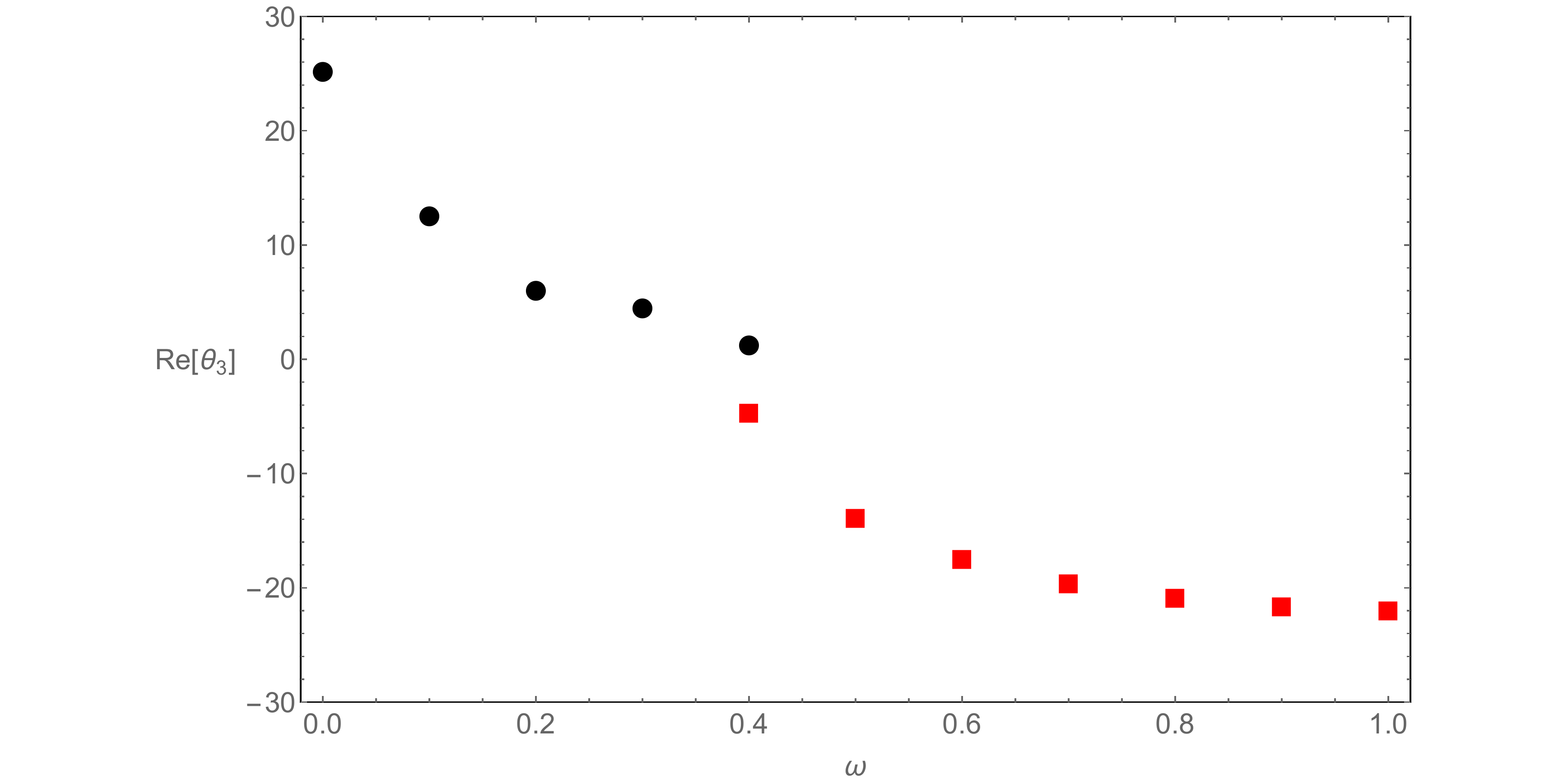}
    \caption{Real part of the critical exponent $\theta_{3}$.}
  \end{subfigure}  
  \caption{$R^2$ truncation in the $\beta=0$ gauge.}
  \label{lambdagR2beta0}
\end{figure}

For $N=2$ truncation, one can see how the values for the fixed point $(g^\ast_0,g^\ast_1,g^\ast_2)$ change with respect to the parameter $\omega$ in Fig.~\ref{R2beta0}. For these plots we use two types of plot markers, namely, black dots and red squares. The reason is the following: For $0\le\omega < 0.4$ the UV fixed point has three relevant directions. At $\omega = 0.4$ a second fixed point with two relevant directions shows up and for $0.4 < \omega \le 1$ the only viable fixed point we obtain has two relevant directions. The fixed points with three relevant directions are identified by black dots while those with two relevant directions, by red squares. We see thus that it is not possible to continuously deform a fixed point for different values of $\omega$ at $N=2$ order. As particular examples, the linear $(\omega = 0)$ and the exponential $(\omega = 1/2)$ parametrizations have fixed points with three and two relevant directions, respectively. Such a difference was already detected in previous works~\cite{Ohta:2015fcu}. For each class of fixed points (black dots and red squares) the numerical values for the fixed points are relatively stable under changes on $\omega$. In Fig.~\ref{lambdagR2beta0} we show the product $\tilde{\Lambda}^\ast \tilde{G}^\ast$ in $N=2$ truncation. We emphasize that at $\omega=0.4$, the black dot and the red square almost coincide. 
We also show how the critical exponents change under variations of $\omega$ in Fig.~\ref{lambdagR2beta0}.
As pointed out before, for each class of fixed points, the results do not show strong $\omega$-dependence.

\begin{figure}[t!]
  \centering
  \begin{subfigure}[b]{0.49\linewidth}
    \includegraphics[width=\linewidth]{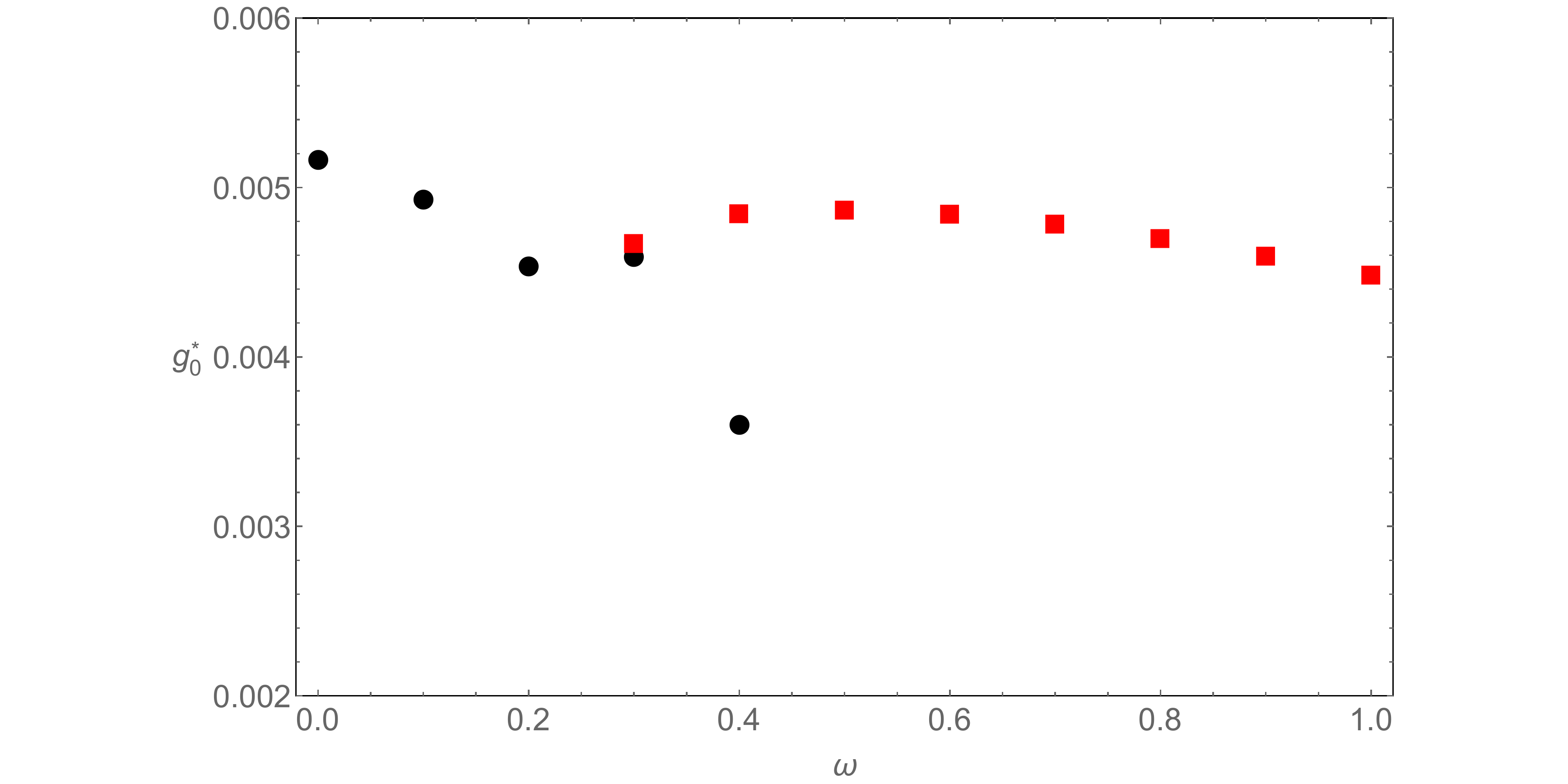}
  \end{subfigure}
  \begin{subfigure}[b]{0.49\linewidth}
    \includegraphics[width=\linewidth]{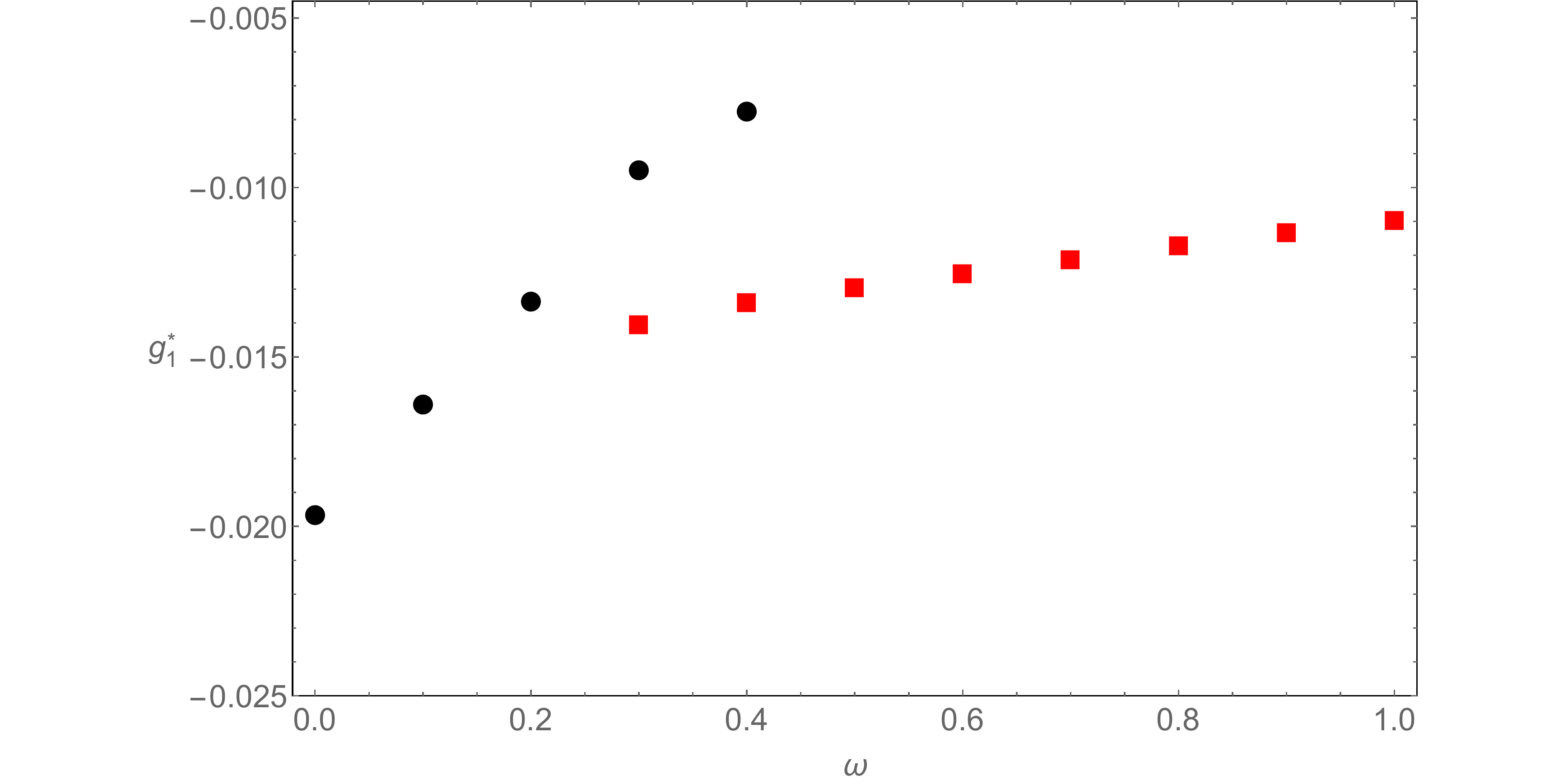}
  \end{subfigure}
    \begin{subfigure}[b]{0.49\linewidth}
    \includegraphics[width=\linewidth]{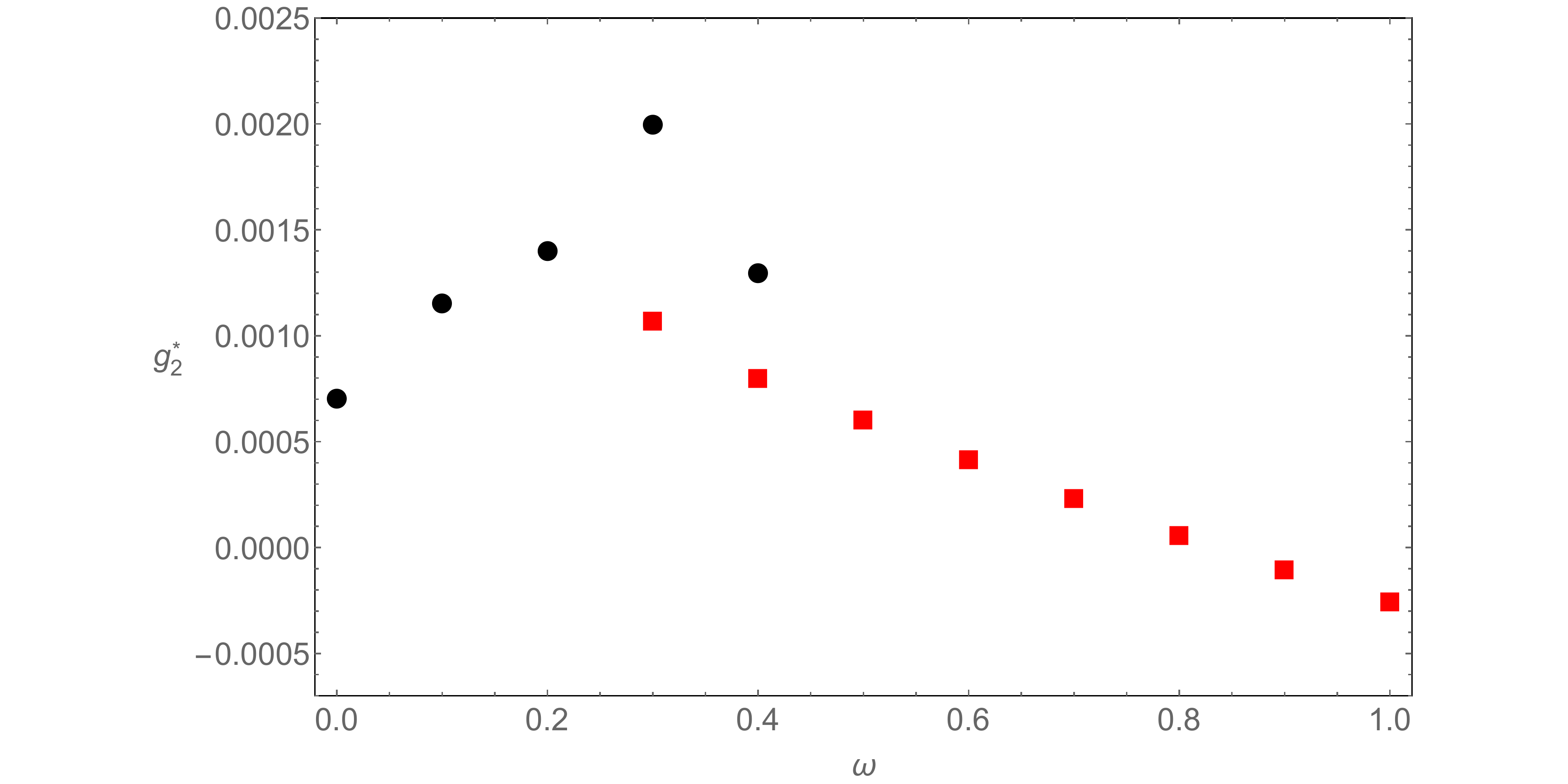}
  \end{subfigure}
  \begin{subfigure}[b]{0.49\linewidth}
    \includegraphics[width=\linewidth]{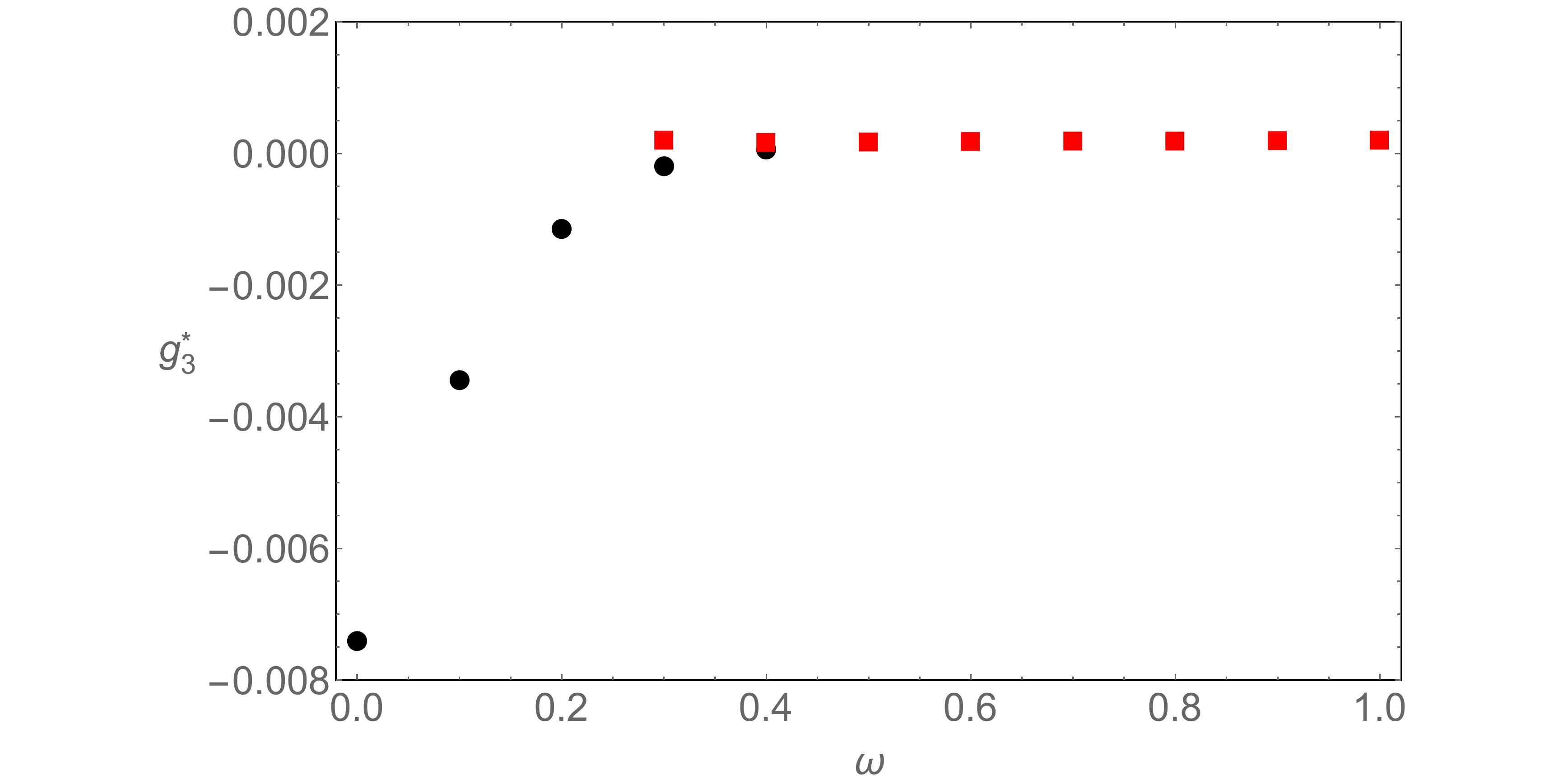}
  \end{subfigure}
  \caption{Fixed point values for the couplings $g_0$, $g_1$, $g_2$ and $g_3$ in the $R^3$ truncation in the $\beta=0$ gauge.}
  \label{R3beta0}
\end{figure}

\begin{figure}[t!]
  \centering
  \begin{subfigure}[b]{0.49\linewidth}
    \includegraphics[width=\linewidth]{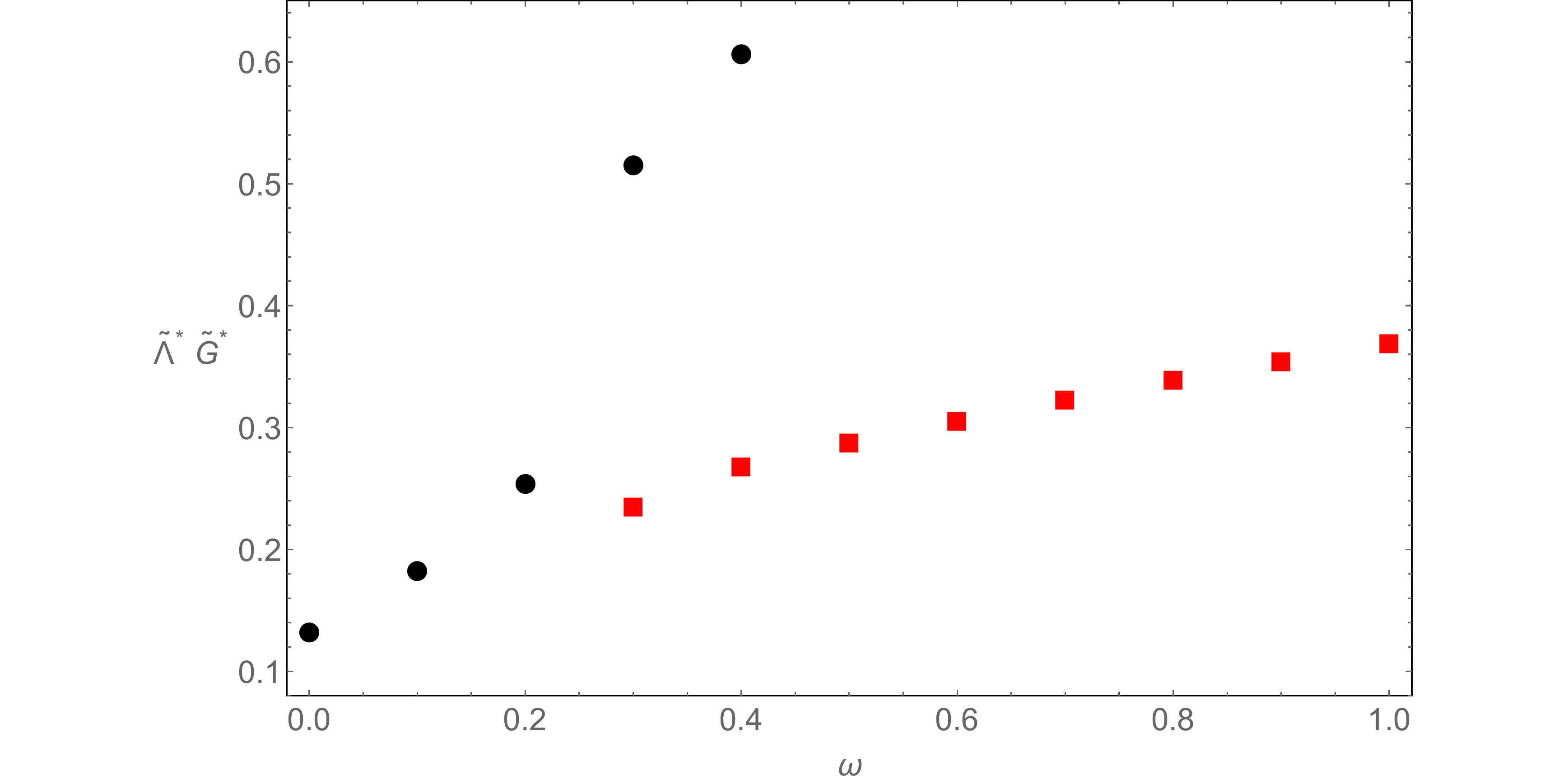}
    \caption{Product $\tilde{\Lambda}^\ast \tilde{G}^\ast$.}
  \end{subfigure}
  \begin{subfigure}[b]{0.49\linewidth}
    \includegraphics[width=\linewidth]{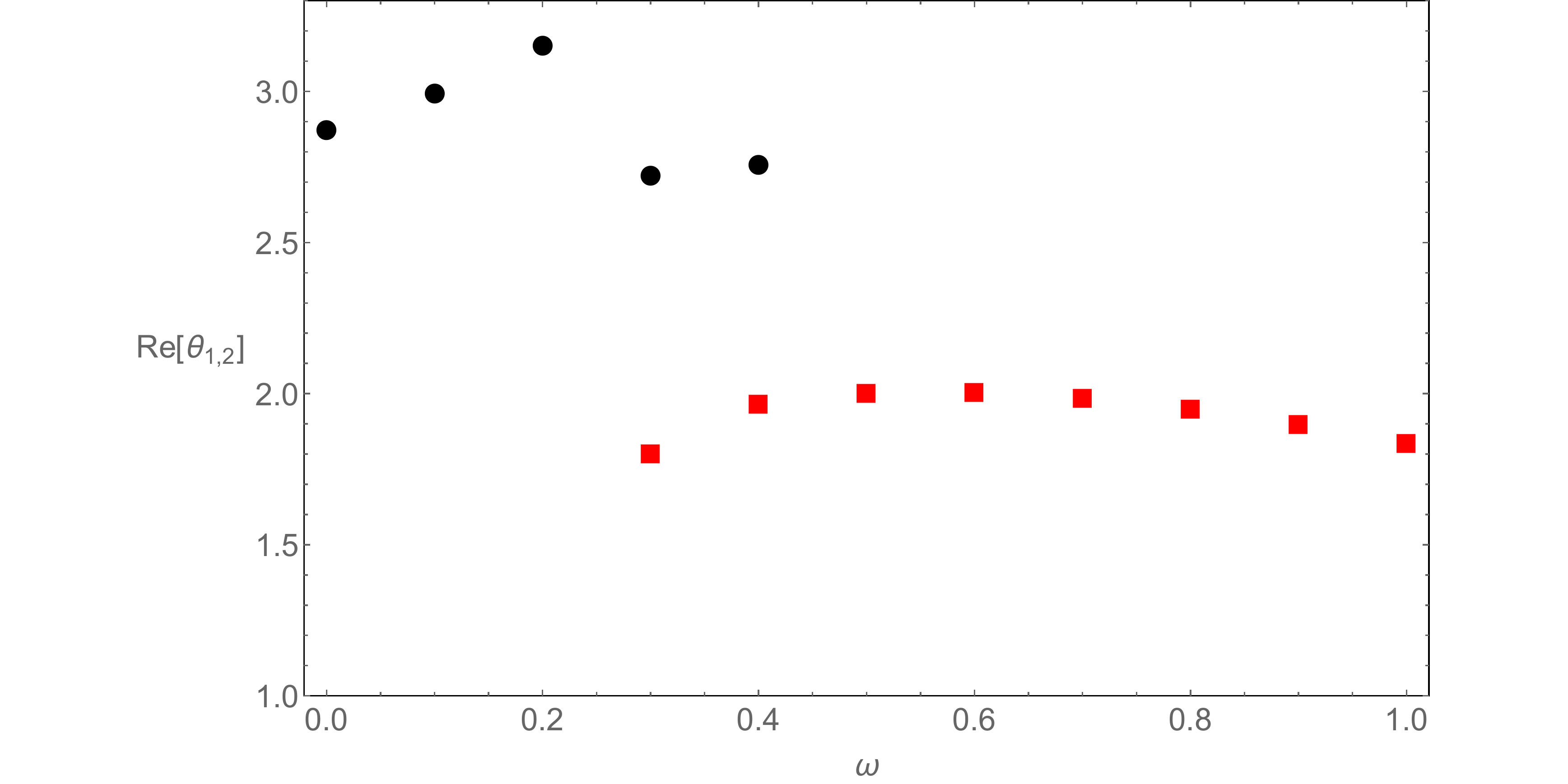}
    \caption{Real part of the critical exponent $\theta_{1,2}$.}
  \end{subfigure}
    \begin{subfigure}[b]{0.49\linewidth}
    \includegraphics[width=\linewidth]{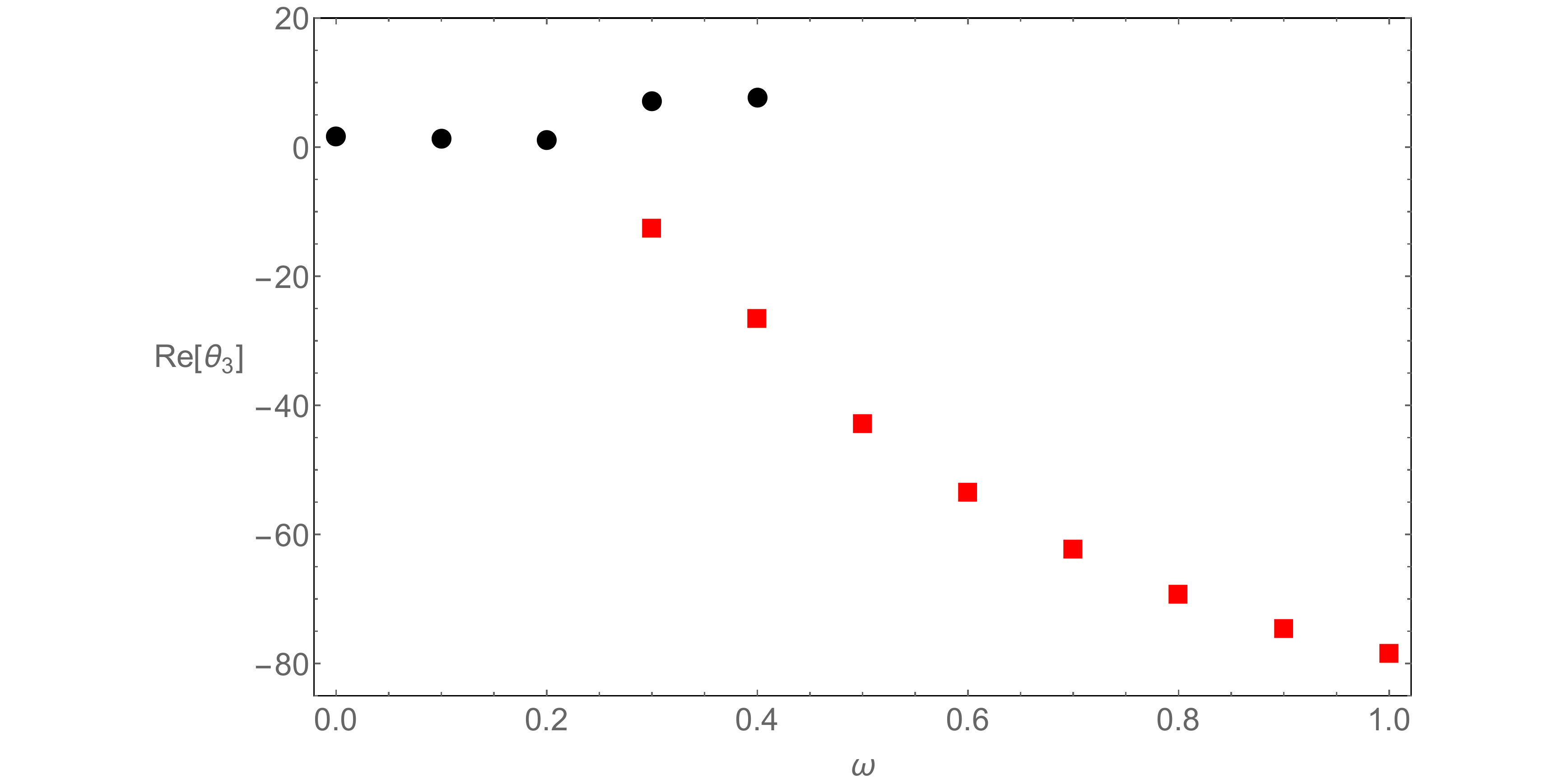}
    \caption{Real part of the critical exponent $\theta_{3}$.}
  \end{subfigure}
      \begin{subfigure}[b]{0.49\linewidth}
    \includegraphics[width=\linewidth]{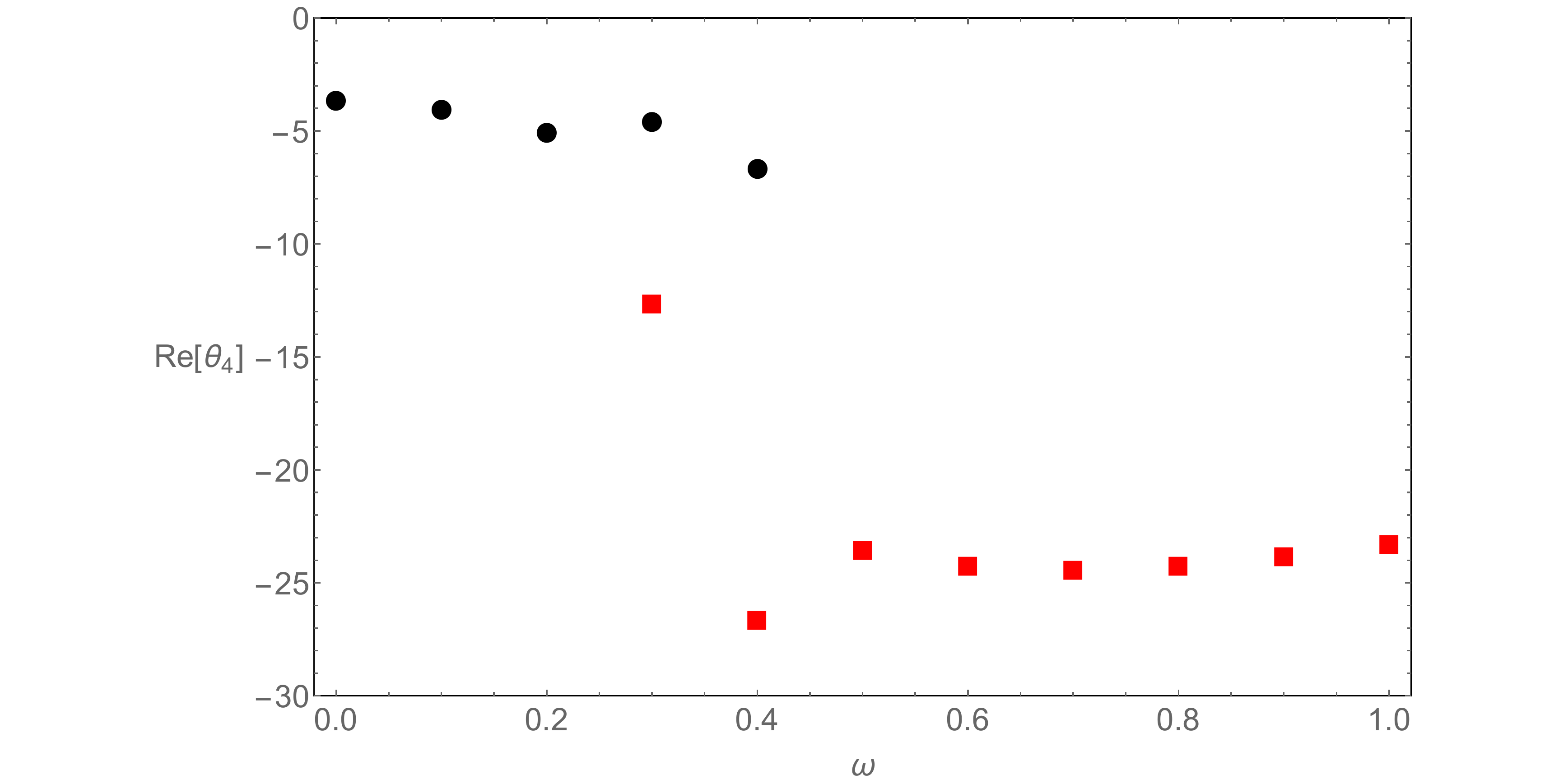}
    \caption{Real part of the critical exponent $\theta_{4}$.}
  \end{subfigure}
  \caption{$R^3$ truncation in the $\beta=0$ gauge.}
  \label{lambdagR3beta0}
\end{figure}

The discrepancy in the number of relevant directions for different choices of parametrization persists at $N=3$ order. In Fig.~\ref{R3beta0} we collect the values of the fixed points for each coupling for different choices of $\omega$. As  in the $N=2$ truncation, we find two types of fixed points: For $0\le\omega < 0.3$ we obtain fixed points with three relevant directions; at $\omega = 0.3$ and $\omega = 0.4$ two fixed points are viable, one with three and other with two relevant directions; for $0.4 < \omega \le 1$ the fixed point obtained has two relevant directions. This is the same qualitative behavior observed in the $N=2$ truncation. 
The $\omega$ dependence of the product $\tilde{\Lambda}^\ast \tilde{G}^\ast$ as well as of the critical exponents calculated at $N=3$ are collected in Fig.~\ref{lambdagR3beta0}. 

\begin{figure}[t!]
  \centering
  \begin{subfigure}[b]{0.49\linewidth}
    \includegraphics[width=\linewidth]{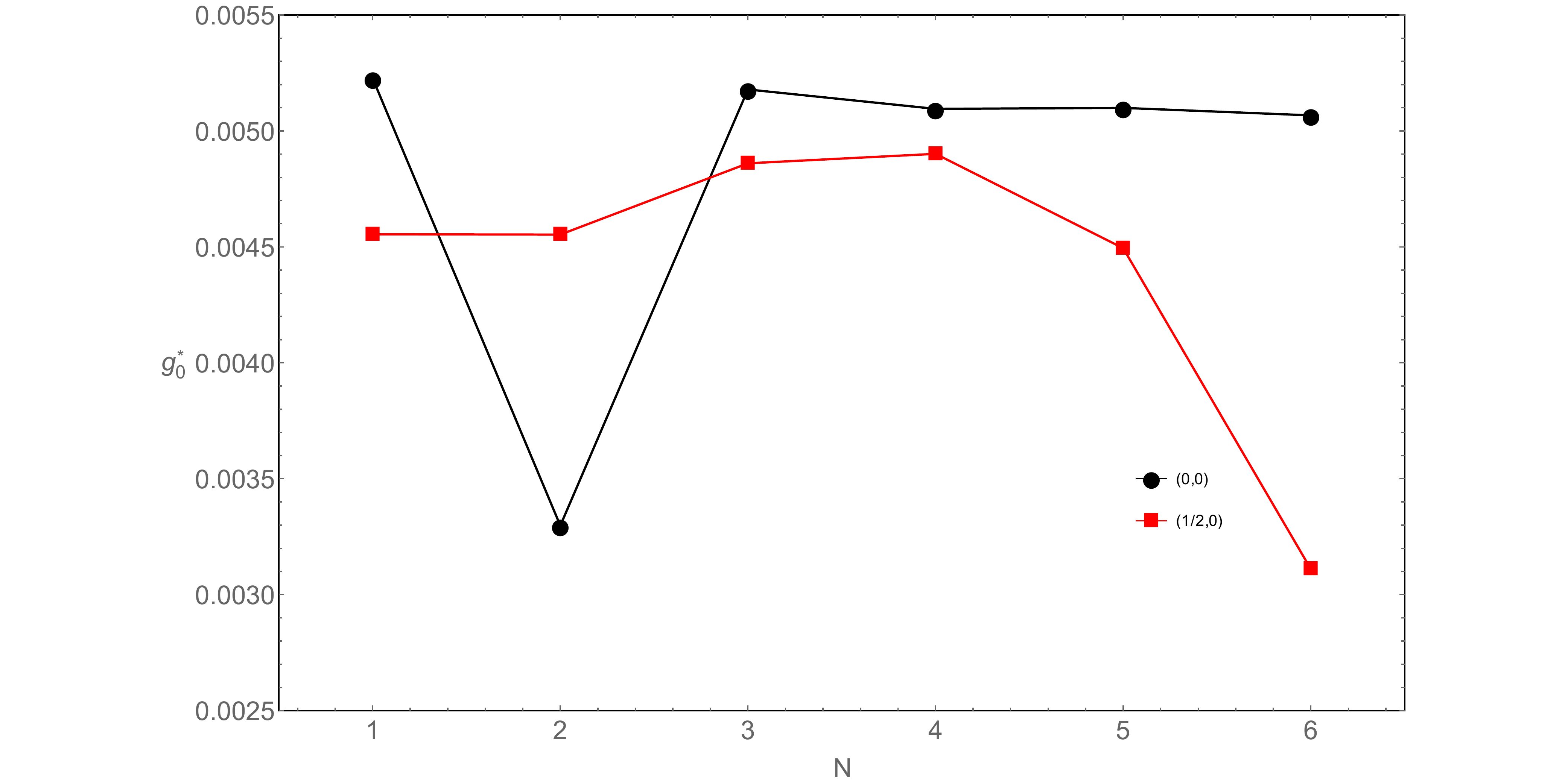}
  \end{subfigure}
  \begin{subfigure}[b]{0.49\linewidth}
    \includegraphics[width=\linewidth]{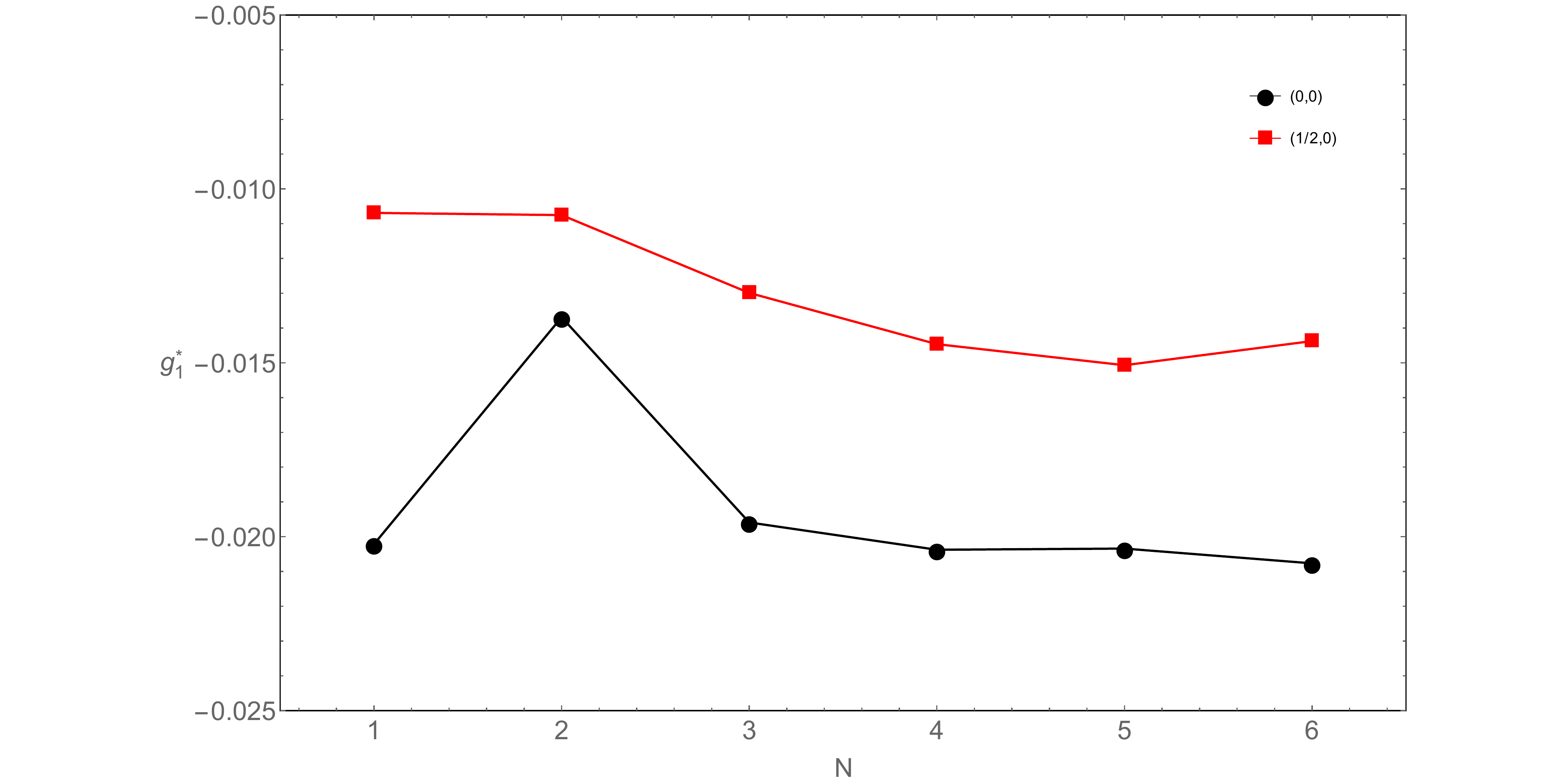}
  \end{subfigure}
    \begin{subfigure}[b]{0.49\linewidth}
    \includegraphics[width=\linewidth]{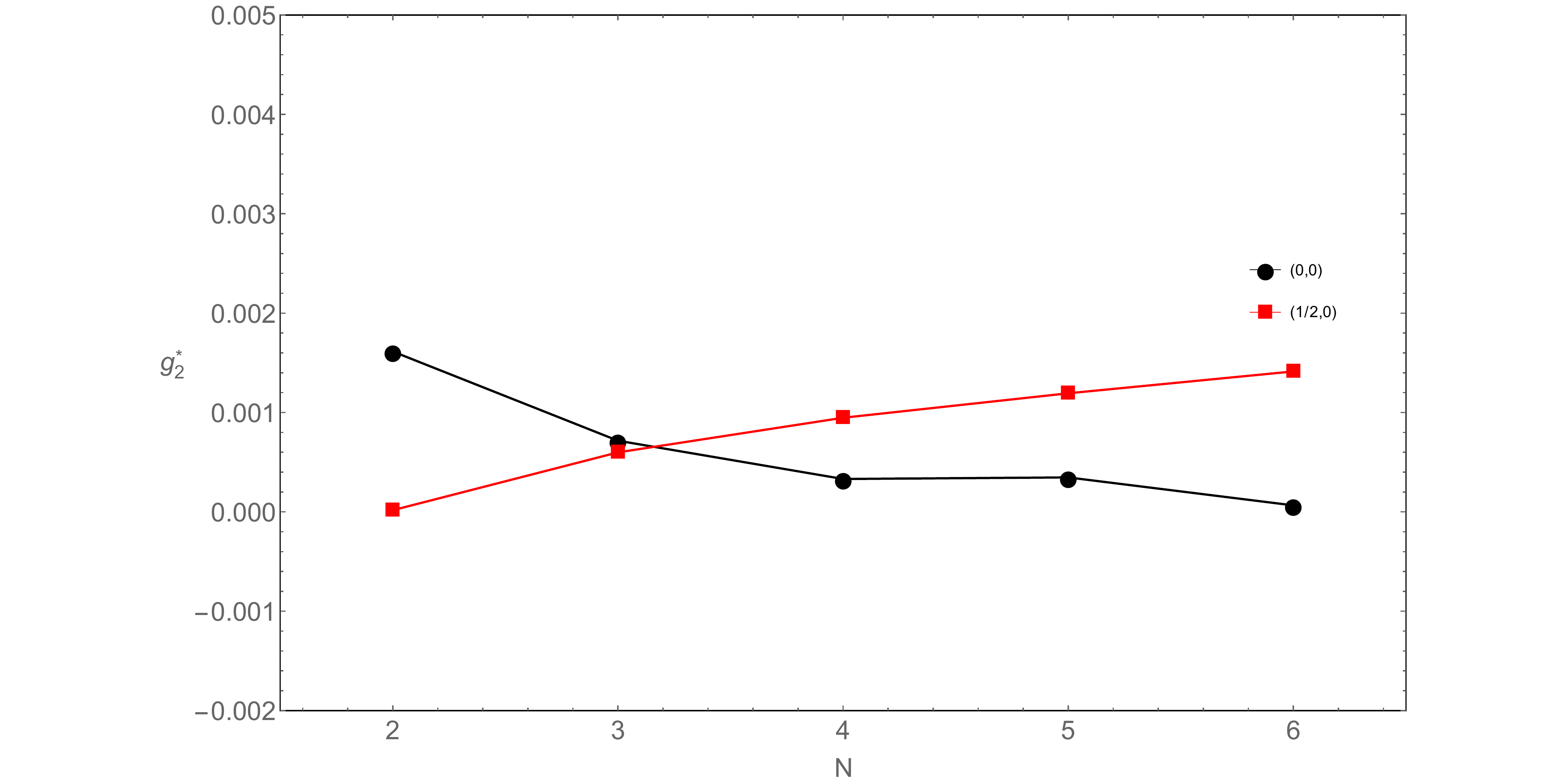}
  \end{subfigure}
      \begin{subfigure}[b]{0.49\linewidth}
    \includegraphics[width=\linewidth]{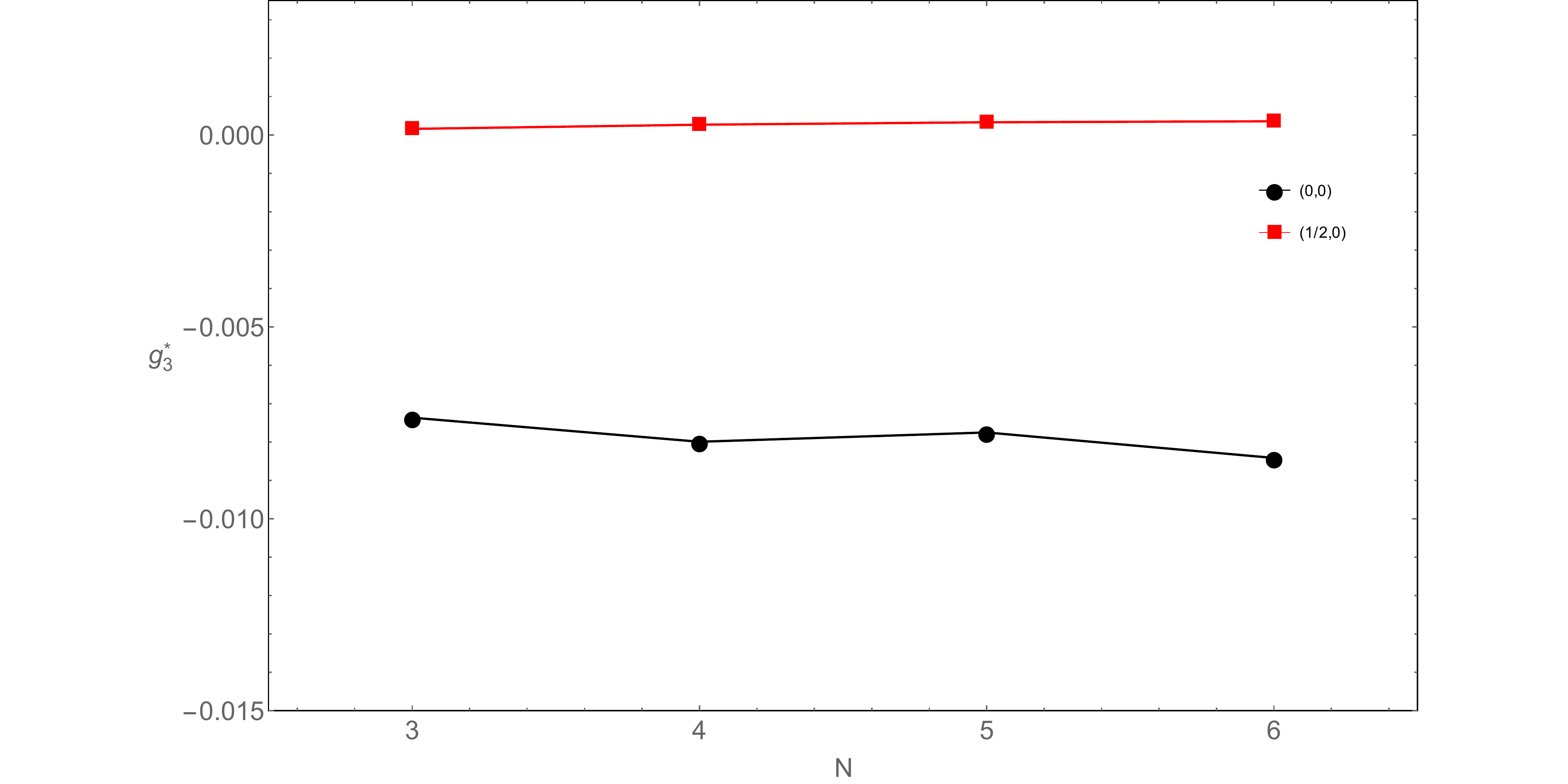}
  \end{subfigure}
  \caption{Values of the fixed point couplings $g^\ast_0$, $g^\ast_1$, $g^\ast_2$ and $g^\ast_3$ for different truncations in the $\beta=0$ gauge for $(\omega,m)=(0,0)$ and $(1/2,0)$.}
  \label{truncdepbeta0}
\end{figure}

In order to understand how, in a given parametrization, the fixed point values change upon enlargement of the truncation, we plot the values of the fixed points against the curvature power $N$ in Fig.~\ref{truncdepbeta0}. We consider truncations up to $N=6$ and we plot the couplings $(g^\ast_0,g^\ast_1,g^\ast_2,g^\ast_3)$. We show the cases of linear and exponential parametrizations. It is known in the literature that for both parametrizations, it is possible to find very good convergence for the fixed point values upon truncation improvements, see \cite{Falls:2014tra,Ohta:2015fcu,Alkofer:2018fxj}. The qualitative picture we have discussed before, namely, that in the linear split one finds a fixed point with three relevant directions and in the exponential parametrization, a fixed point with two relevant directions, persists in larger truncations. However, in the exponential parametrization, the fixed point structure is not too much stable up to $N=6$. In particular, using the prescriptions described in the previous sections, we typically find more than one fixed point candidate for a given $N$ which fulfills the basic requirement of stability under truncation improvement.\footnote{For this choice of gauge parameter, we found two viable fixed points for $3\le N \le 6$. For $N=3$ both display two relevant directions. For $N=4,5,6$, there is an oscillation on the number of relevant directions. One of the candidate fixed point oscillates from two to three relevant directions at $N=4$ but for $N=5,6$ it returns to two relevant directions. The other one remains with two relevant directions at $N=4$ but for $N=5,6$ the number grows to four relevant directions. In order to completely distinguish which fixed point is stable (if any) and which is truncation artifact, one needs to enlarge the truncation for larger values of $N$.} Thus in the exponential parametrization, the fixed point structure is not so stable up to $N=6$. Ultimately one needs to go beyond $N=6$ in order to check if the convergence property improves or not in this case. For the construction of the plots, we have chosen a particular set of fixed points for illustration. To make this point clear and avoid too many plots, we collect these results in tables in Appendix~\ref{numres}.

\subsubsection{\texorpdfstring{$m$}{TEXT}-dependence}

Let us briefly discuss the $m$-dependence of our results on fixed point values and number of relevant directions.
For simplicity, we consider the well-studied cases of $\omega=0$ (the linear split of the metric) and $\omega=1/2$ (the exponential split). However, for $\omega=1/2$, the $m$-dependence just cancels out since $m$ always appears in the combination $(1-2\omega)(1+dm)$ or as an overall factor in the Hessians, see Appendix~\ref{hessians&Cutoffs}.
In this connection, we also note that the $\omega$-dependence also disappears for $m=-1/d$, but this case, corresponding to unimodular gravity, is singular~\cite{Ohta:2016npm}. Nevertheless we may consider that case by extrapolating our results~\cite{Ardon:2017atk}.

We find the following results for $\omega=0$:
In the Einstein-Hilbert truncation, the coupling $g_0$ corresponding to the cosmological constant starts from
the value $0.00523$ at $m=0$, decreases to 0.00211 at $m=-1/4$ and increases to 0.00365 at $m=-1/2$,
whereas $g_1$ for the Einstein-Hilbert term starts from $-0.0202$, increases to $-0.00875$ at $m=-1/4$
and decreases to $-0.0117$ at $m=-1/2$. Both operators are relevant for those choices of $m$.
In the $N=2$ truncation, 
$g_0=0.00330,\, g_1=-0.137,\, g_2=0.00161$ with three relevant operators at $m=0$;
$g_0=0.00211,\, g_1=-0.0102,\, g_2=-0.00283$ with three relevant operators at $m=-1/4$;
and $g_0=0.00370,\, g_1=-0.00778,\, g_2=-0.000680$ with two relevant operators at $m=-1/2$.
Thus the number of relevant operators changes from three to two here.
When we go to $N=3$ truncation,
$g_0=0.00518,\, g_1=-0.0196,\, g_2=0.000716,\, g_3=-0.00737$ with three relevant operators at $m=0$;
$g_0=0.00211,\, g_1=-0.0767,\, g_2=0.00157,\, g_3=0.000998$ with three relevant operators at $m=-1/4$;
and $g_0=0.00262,\, g_1=-0.0101,\, g_2=-0.0838,\, g_3=-0.0532$ with two relevant operators at $m=-1/2$.
Here again the number of relevant operator changes from three to two.
We thus find that the dimensionality of critical surface also changes according to $m$, just as for $\omega$.



\subsection{\texorpdfstring{$\beta=-\infty$}{TEXT}} \label{betainftysub}

Recently, different works in the context of asymptotic safety have employed the gauge choice $\beta=-\infty$, see \cite{Percacci:2015wwa,Gies:2015tca,Ohta:2015efa,Ohta:2015fcu,Ohta:2016npm,Ohta:2016jvw,Alkofer:2018fxj}. The combination of this choice with the exponential parametrization (in our notation $\omega=1/2$) minimizes the dependence of the beta functions (or divergences) on other free parameters as discussed in \cite{Gies:2015tca,Ohta:2016npm}. Frequently, this gauge choice is called ``physical gauge''. Generically it amounts to setting $\xi_\mu = 0$ and $h=0$. As a consequence, the flow equation analysis is simpler~\cite{Percacci:2015wwa,Ohta:2015efa,Ohta:2015fcu,Alkofer:2018fxj}. In the following, we collect our results for this choice in the same style we did in the previous subsection and comment on the comparison with the existent results in the literature afterwards. Here we restrict our discussions to the most studied case of $m=0$.

\begin{figure}[t!]
  \centering
  \begin{subfigure}[b]{0.49\linewidth}
    \includegraphics[width=\linewidth]{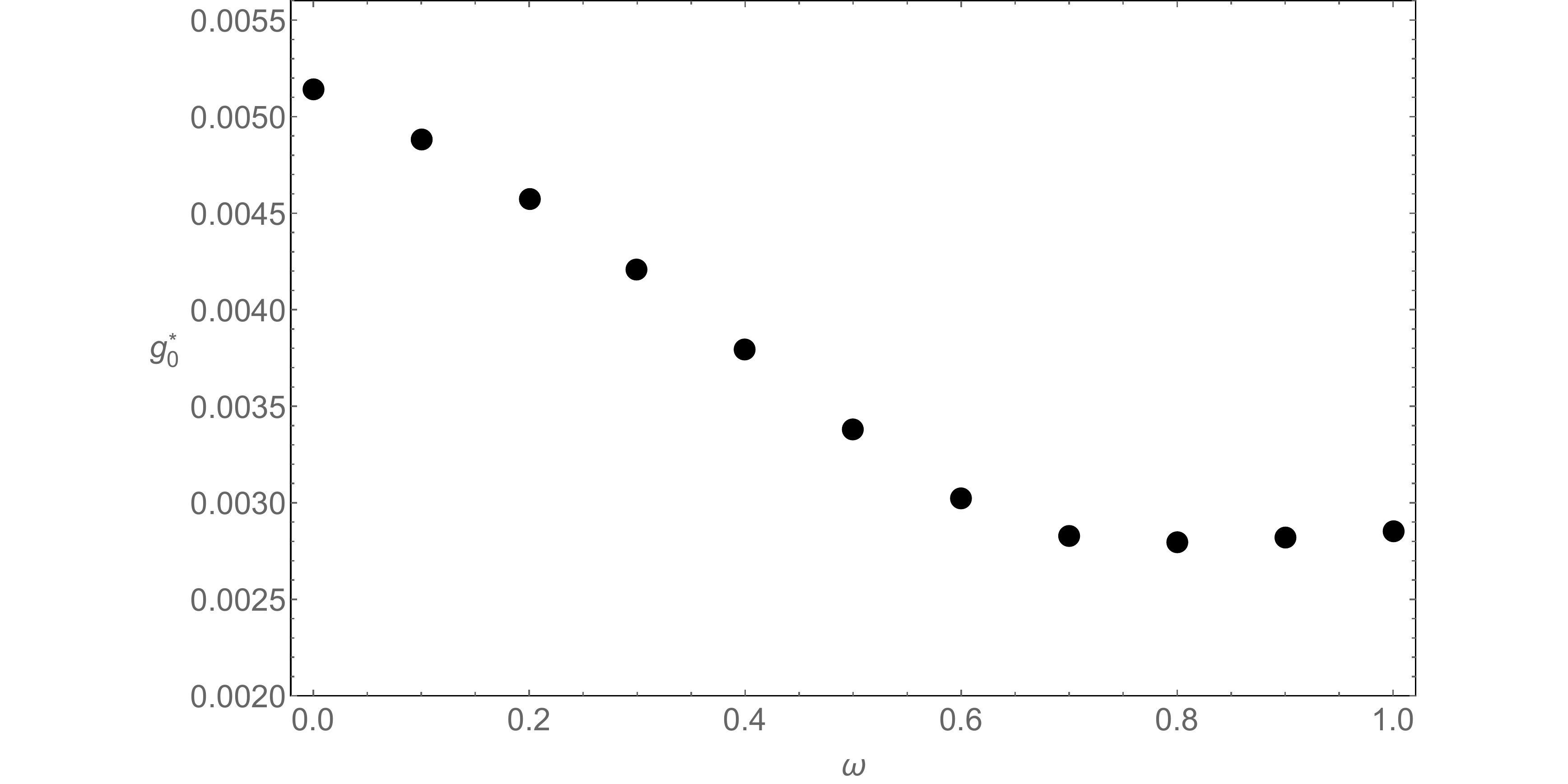}
  \end{subfigure}
  \begin{subfigure}[b]{0.49\linewidth}
    \includegraphics[width=\linewidth]{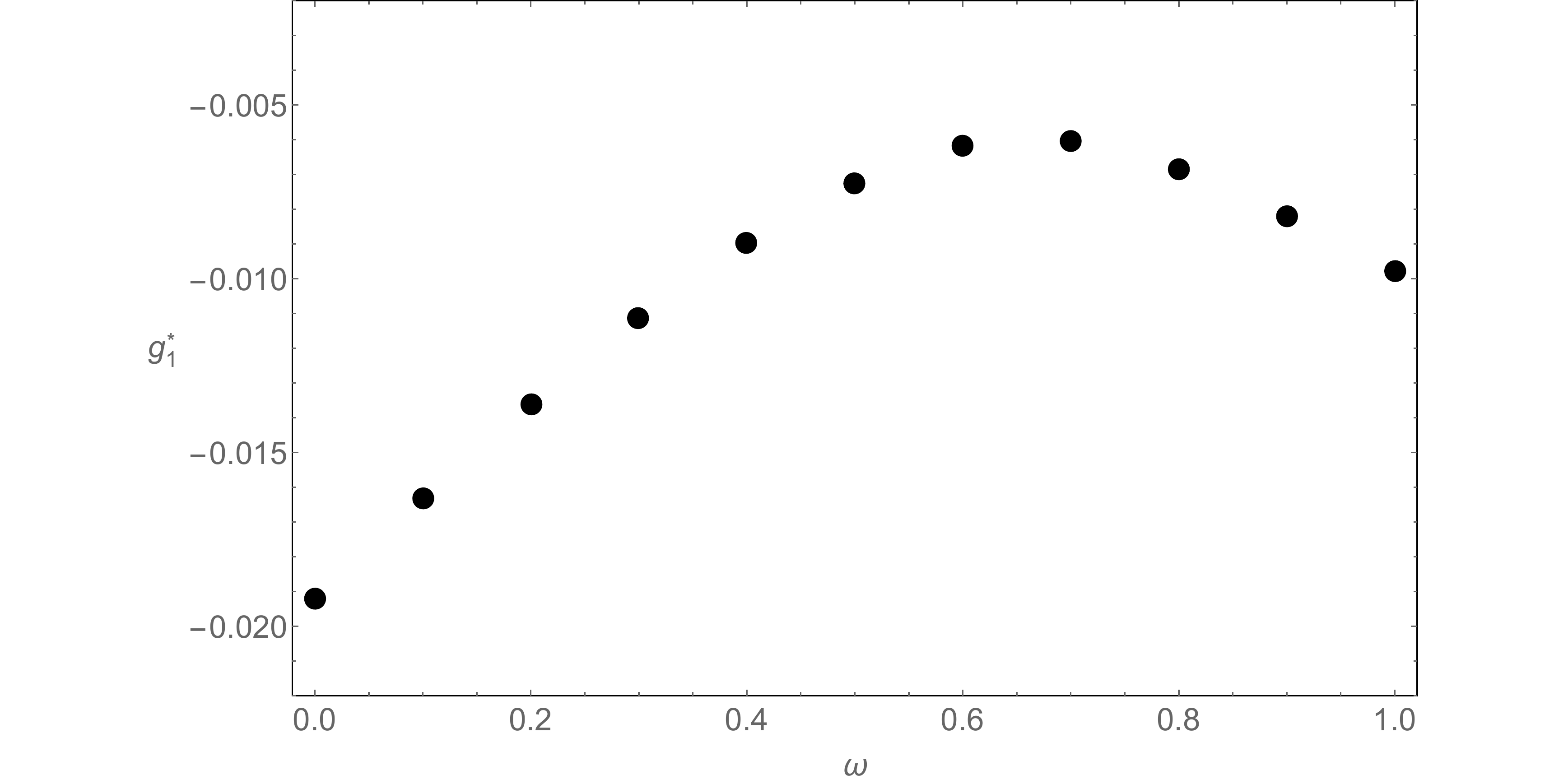}
  \end{subfigure}
\caption{Fixed point values for the couplings $g_0$ and $g_1$ in the Einstein-Hilbert truncation in the physical gauge.}
  \label{EHbetainfty}
\end{figure}

\begin{figure}[t!]
  \centering
  \begin{subfigure}[b]{0.49\linewidth}
    \includegraphics[width=\linewidth]{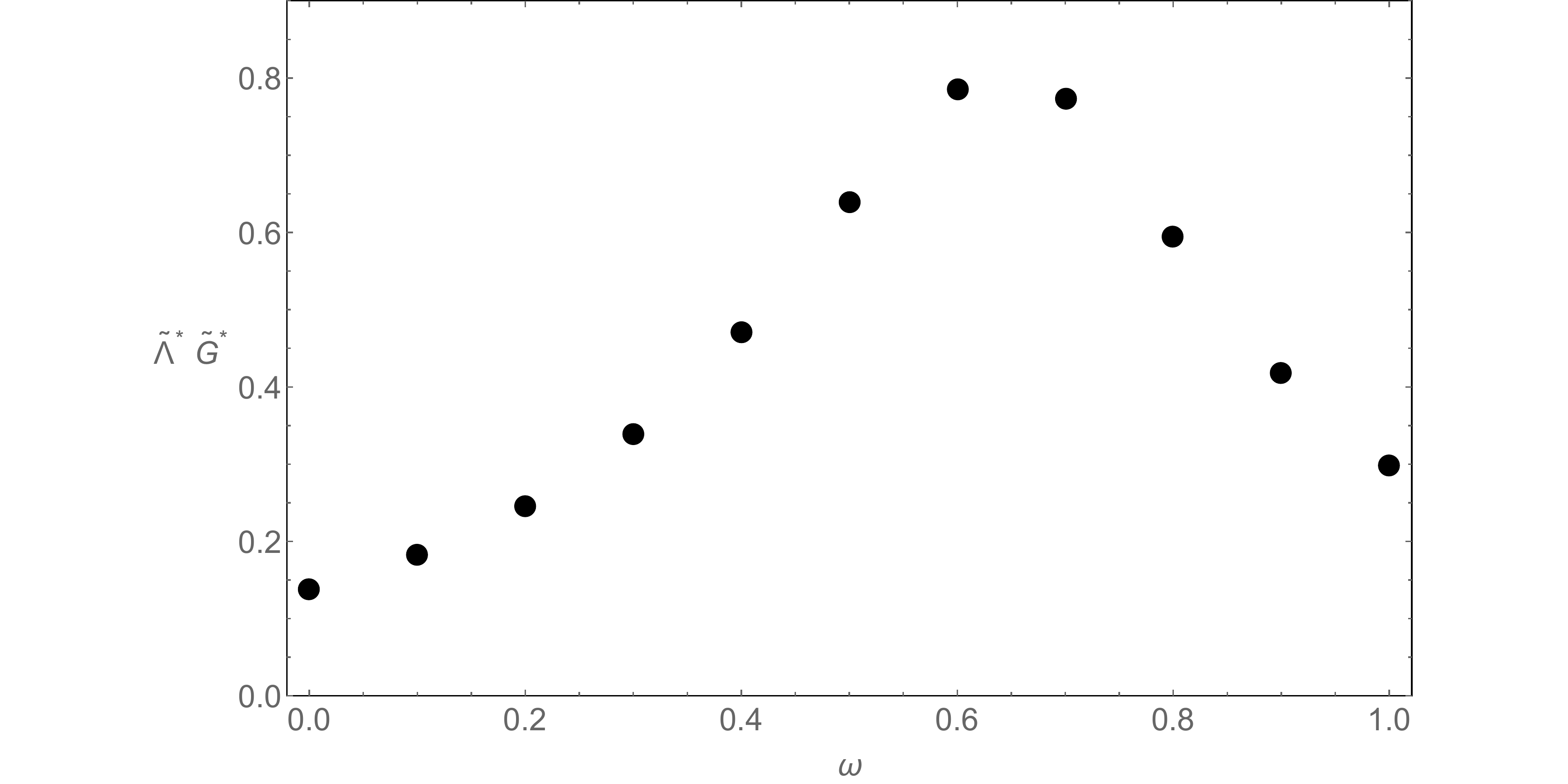}
    \caption{Product $\tilde{\Lambda}^{\ast}\tilde{G}^{\ast}$.}
  \end{subfigure}
  \begin{subfigure}[b]{0.49\linewidth}
    \includegraphics[width=\linewidth]{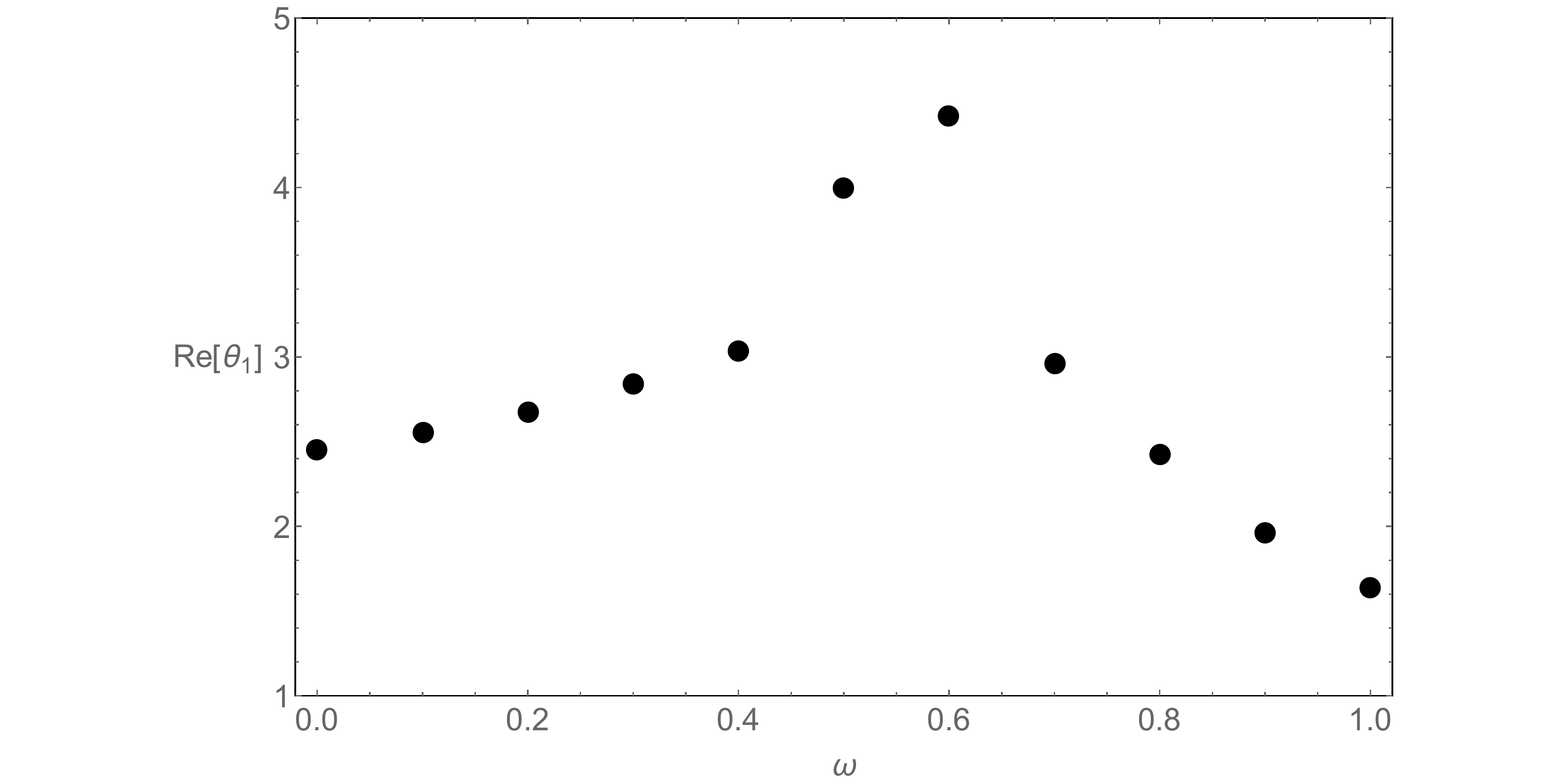}
    \caption{Real part of $\theta_1$.}
  \end{subfigure}
  \begin{subfigure}[b]{0.49\linewidth}
    \includegraphics[width=\linewidth]{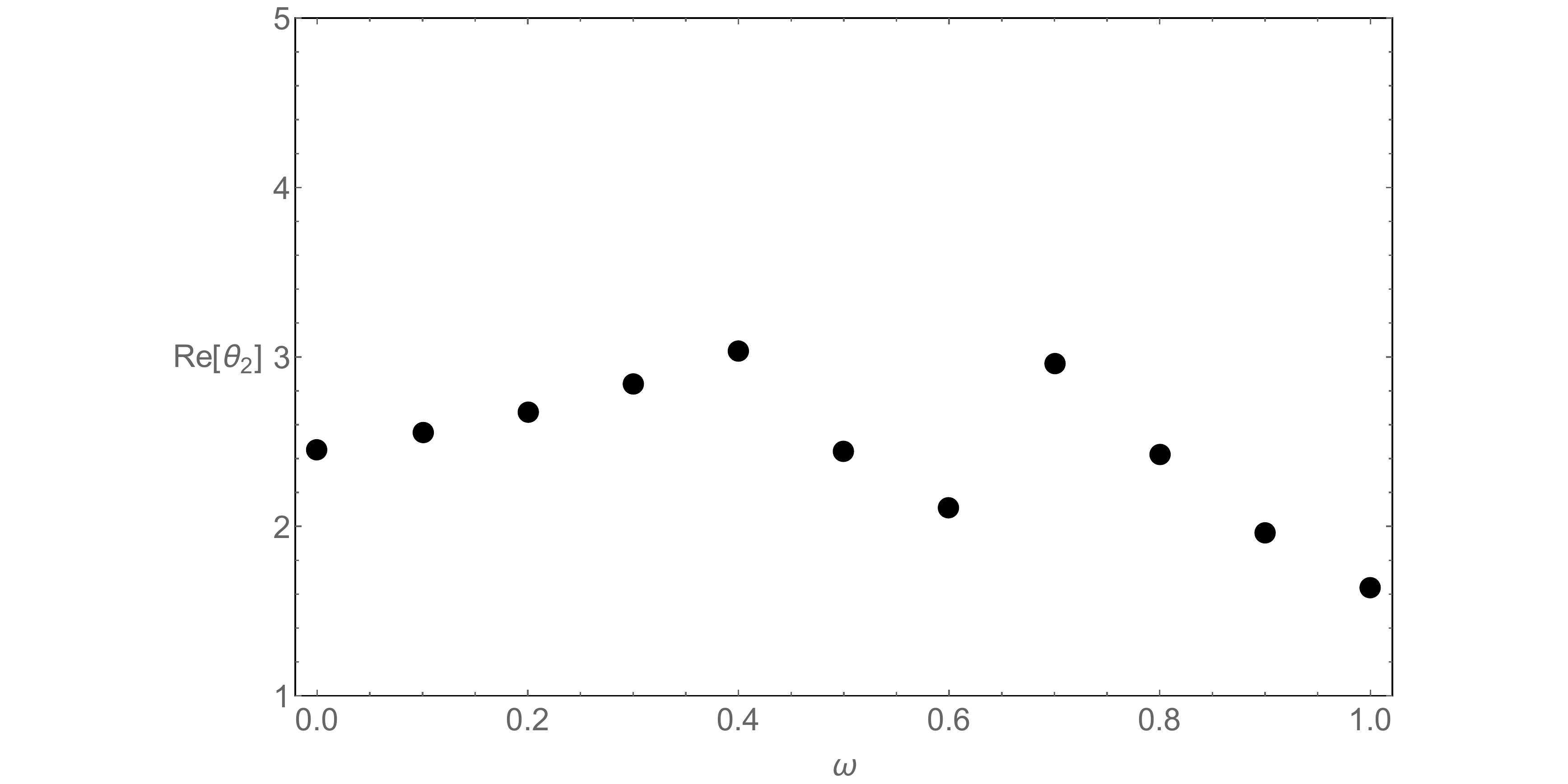}
    \caption{Real part of $\theta_2$.}
  \end{subfigure}  
  \caption{Einstein-Hilbert truncation in the physical gauge.}
  \label{lambdagEHbetainfty}
\end{figure}

In the Einstein-Hilbert truncation, we display how the fixed point values $(g^\ast_0,g^\ast_1)$ change with $\omega$ in Fig.~\ref{EHbetainfty}. We find that not only the dependence on $\omega$ is similar to the one reported for $\beta=0$ in Fig.~\ref{EHbeta0}, but also the numerical values are quite close to those computed for $\beta=0$. One sees that the local maximum for $g^\ast_1$ is located around $\omega = 0.6$ differently from the $\beta = 0$ gauge where it is located near $\omega = 1/2$. Nevertheless, the results are still close.  
In Fig.~\ref{lambdagEHbetainfty} we show the product $\tilde{\Lambda}^{\ast}\tilde{G}^{\ast}$ and (real part of) the critical exponents $\theta_1$ and $\theta_2$. In contrast to the $\beta = 0$ choice, the critical exponents at $\omega=1/2$ and $\omega = 0.6$ are real, but they are both positive, leading to two relevant directions as for $\beta = 0$. Therefore, within the Einstein-Hilbert truncation we see that the results obtained for $\beta = 0$ and $\beta = -\infty$ are qualitatively similar. 

\begin{figure}[t!]
  \centering
  \begin{subfigure}[b]{0.49\linewidth}
    \includegraphics[width=\linewidth]{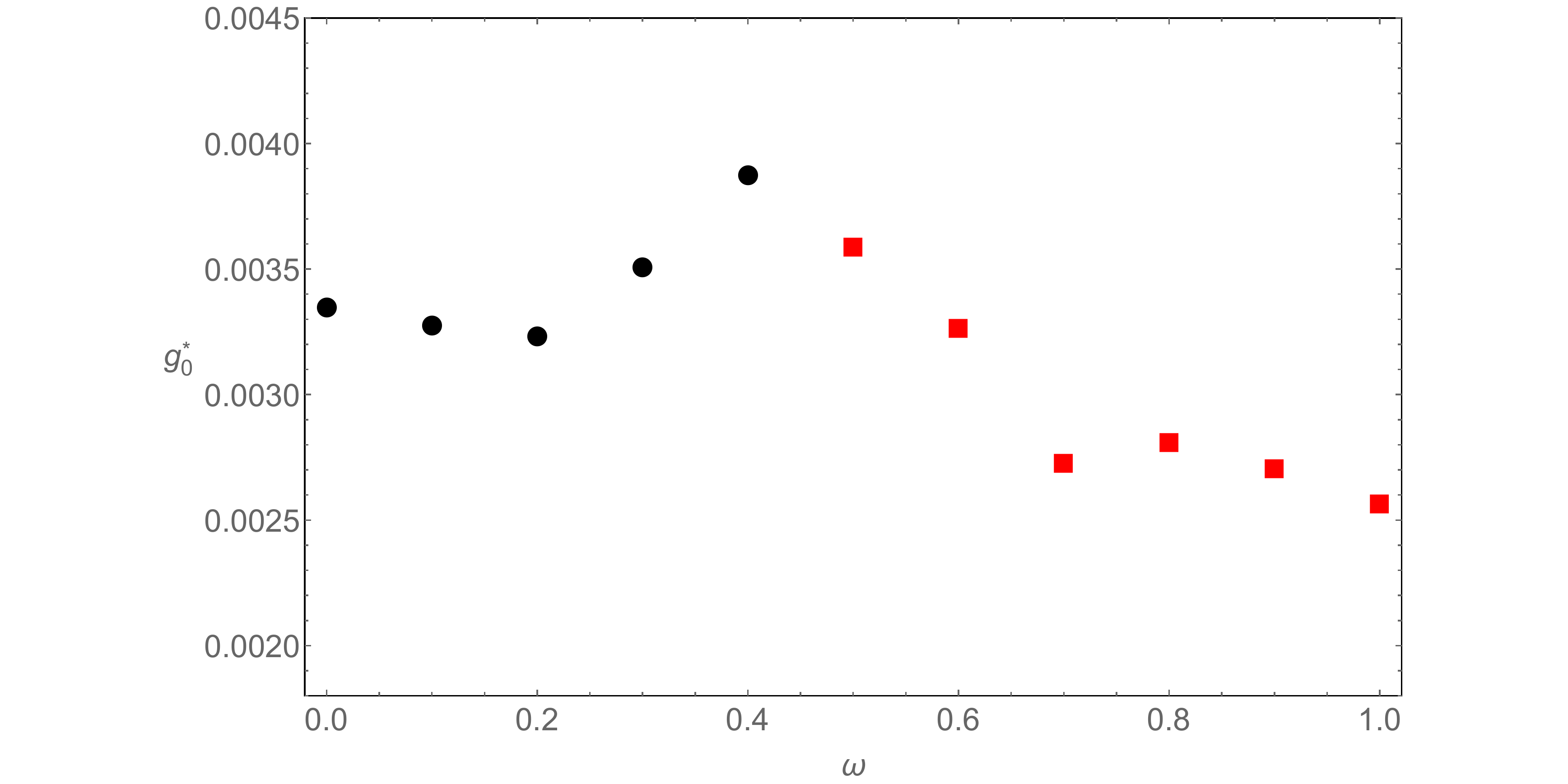}
  \end{subfigure}
  \begin{subfigure}[b]{0.49\linewidth}
    \includegraphics[width=\linewidth]{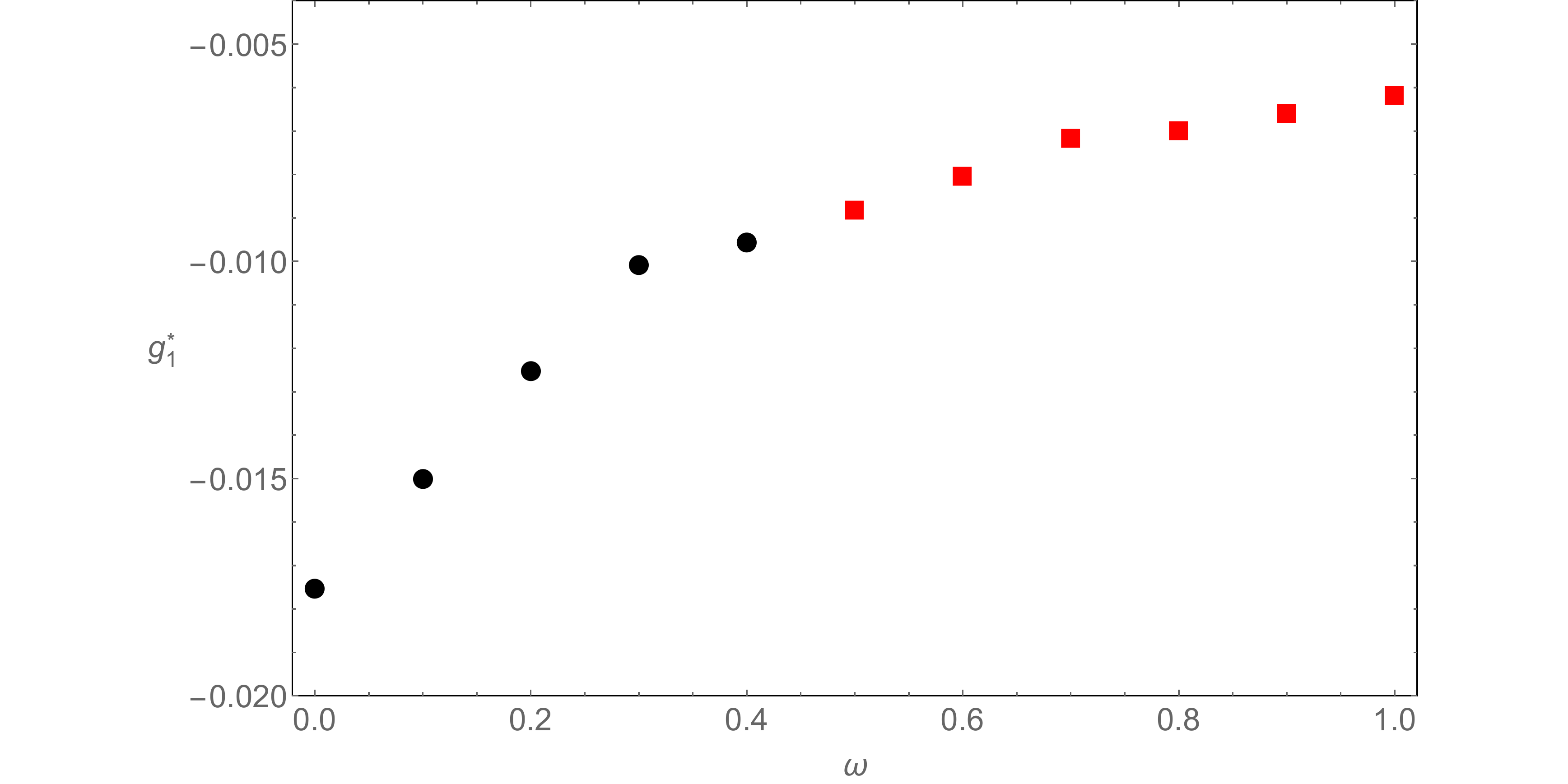}
  \end{subfigure}
    \begin{subfigure}[b]{0.49\linewidth}
    \includegraphics[width=\linewidth]{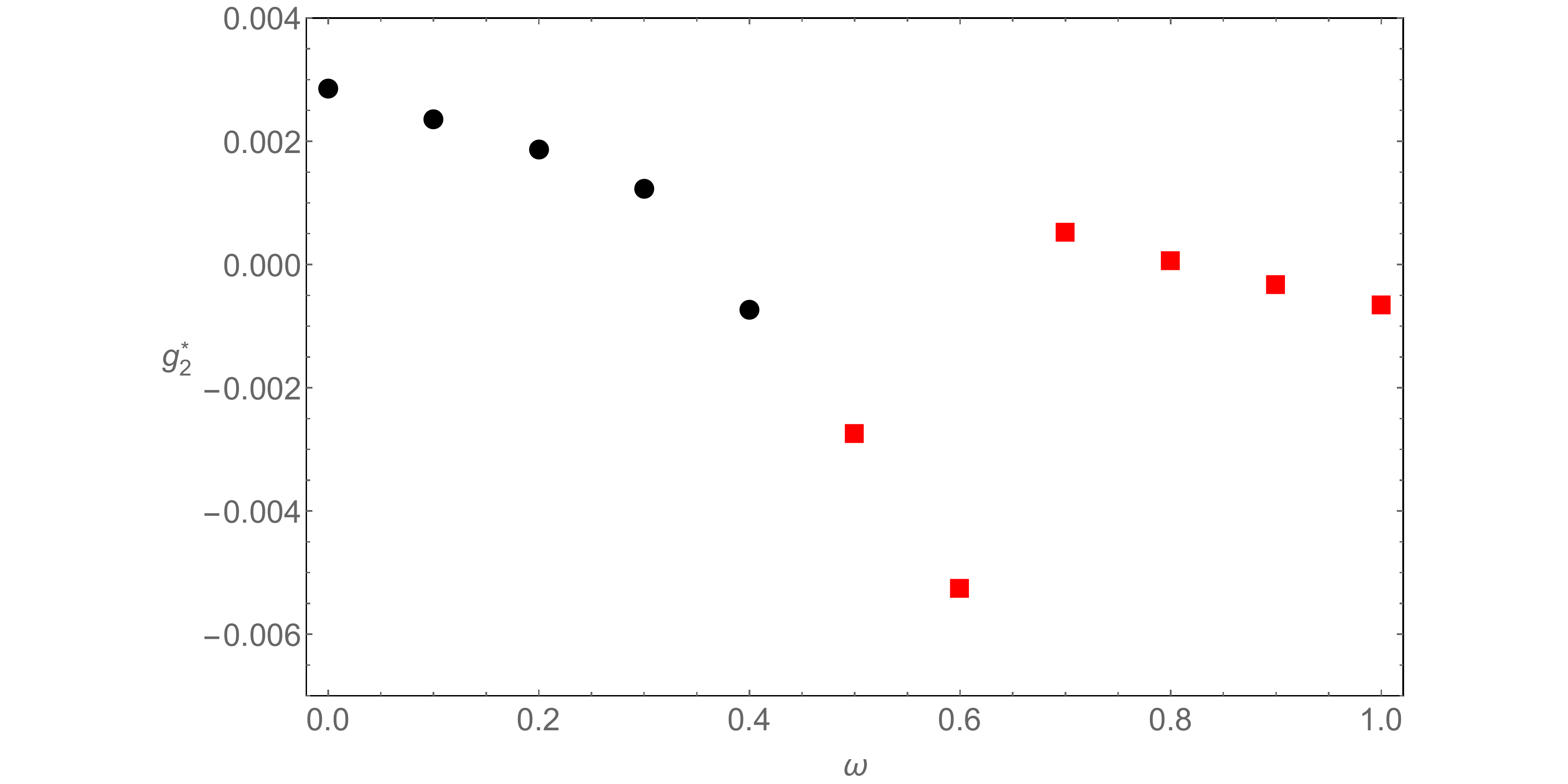}
  \end{subfigure}
  \caption{Fixed point values for the couplings $g_0$, $g_1$ and $g_2$ in the $R^2$ truncation in the physical gauge.}
  \label{R2betainfty}
\end{figure}

\begin{figure}[t!]
  \centering
  \begin{subfigure}[b]{0.49\linewidth}
    \includegraphics[width=\linewidth]{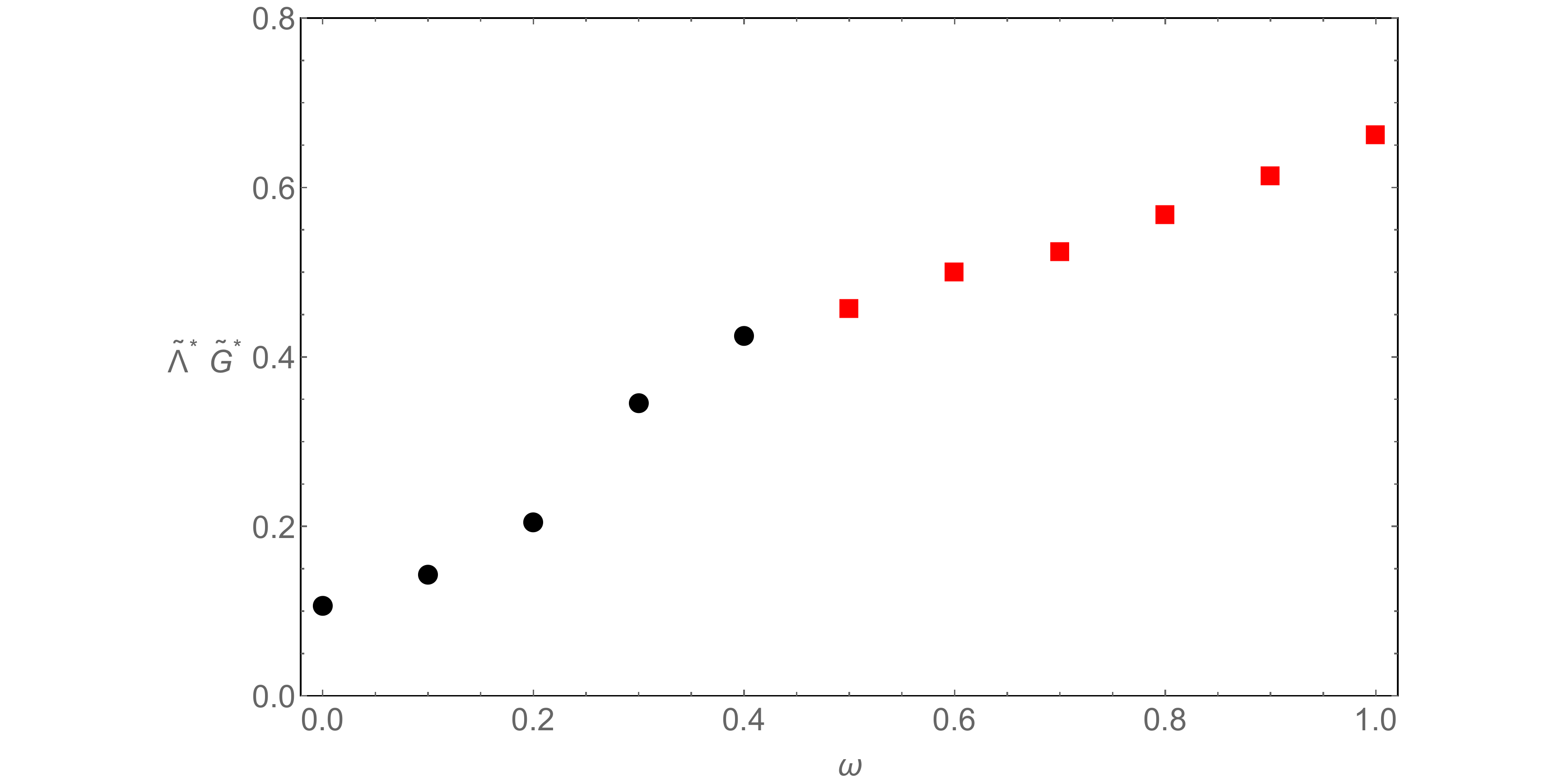}
    \caption{Product $\tilde{\Lambda}^{\ast}\tilde{G}^{\ast}$.}
  \end{subfigure}
  \begin{subfigure}[b]{0.49\linewidth}
    \includegraphics[width=\linewidth]{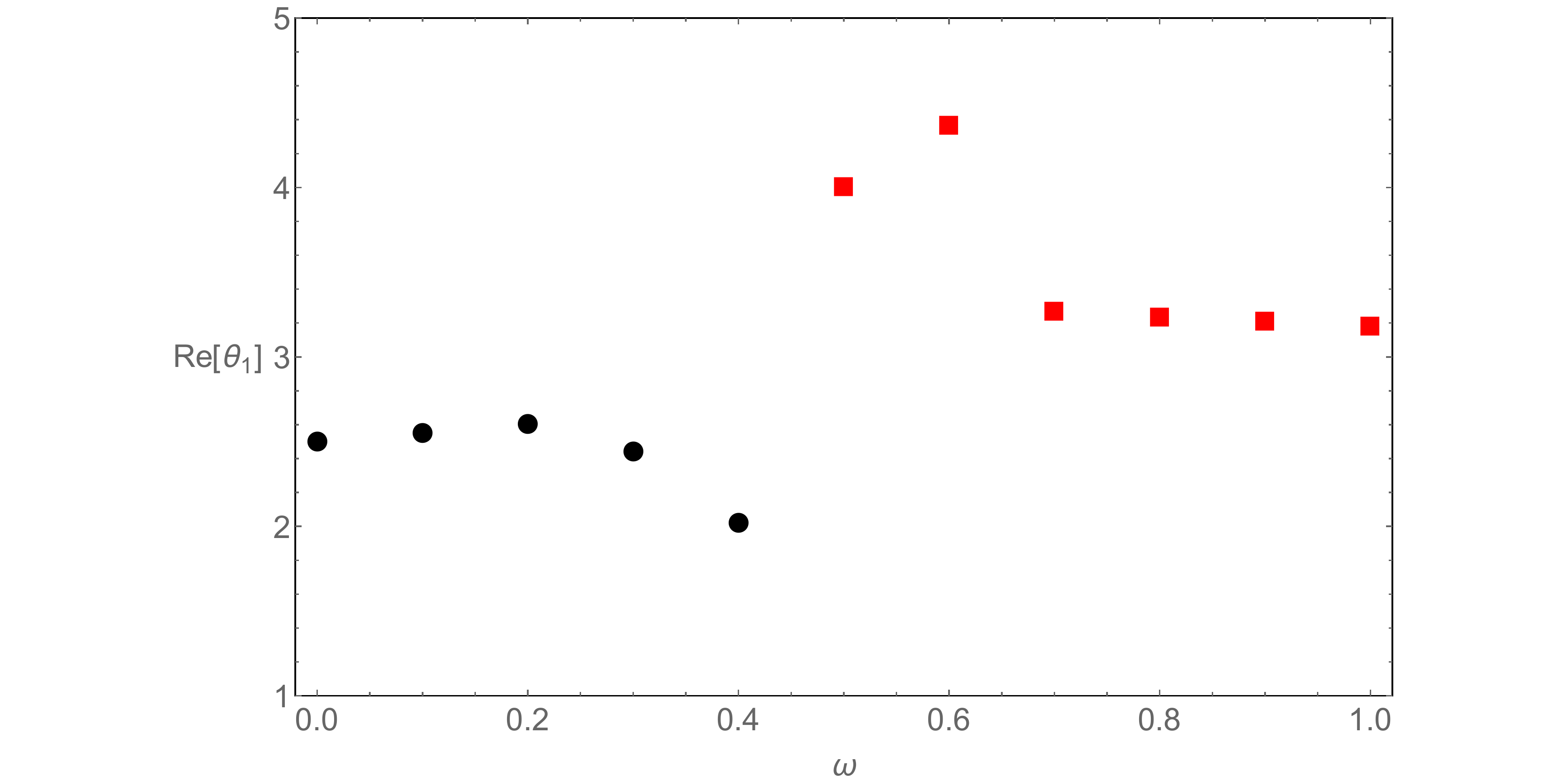}
    \caption{Real part of the critical exponent $\theta_1$.}
  \end{subfigure}
    \begin{subfigure}[b]{0.49\linewidth}
    \includegraphics[width=\linewidth]{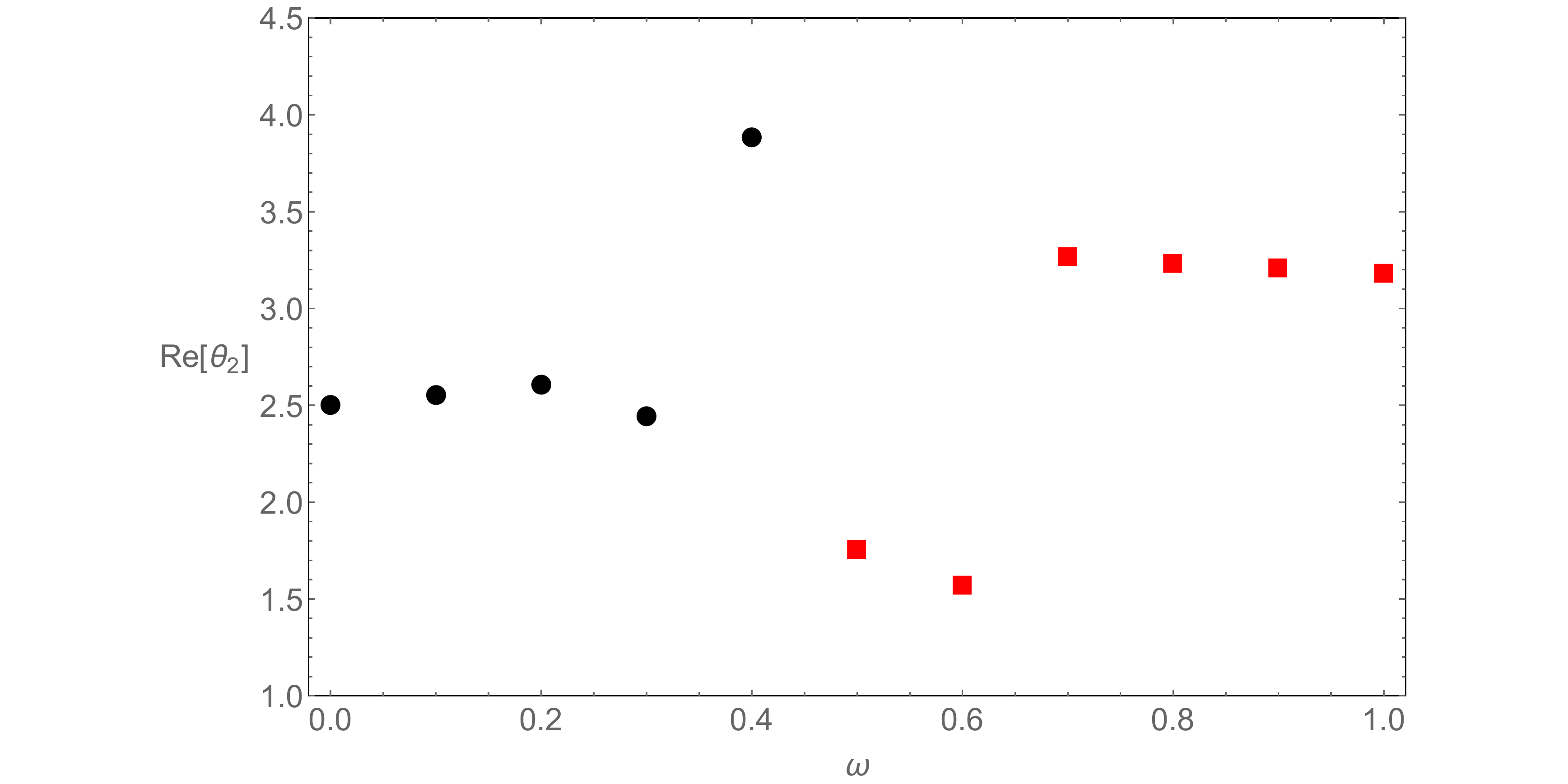}
    \caption{Real part of the critical exponent $\theta_2$.}
  \end{subfigure}
	\begin{subfigure}[b]{0.49\linewidth}
    \includegraphics[width=\linewidth]{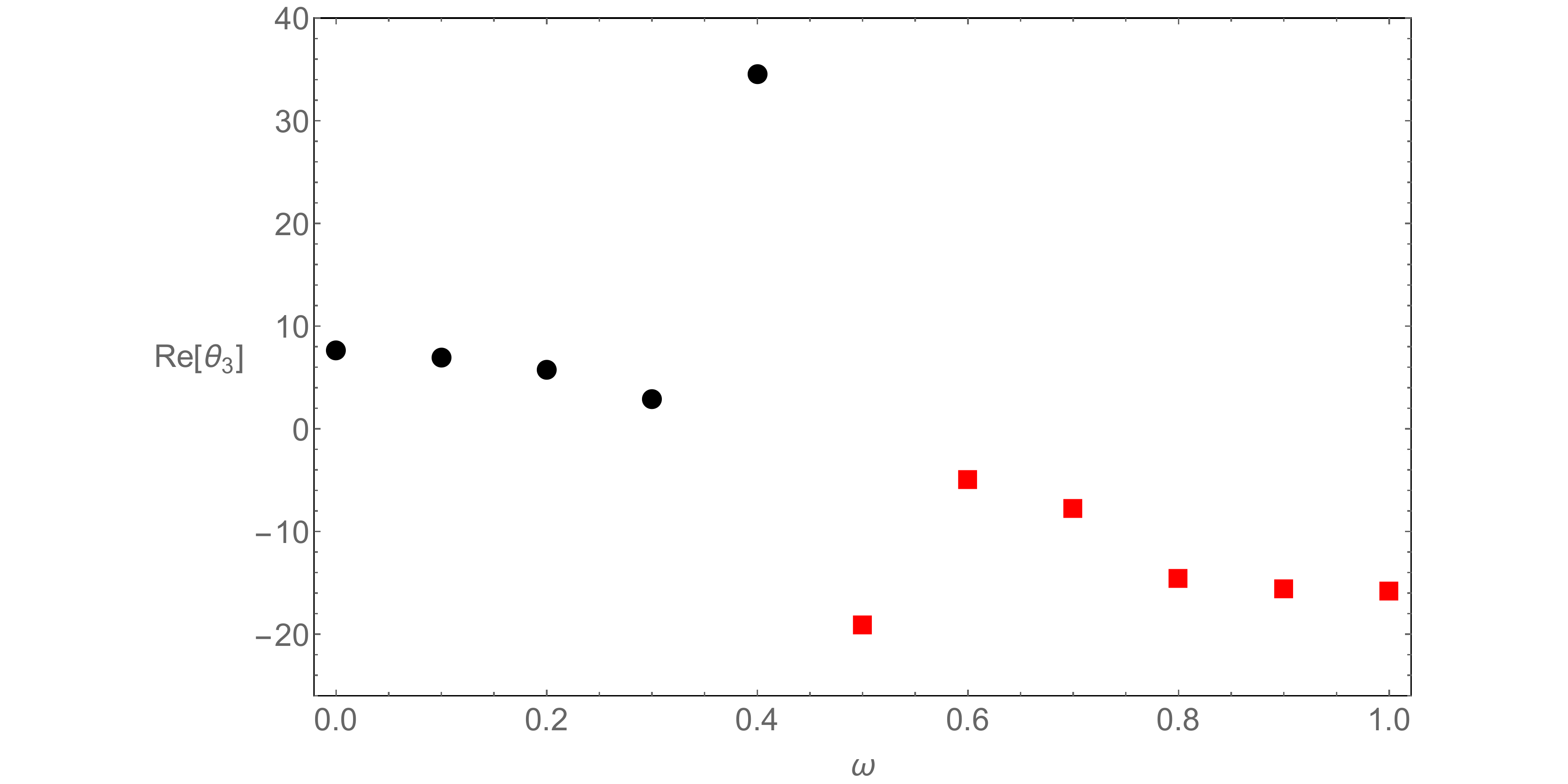}
    \caption{Real part of the critical exponent $\theta_3$.}
  \end{subfigure}  
  \caption{$R^2$ truncation in the physical gauge.}
  \label{lambdagR2betainfty}
\end{figure}

The fixed points values $(g^\ast_0,g^\ast_1,g^\ast_2)$ for different values of $\omega$ in the $N=2$ truncation are shown in Fig.~\ref{R2betainfty}. We employ the same representation as in the $\beta = 0$: black dots represent fixed points with three relevant directions and red squares, fixed points with two relevant directions. As exhibited in Fig.~\ref{R2betainfty}, for $0\le \omega < 1/2$ one finds a nontrivial fixed point with three relevant directions while for $1/2\le\omega\le 1$, a fixed point with just two relevant directions. This is in qualitative agreement with the $\beta=0$ results in Fig.~\ref{R2beta0}. It should be noticed that, typically, for a given $\omega$, one obtains more than one fixed point which satisfies the basic requirements for a viable UV fixed point (sometimes, viable fixed points with different number of relevant directions are found as already pointed out in the $\beta=0$ case). The selection rule we employ is to check which of those fixed points are still present under truncation enlargement.  
The product $\tilde{\Lambda}^\ast \tilde{G}^\ast$ together with the critical exponents in the $N=2$ truncation are collected in Fig.~\ref{lambdagR2betainfty}. As in the Einstein-Hilbert truncation, there are specific choices of $\omega$ where the critical exponents $\theta_1$ and $\theta_2$ are real. This happens, in particular, for $\omega=\left\{0.4,0.5,0.6\right\}$. As also happens in the case $\beta=0$, the critical exponents in this truncation are rather large.

\begin{figure}[t!]
  \centering
  \begin{subfigure}[b]{0.49\linewidth}
    \includegraphics[width=\linewidth]{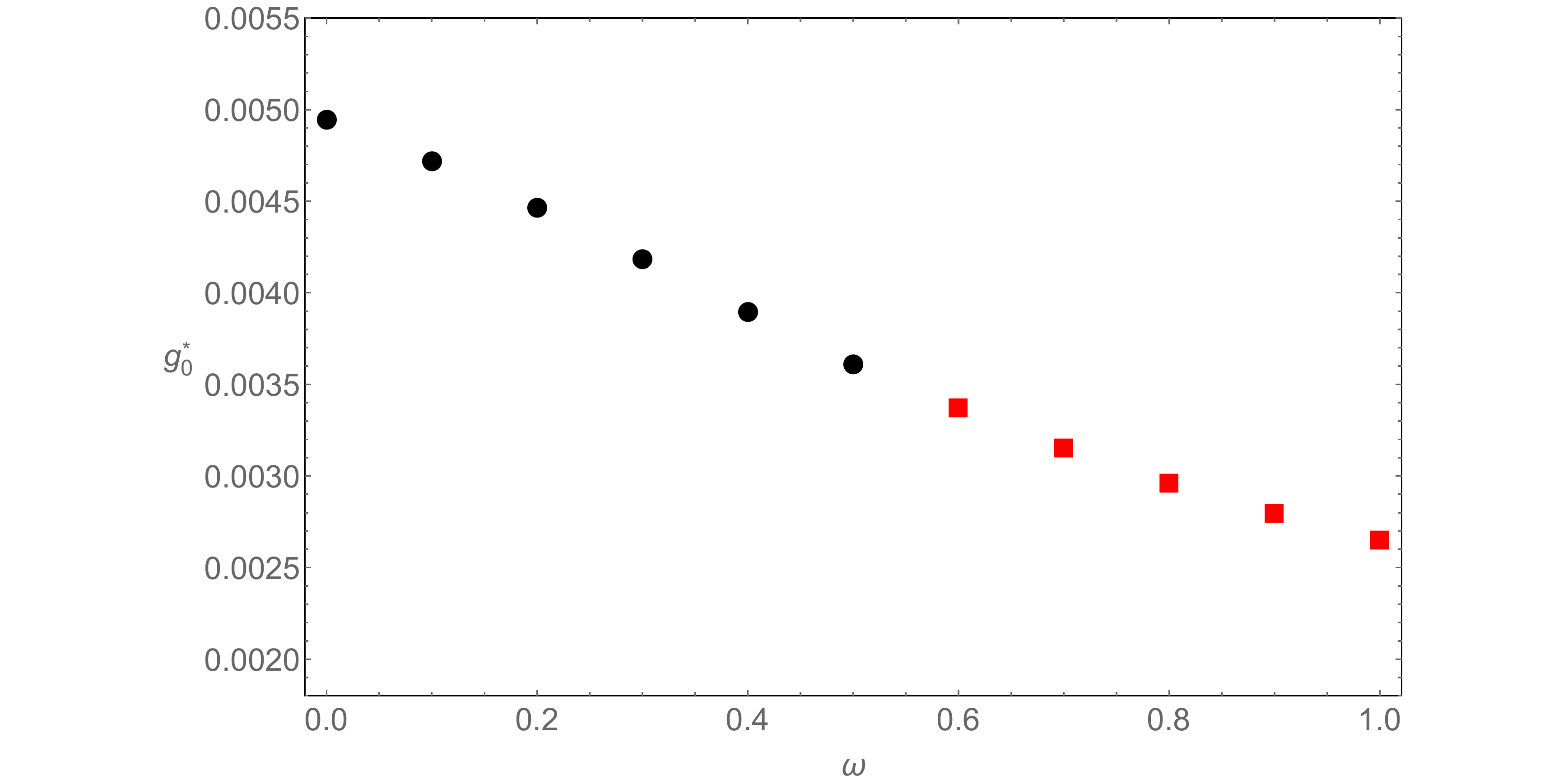}
  \end{subfigure}
  \begin{subfigure}[b]{0.49\linewidth}
    \includegraphics[width=\linewidth]{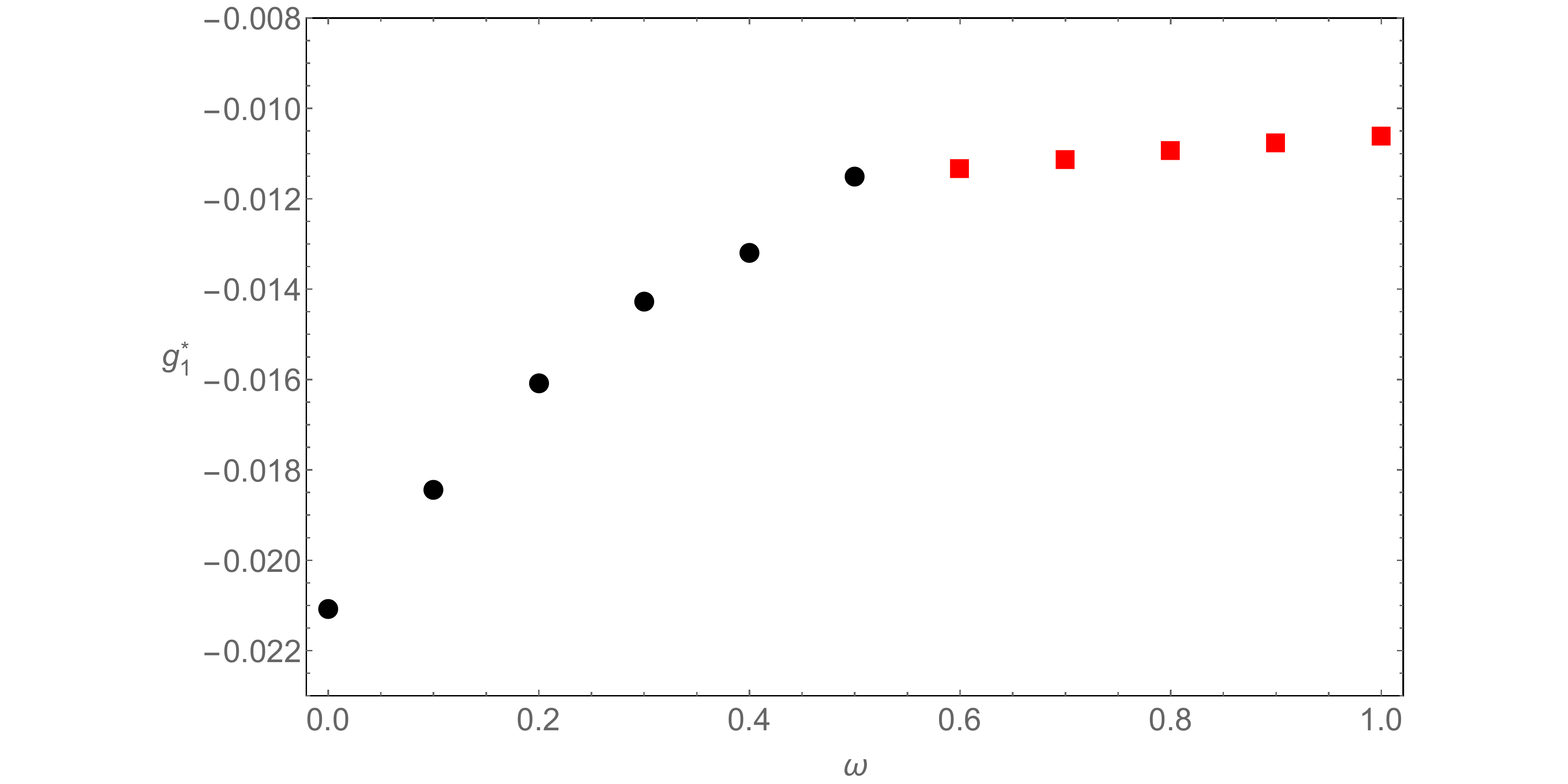}
  \end{subfigure}
    \begin{subfigure}[b]{0.49\linewidth}
    \includegraphics[width=\linewidth]{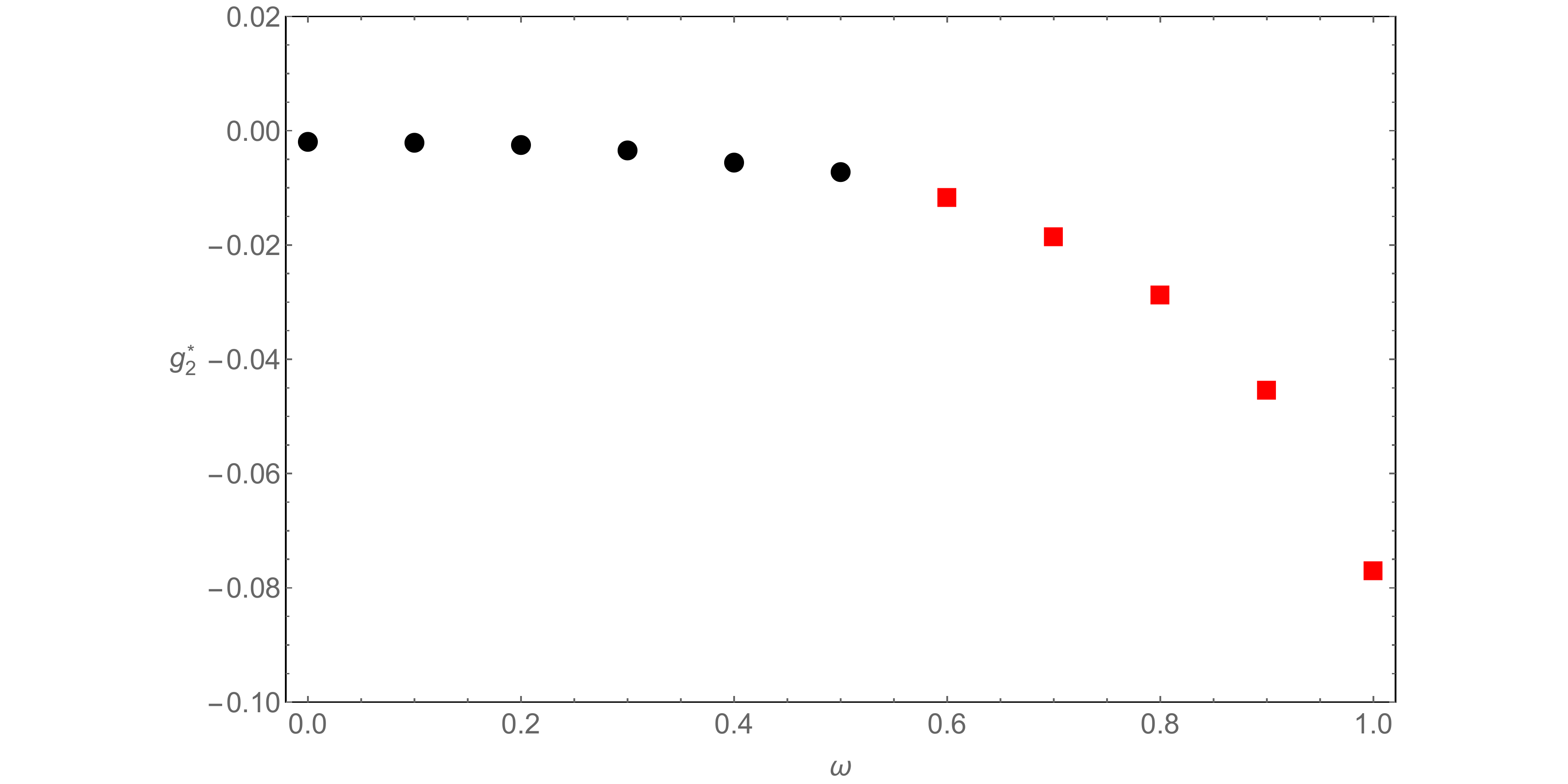}
  \end{subfigure}
    \begin{subfigure}[b]{0.49\linewidth}
    \includegraphics[width=\linewidth]{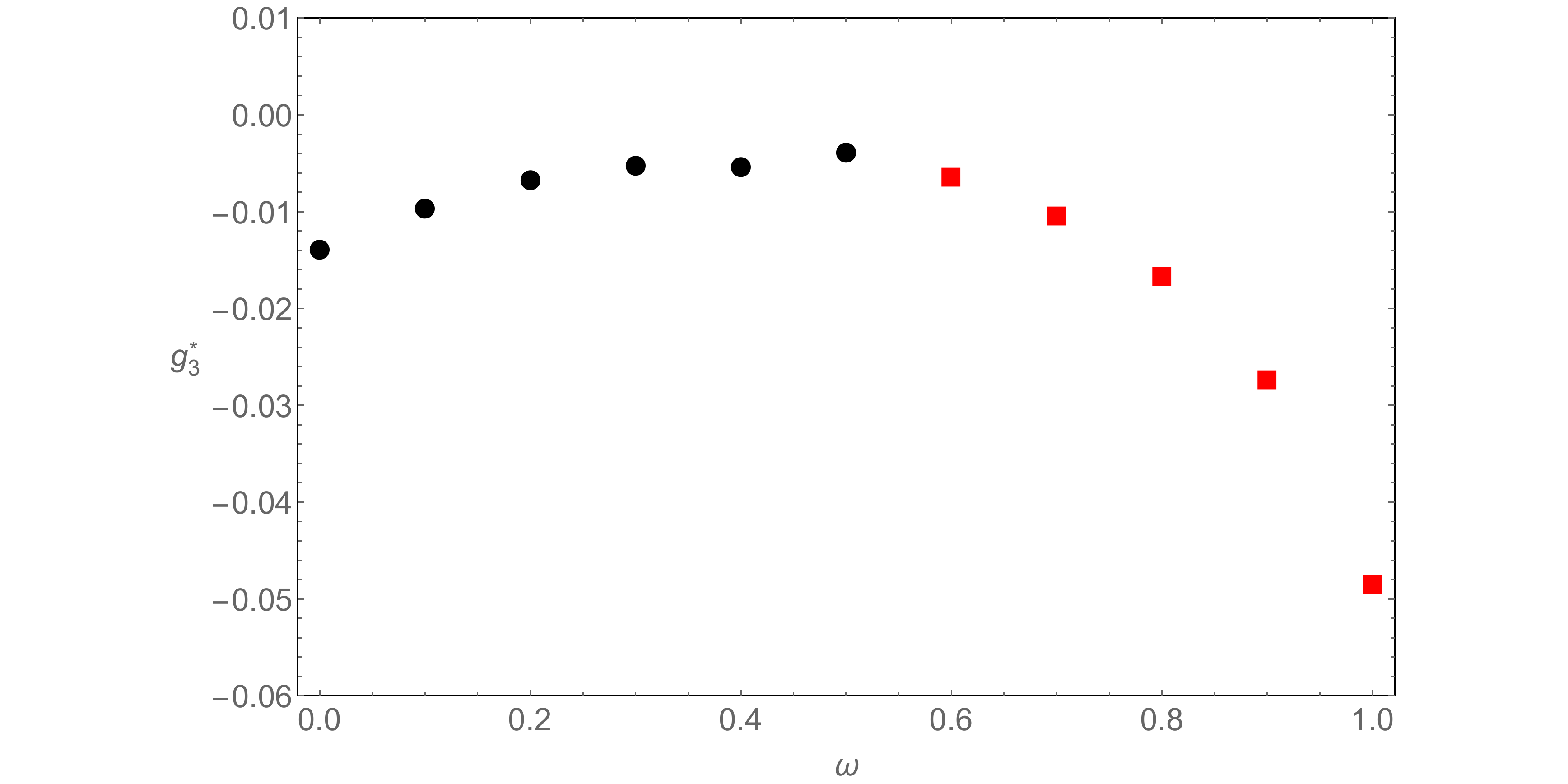}
  \end{subfigure}  
  \caption{Fixed point values for the couplings $g_0$, $g_1$, $g_2$ and $g_3$ in the $R^3$ truncation in the physical gauge.}
  \label{R3betainfty}
\end{figure}

\begin{figure}[t!]
  \centering
  \begin{subfigure}[b]{0.49\linewidth}
    \includegraphics[width=\linewidth]{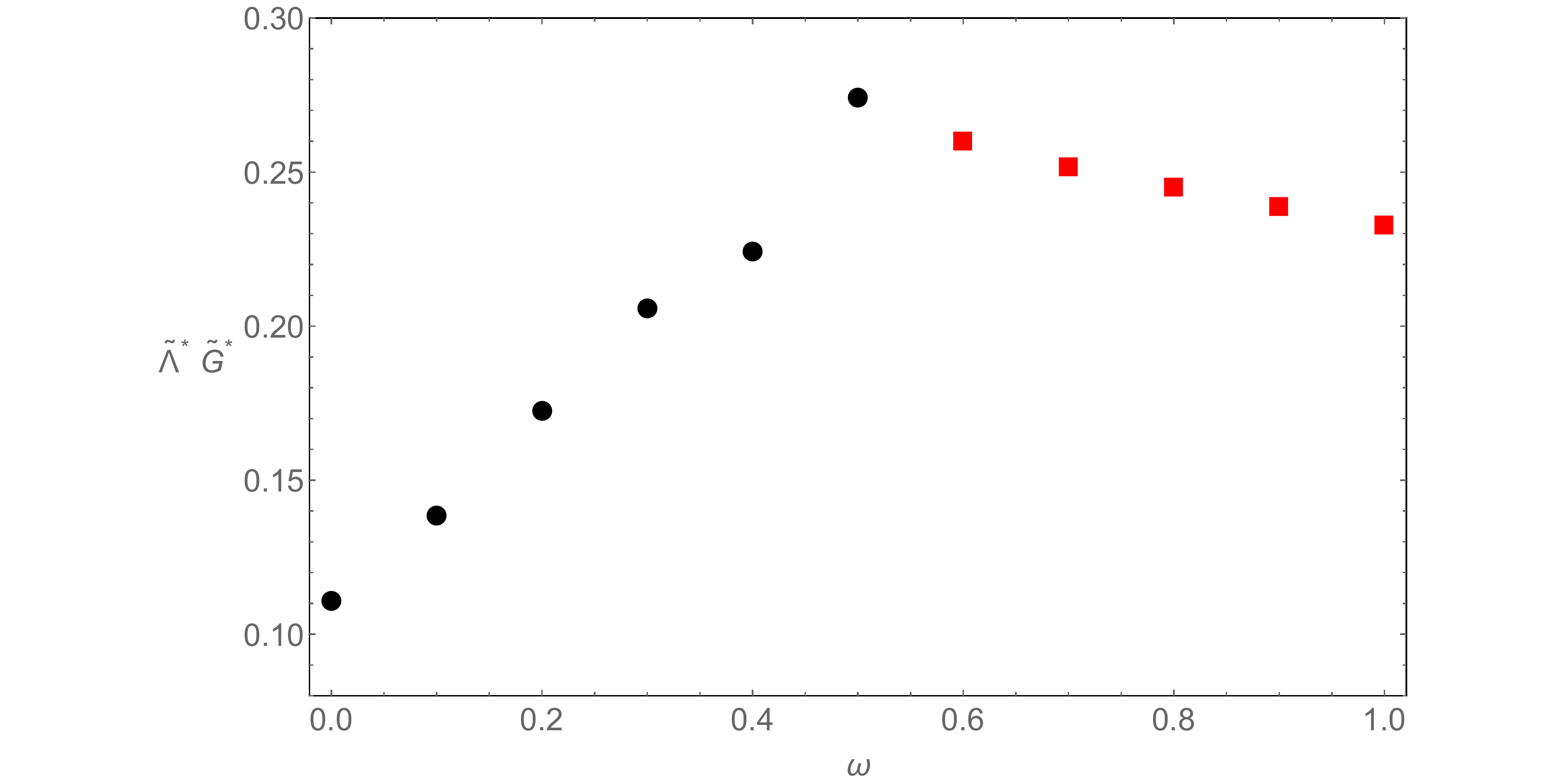}
    \caption{Product $\tilde{\Lambda}^{\ast}\tilde{G}^{\ast}$.}
  \end{subfigure}
  \begin{subfigure}[b]{0.49\linewidth}
    \includegraphics[width=\linewidth]{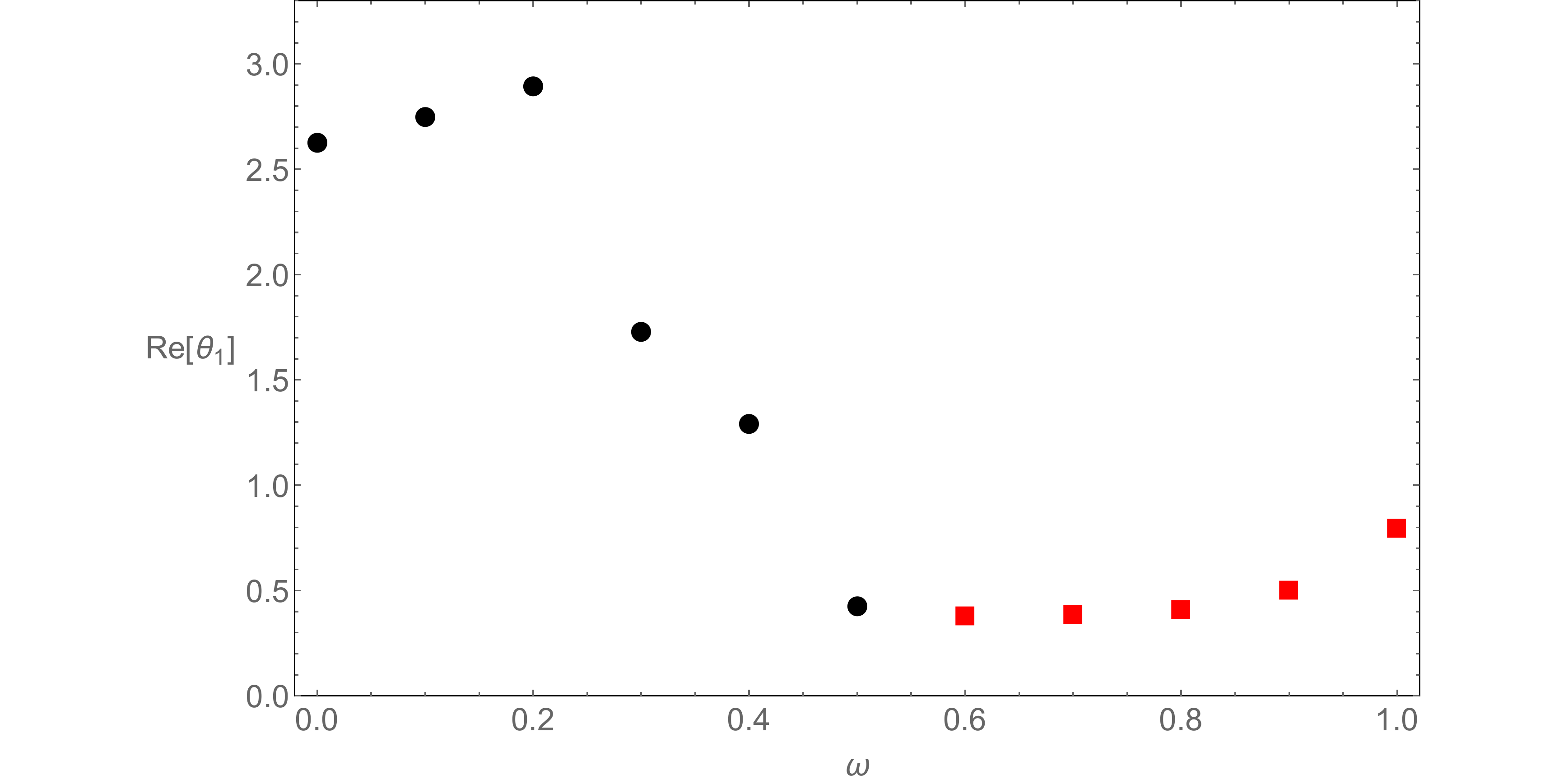}
    \caption{Real part of the critical exponent $\theta_1$.}
  \end{subfigure}
    \begin{subfigure}[b]{0.49\linewidth}
    \includegraphics[width=\linewidth]{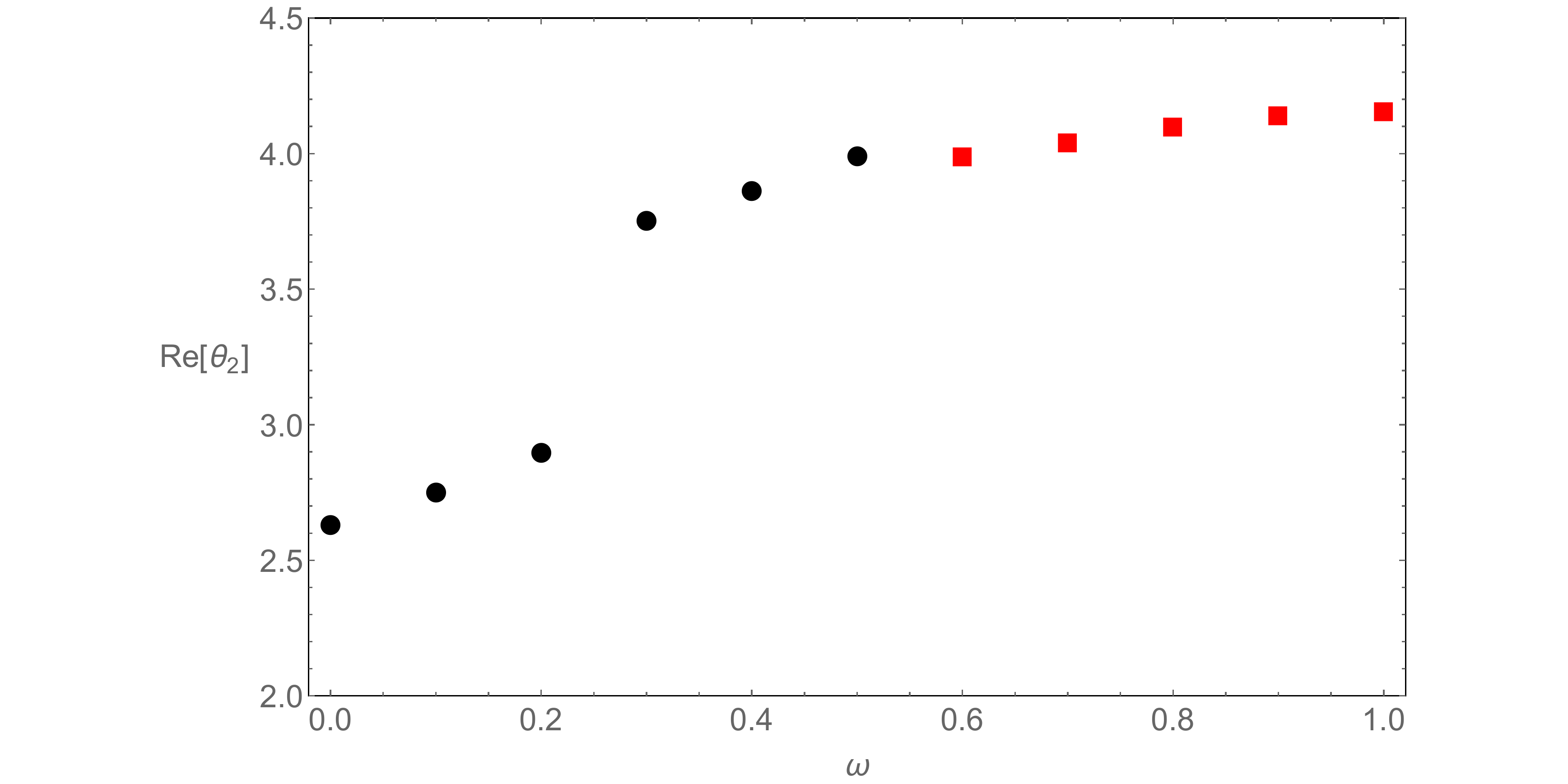}
    \caption{Real part of the critical exponent $\theta_2$.}
  \end{subfigure}
	\begin{subfigure}[b]{0.49\linewidth}
    \includegraphics[width=\linewidth]{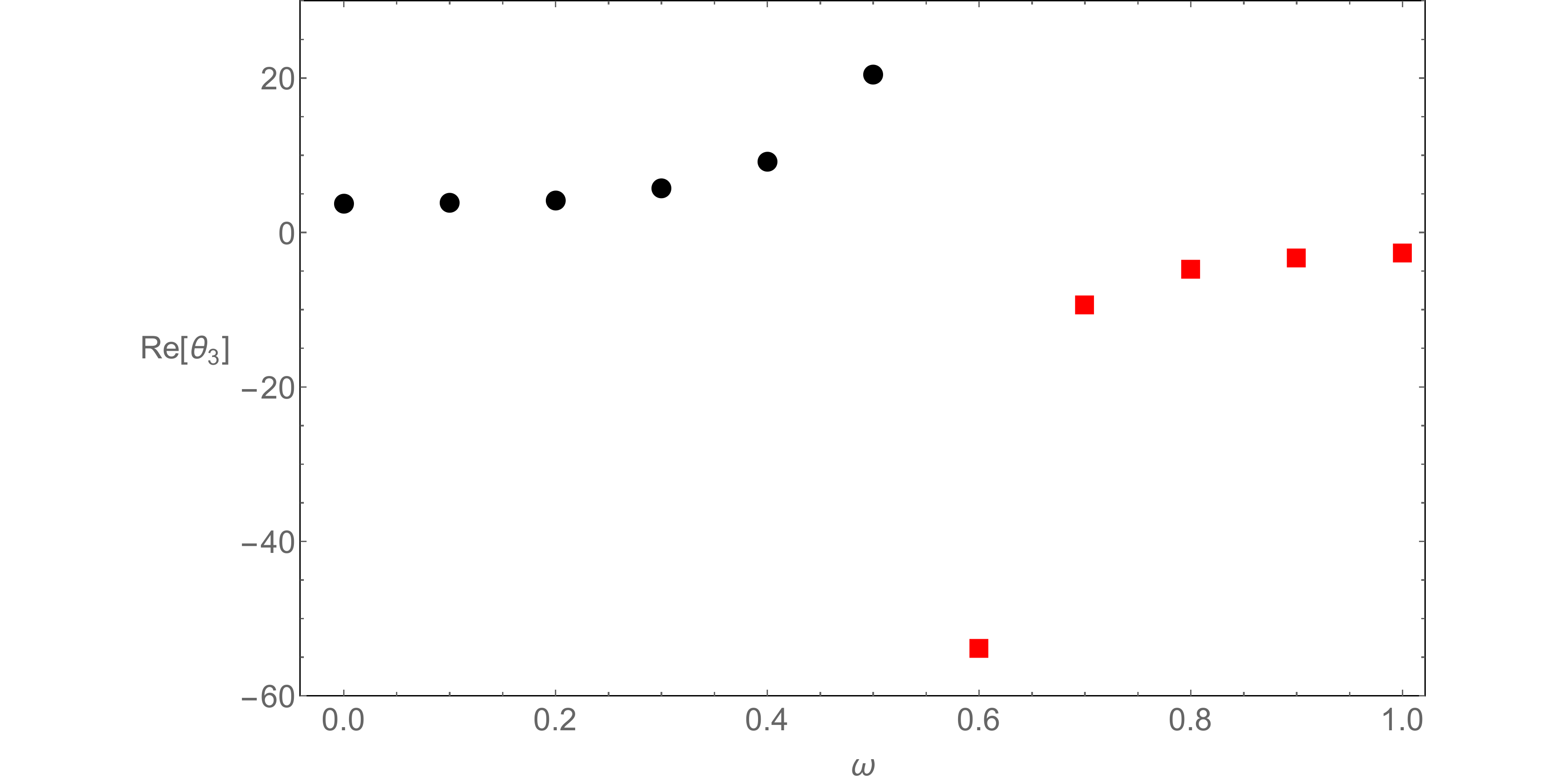}
    \caption{Real part of the critical exponent $\theta_3$.}
  \end{subfigure}  
	\begin{subfigure}[b]{0.49\linewidth}
    \includegraphics[width=\linewidth]{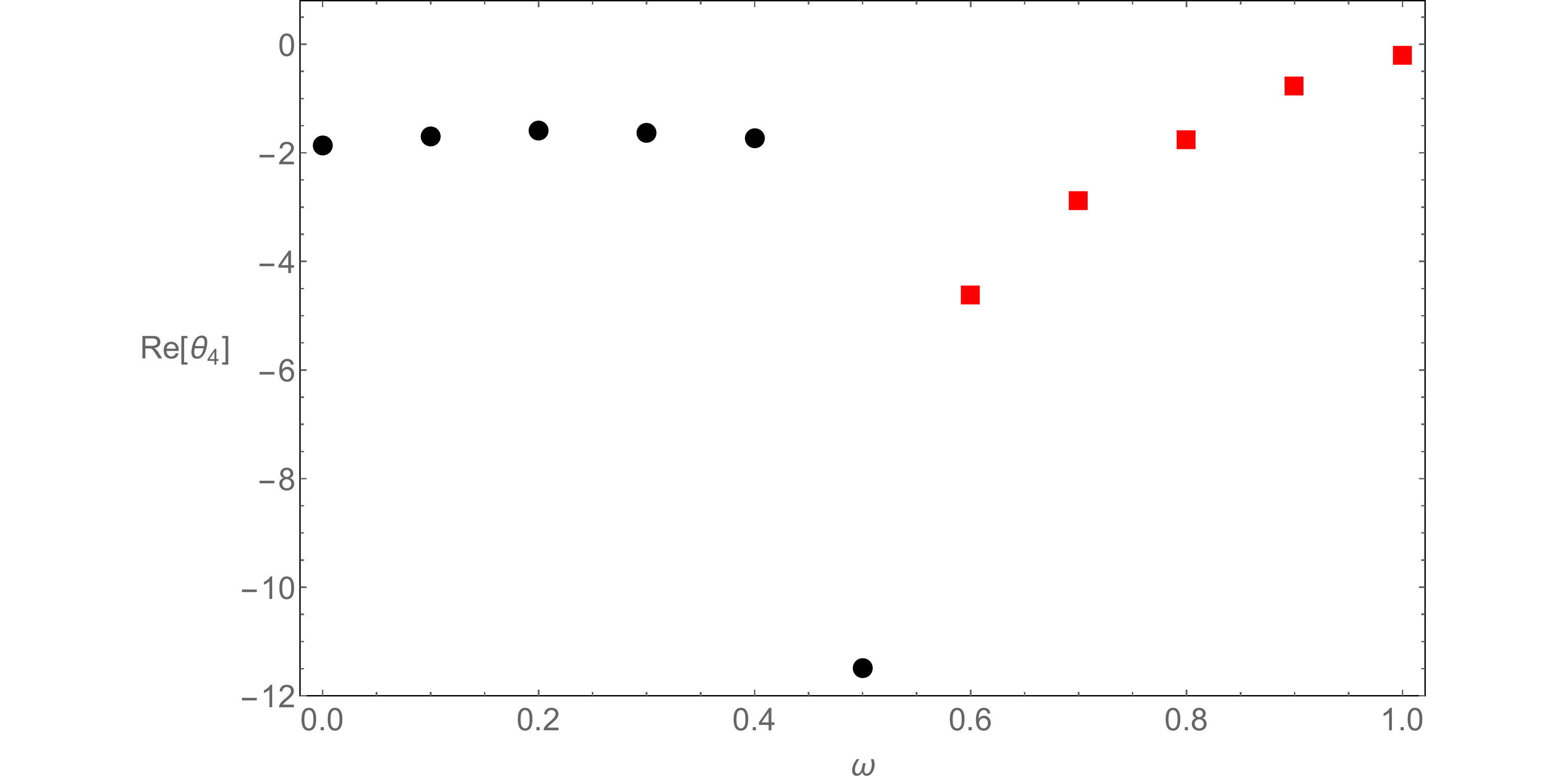}
    \caption{Real part of the critical exponent $\theta_4$.}
  \end{subfigure}    
  \caption{$R^3$ truncation in the physical gauge.}
  \label{lambdagR3betainfty}
\end{figure}

For the $N=3$ truncation, we show the fixed points values $(g^\ast_0,g^\ast_1,g^\ast_2,g^\ast_3)$ as a function of $\omega$ in Fig.~\ref{R3betainfty}. The existence of two different classes of fixed points (with three and two relevant directions, respectively) persists. In particular, one identifies that the ``transition'' from three to two relevant directions occurs at $\omega>1/2$. This is different from the results reported in \cite{Ohta:2015fcu,Alkofer:2018fxj}, for instance, where a fixed point with two relevant directions for pure gravity in the exponential parametrization $(\omega=1/2)$ in the physical gauge was found. On the other hand, it agrees with the result reported in \cite{Eichhorn:2015bna} which also displays a fixed point with two relevant directions, but in the unimodular setting (i.e. the cosmological constant is not a coupling of the theory). Although there is an apparent clash between these results, we argue in the next section that they arise due to different choices of regularization schemes.  
We collect the product $\tilde{\Lambda}^{\ast}\tilde{G}^\ast$ as well as the critical exponents in the $N=3$ truncation in Fig.~\ref{lambdagR3betainfty}. 

\begin{figure}[t!]
  \centering
  \begin{subfigure}[b]{0.49\linewidth}
    \includegraphics[width=\linewidth]{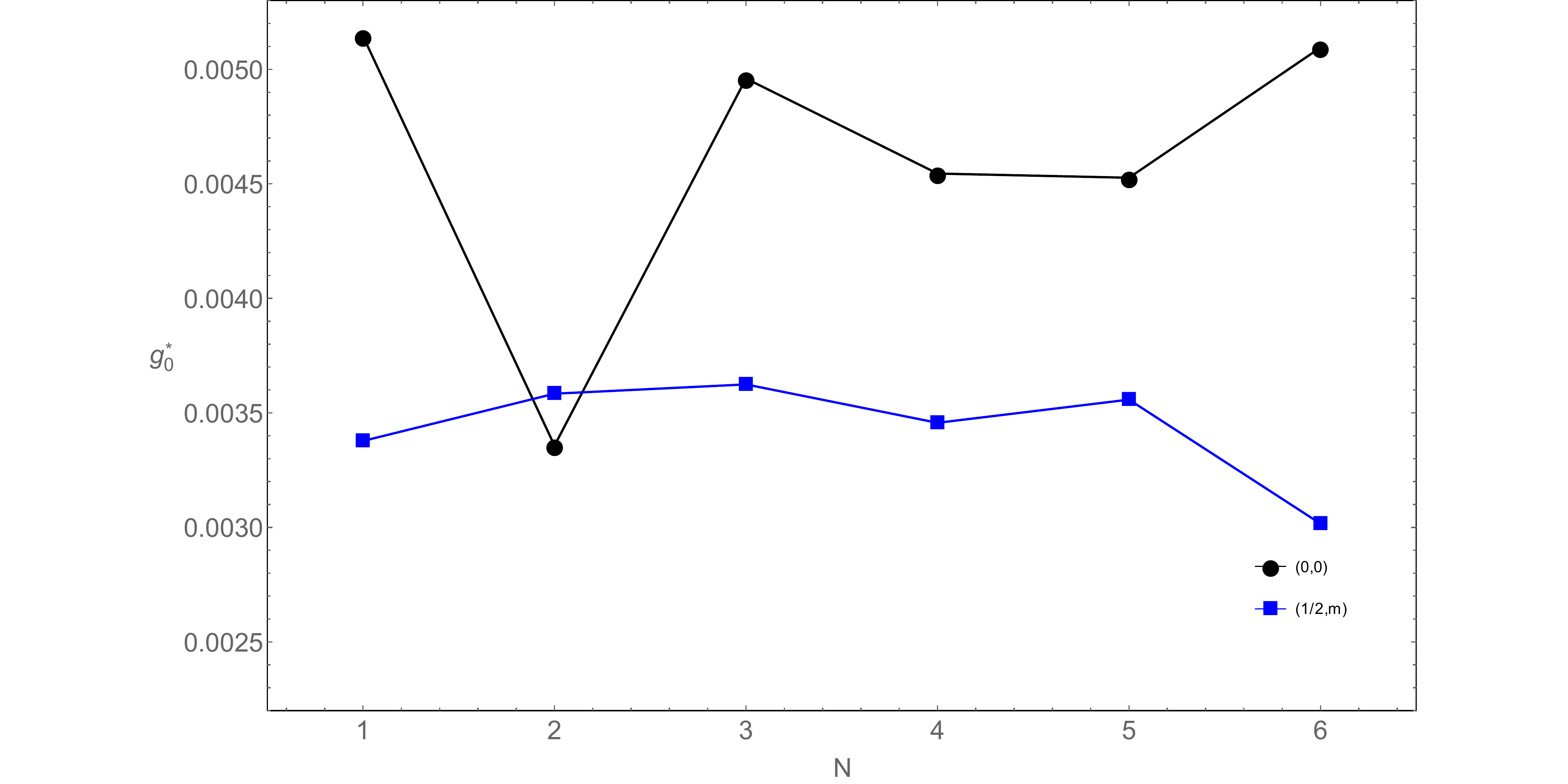}
  \end{subfigure}
  \begin{subfigure}[b]{0.49\linewidth}
    \includegraphics[width=\linewidth]{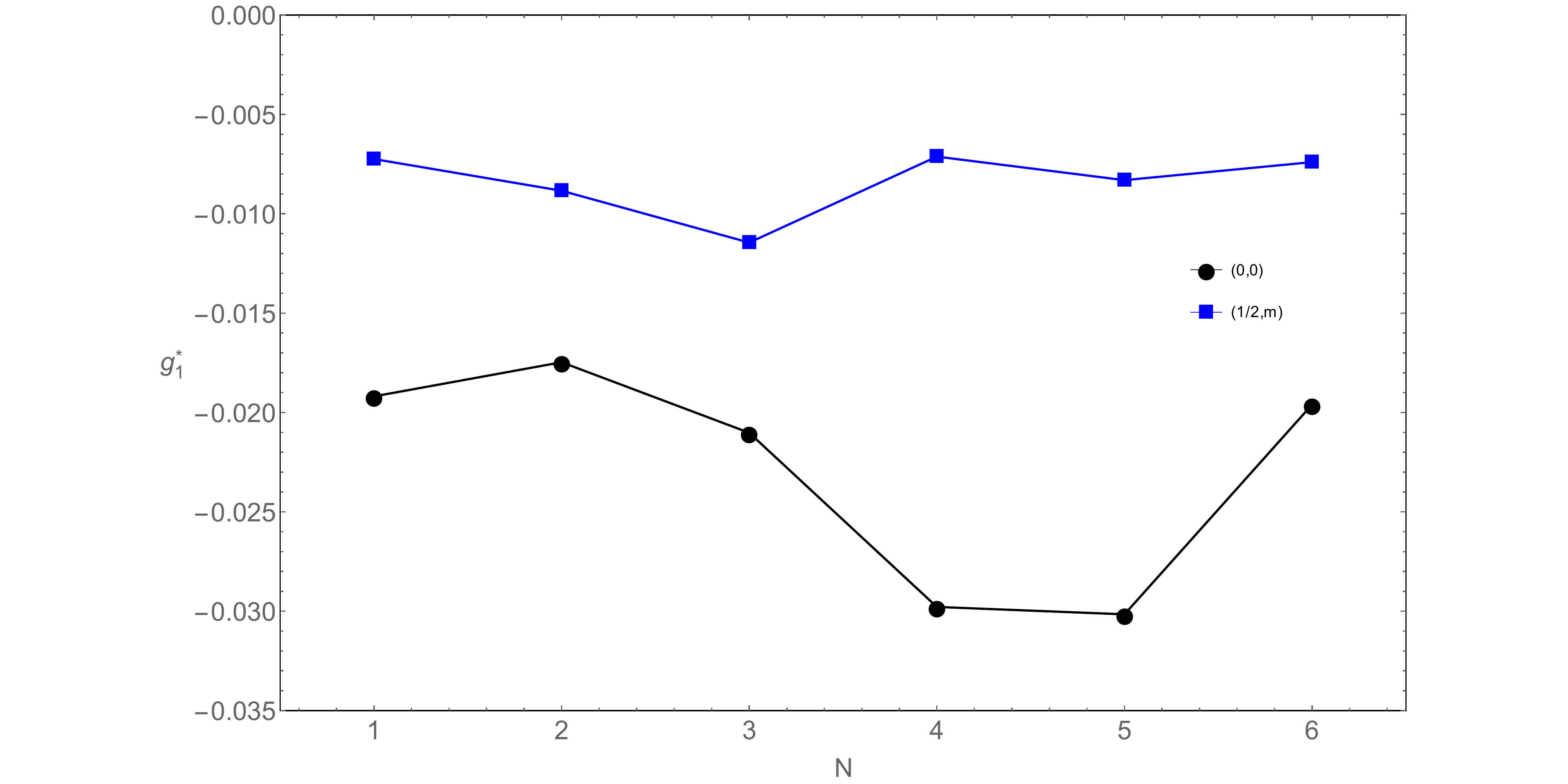}
  \end{subfigure}
    \begin{subfigure}[b]{0.49\linewidth}
    \includegraphics[width=\linewidth]{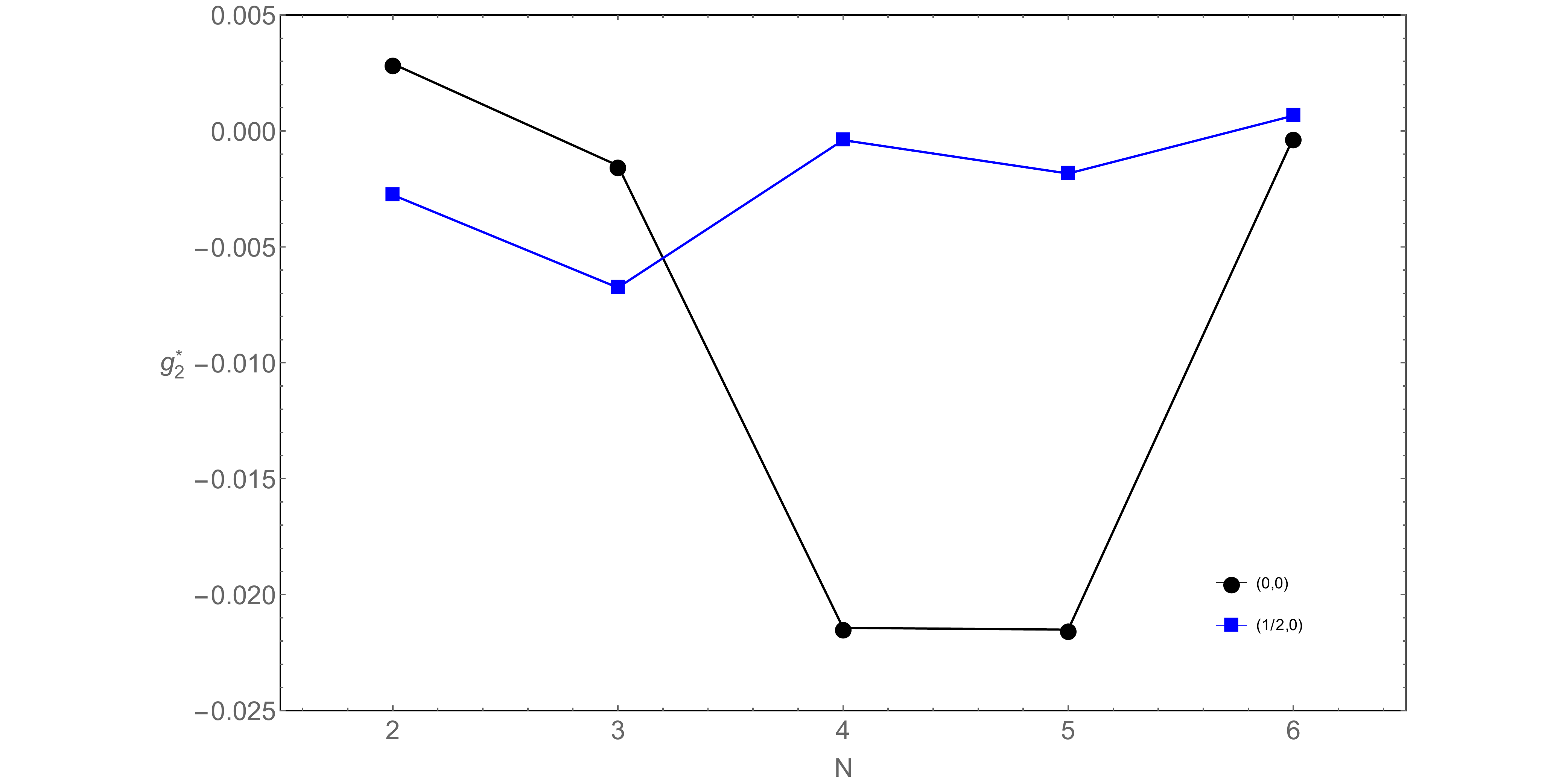}
  \end{subfigure}
      \begin{subfigure}[b]{0.49\linewidth}
    \includegraphics[width=\linewidth]{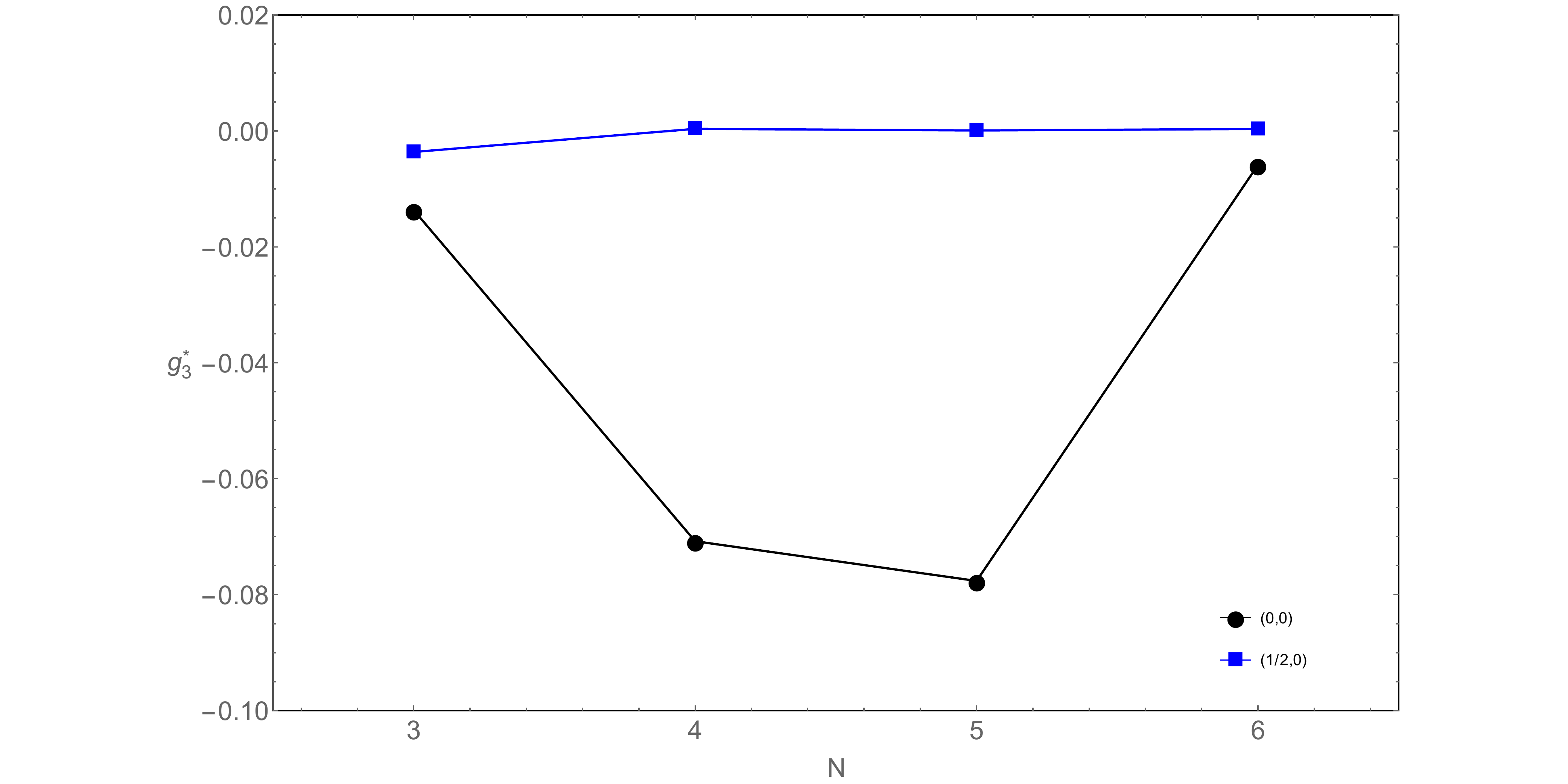}
  \end{subfigure}
  \caption{Values of the fixed point couplings $g^\ast_0$, $g^\ast_1$, $g^\ast_2$ and $g^\ast_3$ for different truncations in the physical gauge for $(\omega,m)=(0,0)$and $(1/2,0)$.}
  \label{truncdepbetainfty}
\end{figure}

It was reported for the linear \cite{Falls:2014tra} and exponential \cite{Ohta:2015fcu,Alkofer:2018fxj} parametrizations that it is possible to identify stability (for numerical convergence, one has to go beyond the $N=6$ truncation) of the fixed point values and critical exponents under truncation enlargement. However, for a fixed truncation, we have seen explicitly that the results are not quite stable against changes in the parameter $\omega$ apart from the Einstein-Hilbert truncation, a result already reported in \cite{Gies:2015tca}. In the next section we comment on this fact and point out some observations about such a dependence. However, before moving to that, we show how the fixed point values change under truncation enlargement in Fig.~\ref{truncdepbetainfty}.
We employ two different plot markers (dots and squares) and two different colors (black and blue) to distinguish between the linear (black dots) and the exponential (blue squares) splits. We note that, in the physical gauge, both parametrizations display a fixed point with three relevant directions.\footnote{This statement should be taken with a grain of salt: as one can see in Appendix \ref{numres}, the number of relevant directions in the exponential parametrization oscillates from three to two under truncation enlargements. For $N=5,6$ it has three relevant directions and in this sense we say the fixed point displays three relevant directions. Clearly, a definite answer requires further studies.}
It should be noticed that, typically, in a given truncation, one finds more than one fixed point with the ``minimum requirement" of positive Newton constant. On top of such requirement, one can demand some selection rules as, for instance, the stability of the fixed point under truncation enlargement or under changes of parametrization and/or gauge parameter values. Since we find different numbers of relevant directions for different choices of $\omega$, the selection of one single fixed point is not always obvious and one has to be careful in the selection of fixed points around the value of $\omega$ where the transition from three to two relevant directions occurs. Ultimately, sufficiently large truncations will ensure whether the selection was properly done (this is particularly important for finding fixed points in large truncations where one typically has to use some input value to look for the numerical solutions). Also, the amount of fixed points one finds in a given truncation varies with the prescription used. For instance, if one uses the prescription\footnote{This prescription, for $\beta=-\infty$ and $\omega=1/2$ has the curious feature that for $N=2$ truncation no fixed points are found. However, at $N=3$ order, one recovers one single fixed point with three relevant directions.} discussed in the beginning of Subsect.~\ref{beta0sub}, namely, the elimination of an extra mode from the transverse ghost sector, there is a reduction in the number of fixed point solutions per truncation, in general. This helps the selection of fixed points.

\section{Discussion} \label{Discussion} 
\subsection{Scheme dependence} \label{ParamVsRegulator}

In this work, we took the point of view that one is allowed to parametrize the quantum fluctuations in different forms, as represented in Eqs.~\eqref{ap11} and \eqref{ap111}. In particular, we have used this freedom to introduce two interpolating parameters $\omega$ and $m$. Then, we employ the Wetterich equation to compute beta functions associated with different couplings in a $f(R)$ truncation. In addition to the $f(R)$ truncation, we also made use of the so-called background approximation which consists in expanding the effective average action up to second order in the fluctuation $h_{\mu\nu}$. After the computation of the Hessian, one turns off the field $h_{\mu\nu}$. Therefore, in the background approximation, there are two pieces that are crucial for the computation: The effective average action projected on the background i.e. $\Gamma_k [\bar{g};h=0]$ and the quadratic terms on the fluctuation $h_{\mu\nu}$. All the other terms will not contribute to the flow equation when the approximation $h_{\mu\nu}=0$ is taken. As is clear from Eqs.~\eqref{ap11} and \eqref{ap111}, different choices of $(\omega,m)$ will certainly modify the quadratic part of $\Gamma_k$. Also, from the Hessians in Appendix~\ref{hessians&Cutoffs}, one sees that parametrization dependence always appears either as terms proportional to the equations of motion or as overall factors. This suggests that working with a background metric $\bg_{\mu\nu}$ which satisfies the equations of motion would remove parametrization dependence from the computations. For an off-shell background, different choices of parametrization lead to different background dependence of the regularized Hessian. This changes the right-hand side of the FRGE leading, typically, to different results in different parametrizations. In the background approximation one might change the background dependence of the regularized Hessian coming from different choices of parametrization by a suitable modification of the regulator. However, it is known that modifying background dependence of the regulator leads to non-universal results if no further constraints are imposed, see \cite{Litim:2002hj,Bridle:2013sra}.

Let us also comment on some other type of freedom in the implementation of the regulator which also brings in ``scheme dependence'' to the results. As discussed in Subsect.~\ref{betainftysub}, there are results in the literature that point to a fixed point in the exponential parametrization (in the physical gauge) with two relevant directions. In the present work, for those choices of parametrization and gauge, we obtain a fixed point with three relevant directions. The difference arises due to different regularization schemes. For comparison, we note that the Hessians reported in Appendix~\ref{hessians&Cutoffs} reduce to those given in \cite{Ohta:2015fcu}. Also, in \cite{Ohta:2015fcu}, the scalar sector $(\sigma,h)$ is rewritten in terms of $\chi$ and $s$ defined by
\begin{equation}
\chi = \frac{[(d-1)\bnabla^2+\br]\sigma -\beta h}{(d-1-\beta)\bnabla^2+\br}\qquad\mathrm{and}\qquad s = h-\bnabla^2\sigma\,.
\label{discussion1}
\end{equation}
Such a change of variables has trivial Jacobian. The pure gravitational part (namely, the one which comes from the $f(R)$ action and not from gauge-fixing terms) contains a diagonal term in $s$ which is expressed as
\begin{equation}
\frac{d-1}{4d}s\left[\frac{2(d-1)}{d}f^{\prime\prime}(\br)\left(-\bnabla^2-\frac{\br}{d-1}\right)+\frac{d-2}{d}f^\prime (\br)\right]\left(-\bnabla^2-\frac{\br}{d-1}\right)s\,.
\label{discussion2}
\end{equation}
In \cite{Ohta:2015fcu}, the second operator that appears in parentheses in eq.\eqref{discussion2} is canceled against the contribution that comes from the Jacobian due to the York decomposition. Consequently, the scalar contribution arising from the pure gravitational sector is encoded in the operator 
\begin{equation}
\frac{2(d-1)}{d}f^{\prime\prime}(\br)\left(-\bnabla^2-\frac{\br}{d-1}\right)+\frac{d-2}{d}f^\prime (\br)\,.
\label{discussion2a}
\end{equation}
Then, one can introduce the regulator for this operator and compute its contribution to the FRGE. Another procedure that leads to the same contribution is to perform the redefinition of the field $s$ by $\tilde{s}=\sqrt{-\bnabla^2-\br/(d-1)}s$. The scalar operator simplifies to \eqref{discussion2a} and the Jacobian generated by such a redefinition is canceled by the York decomposition Jacobian. This was employed in \cite{Alkofer:2018fxj}. They lead to the same results. In contrast, we simply regularize the full operator \eqref{discussion2} as well as the contributions from the Jacobian of the York decomposition. One can show that after regularization, these contributions do not cancel exactly and the resulting contribution to the FRGE differs from those reported in \cite{Ohta:2015fcu,Alkofer:2018fxj}. In particular, this leads to the difference in the number of relevant directions obtained in \cite{Ohta:2015fcu,Alkofer:2018fxj} and in the present paper. Also, since the results reported in \cite{Ohta:2015fcu,Alkofer:2018fxj} are stabler under truncation enlargement than those studied here, one might infer that they are less contaminated by truncation artifacts.

Finally, let us recall that in \cite{Eichhorn:2015bna}, the $f(R)$ truncation was analyzed in the unimodular setting with the exponential parametrization. A fixed point with two relevant directions is obtained. This is consistent with our findings, given that in the unimodular case, the cosmological constant is not a coupling (hence, one would have an extra relevant direction in the standard case). This also suggests that, in the background approximation, different schemes in the (field) basis one uses for the FRGE, the modes one subtracts or not from the flow and the choice of regulator can lead to different numbers of relevant directions, a known fact in the functional renormalization group literature. Therefore, if one wants to capture the effects of more sophisticated computations using the background approximation, a possible consistent way is to constrain such a freedom on the scheme used for the computations.

\subsection{Graviton propagator poles and parametrization dependence}

As we have seen in the previous section, the number of relevant directions changes under the variation of the parameters $\omega$ and $m$.
In order to understand $\omega$-dependence of the critical exponents, we investigate the structure of the beta functions.
To this end, we consider the $R^2$ truncation in the Landau gauge $\alpha\to 0$ with the choice $\beta=0$ and $m=0$.
In this case, the graviton propagator contains the following structures for the traceless-transverse sector and the trace mode sector, respectively: 
\begin{equation}
\frac{1}{(1-2\omega){\tilde g}_0+{\tilde g}_1}\qquad\mathrm{and}\qquad
\frac{1}{2(1+2\omega){\tilde g}_0+3{\tilde g}_1+18{\tilde g}_2}.
\label{graviton propagator}
\end{equation}
For a fixed point ${\tilde g}_0^*$, ${\tilde g}_1^*$, ${\tilde g}_2^*$,  the values $\omega=({\tilde g}_0^*+{\tilde g}_1^*)/2{\tilde g}_0^*$ and $(-2{\tilde g}_0^*-3{\tilde g}_1^*-18{\tilde g}_2^*)/4{\tilde g}_0^*$, become poles of \eqref{graviton propagator}, at which  the graviton fluctuations are enhanced. 
In such a case, the results of the critical exponents tends to be unstable, and then one has to avoid such choices of $\omega$.

\begin{figure}[t!]
  \centering
  \begin{subfigure}[b]{0.49\linewidth}
    \includegraphics[width=\linewidth]{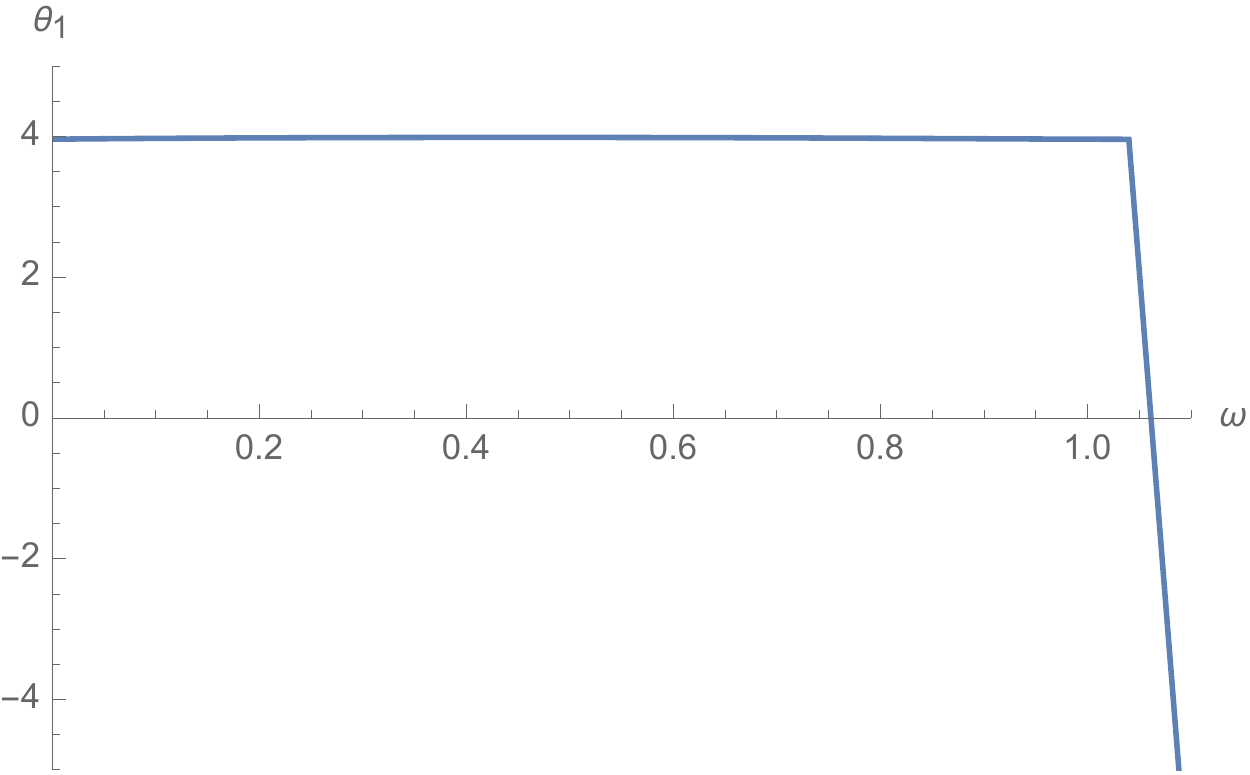}
  \end{subfigure}
  \begin{subfigure}[b]{0.49\linewidth}
    \includegraphics[width=\linewidth]{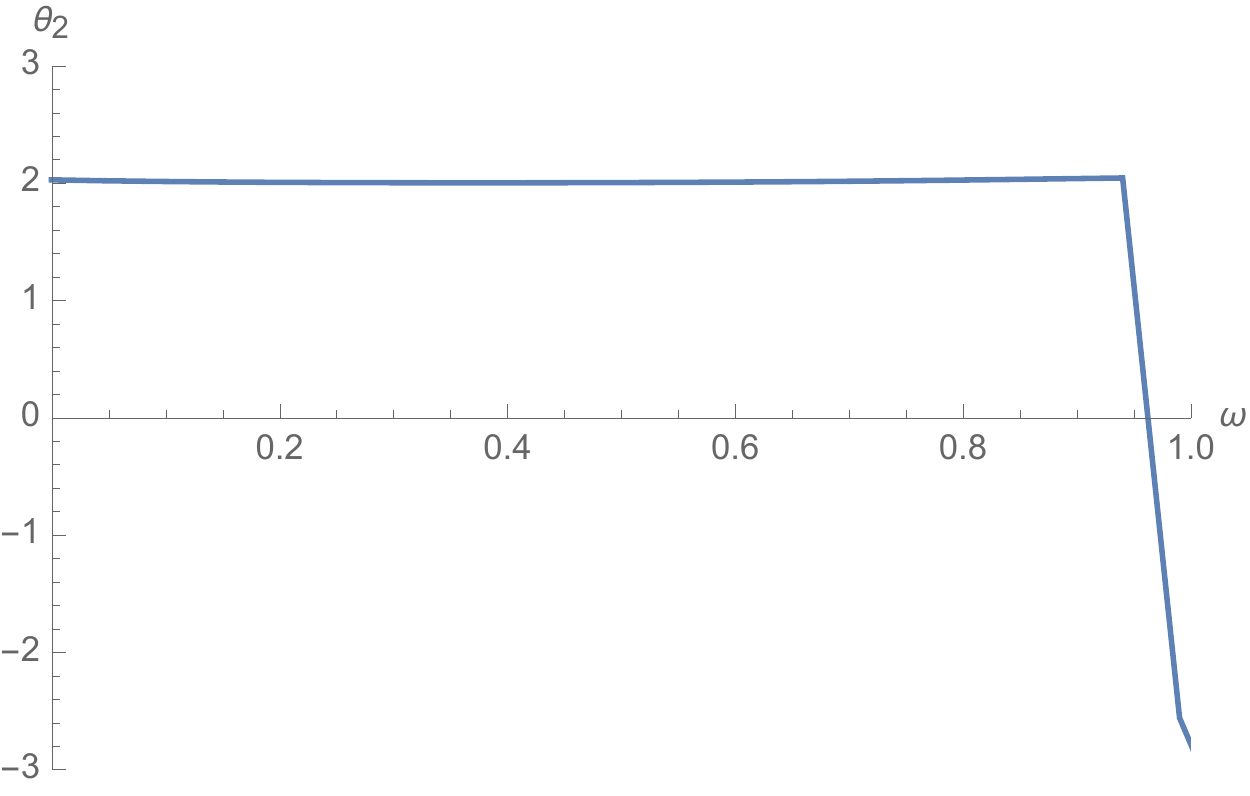}
  \end{subfigure}
    \begin{subfigure}[b]{0.49\linewidth}
    \includegraphics[width=\linewidth]{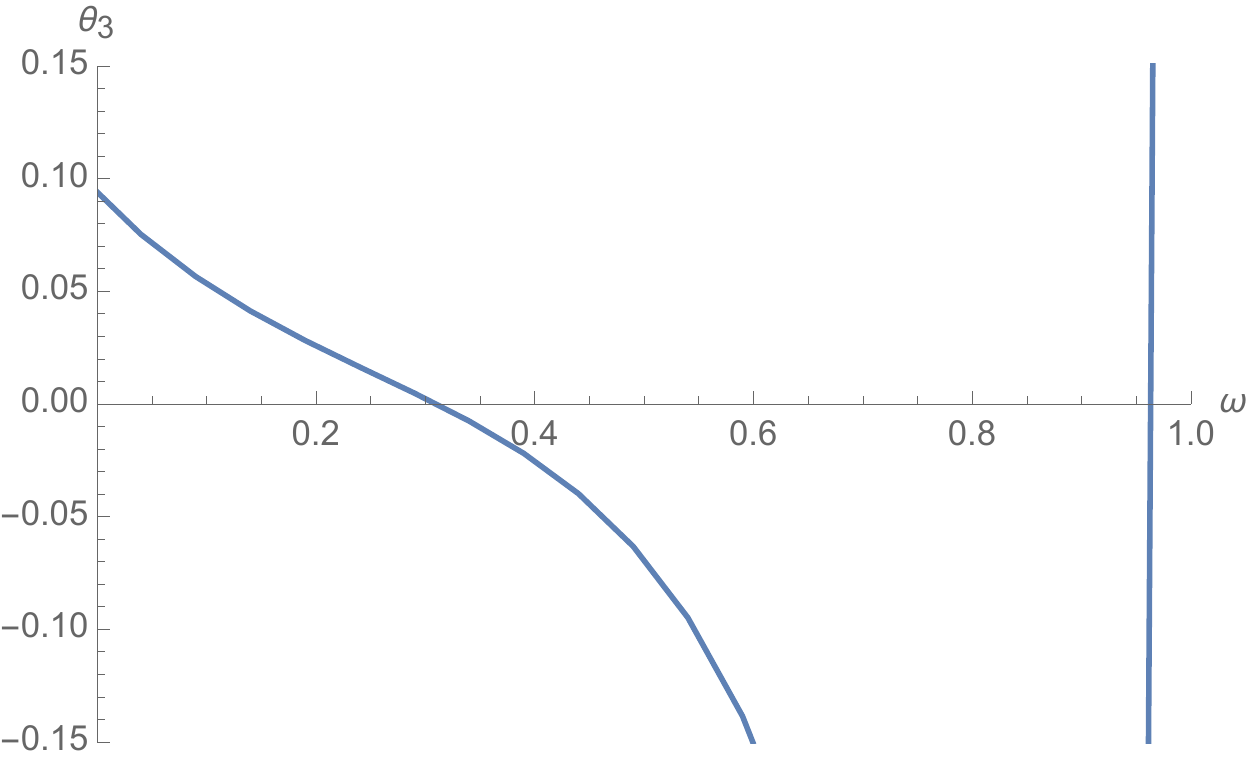}
  \end{subfigure}
  \caption{The $\omega$-dependence of the critical exponents in the $R^2$ truncation with $\alpha\to 0$, $\beta=0$ and $m=0$. The value of the fixed point is set to $g_0^*=1/3$, $g_1^*=-1$, $g_2^*=0.01$.}
  \label{critical exponent with omega}
\end{figure}

Next, let us see the explicit forms of the critical exponents within a certain approximation:
Since the absolute value of the $R^2$ coupling constant ${\tilde g}_2^*$ is smaller than those of the cosmological constant ${\tilde g}_0^*$ and the Newton constant $|{\tilde g}_1^*|$ as we can see from the fixed point analysis in the previous section, we neglect it.
Also, we assume that ${\tilde g}_0^*$ is smaller than $|{\tilde g}_1^*|$ - which is typically the case - and take into account terms up to order $\mathcal{O}(g_0)$.
Although the stability matrix is not diagonal, the main contributions to the critical exponents come from the diagonal part.
Then, assuming that $\theta_i\simeq \partial \beta_i/\partial g_i|_{g=g^*}$, we have
\al{
\theta_1&\simeq 4+\frac{17-26 \omega }{72 \pi ^2 \tilde g_1^*}+ \frac{-49+164 \omega -196 \omega ^2}{108 \pi ^2 \tilde g_1^*{}^2}g_0^*,\\
\theta_2&\simeq 2+\frac{597-1922 \omega+1568 \omega ^2 }{1728\pi^2 g_1^*{}^2}g_0^*,\\
\theta_3&\simeq \frac{-49+79 \omega +24 \omega ^2}{144 \pi ^2 g _1^*}+\frac{5791-19438 \omega+13560 \omega ^2 -4800 \omega ^3}{4320 \pi ^2 g_1^*{}^2}g _0^* .\label{critexp3}
}
We see that due to the dominance of the canonical scaling dimensions, $\theta_1$ and $\theta_2$ would be always positive under variation of $\omega$.
On the other hand, the sign of $\theta_3$ depends on the value of $\omega$.
The first term on the right-hand side of \eqref{critexp3} vanishes at $\omega^*=0.533716$.
With a negative value of $g_1^*$, it could become positive for $\omega$ smaller than $\omega^*$, whereas for a larger
$\omega$, the sign of $\theta_3$ could become negative.
In order to determine the behavior more precisely, we evaluate the critical exponents numerically (without resorting to the approximations discussed above).
In Fig.~\ref{critical exponent with omega}, we show the $\omega$-dependence of the critical exponents.
As illustrative values of the fixed point, we use $g_0^*=1/3$, $g_1^*=-1$, $g_2^*=0.01$ for which $\omega=1.615$ leads to a singularity in the graviton propagator \eqref{graviton propagator}.
We see that $\theta_1$ and $\theta_2$ are stable under variations of $\omega$ except around the singular point.
In contrast, $\theta_3$ strongly depends on $\omega$ and its sign changes at $\omega\simeq 0.3$.
From this fact, we can understand the difference of the number of the relevant directions between the linear split ($\omega=0$) and the exponential one ($\omega=1/2$). It is precisely this type of mechanism that generates different counting on the number of relevant directions.

\section{Conclusions} \label{Conclusions}
In this work, we discussed how different choices of the parametrization for the quantum fluctuations affect the fixed point structure in $f(R)$ truncations. In particular, this study was carried out in the background approximation for the effective average action. The quantum fluctuations were parametrized by two free parameters $(\omega,m)$. All the computations were performed in the Landau gauge $\alpha=0$. For comparison, we analyzed parametrization dependence for two choices of the gauge parameter $\beta$: $0$ and $-\infty$. Qualitatively, the results are similar for both choices  of $\beta$: In the Einstein-Hilbert truncation, the fixed point structure is relatively stable for different values of $\omega$. Starting from the $N=2$ truncation, setting $m=0$ for simplicity, one sees that for $\omega \in [0,1]$ there are two different types of fixed points, namely, one with three relevant directions and the other with two relevant directions. Typically, the transition from one fixed point to the other occurs in the vicinity of the exponential parametrization choice, namely, $\omega=1/2$. For $\beta=0$, this transition happens for $\omega < 1/2$ leading to a fixed point with two relevant directions in the exponential parametrization while for $\beta=-\infty$, the transition takes place for $\omega > 1/2$, which entails a fixed point with three relevant directions in the exponential parametrization. 
We expect the same behavior in other higher derivative truncations such as the $R^2+R_{\mu\nu}^2$-type truncation since the Ricci tensor-squared is a marginal operator in four-dimensional spacetime.
In this connection, we note that three relevant directions are observed in \cite{Benedetti:2009rx,Benedetti:2009gn} whereas four relevant directions are found when a different prescription for the cutoff scheme (as well as the treatment of zero modes) is employed~\cite{Hamada:2017rvn}.

We have found that in the background approximation, changes in the parametrization, i.e. in the background dependence of the regularized Hessian, lead to modifications in the number of relevant directions at the UV fixed point. This result is compatible with those reported in \cite{Alkofer:2018fxj}, where modifications on the endomorphisms present in the regulator can lead to different numbers of relevant directions of the UV fixed point for gravity-matter systems. In our perspective, this is a limitation of the background approximation and further constraints should be imposed on the calculations (e.g. on the choice of regulators, compatibility with Ward identities and so on). See also \cite{Lippoldt:2018wvi} where a new version of the effective average action is put forward leading to novel contributions in the background approximation. On the other hand, one might still argue that albeit being not able to give an unambiguous result for the number of relevant directions, one still finds a nontrivial fixed point with a critical surface dimensionality lower than the truncated theory space dimension. Hence, even in the simple background approximation, it is possible to find a suitable candidate for a nontrivial UV fixed point. Ultimately, in order to obtain more precise informations about the fixed point (namely, how many relevant directions are associated with it), more sophisticated approximations should be employed. 

Of course, the issue whether the fixed point features two, three or other finite number of relevant directions is crucial. In particular, relevant directions count the number of free parameters in the theory that should be fixed, ultimately, by experiments. As a consequence, depending on the number of relevant directions, higher curvature couplings, for example, can be a prediction or not (namely, are free parameters) of asymptotically safe quantum gravity. Therefore, establishing precisely the number of relevant directions associated with the nontrivial fixed point is not simply a formal exercise. The present work suggests that this answer should be given in schemes that go beyond the background approximation or in improved versions of it. 
\section*{Acknowledgments}
We acknowledge Nat\'alia Alkofer, Astrid Eichhorn, Kevin Falls, Stefan Lippoldt, Jan Pawlowski and Roberto Percacci for helpful discussions. We also thanks to C.E. Cede\~{n}o for helpful suggestion on the numerical computations performed in this work.
The research of GPB is supported by CNPq, No. 142049/2016-6. The work of NO was supported in part by the Grant-in-Aid for Scientific Research Fund of the JSPS (C) No. 16K05331.
ADP acknowledges funding by the DFG, Grant Ei/1037-1. The work of MY is supported by the DFG Collaborative Research Centre SFB 1225 (ISOQUANT). 

\appendix

\section{Quantum Fluctuation Parametrization }\label{ap1}
In this work we consider the split of the metric $g_{\mu\nu}$ of the form \cite{Ohta:2016npm,Ohta:2016jvw},
\begin{equation}
g_{\mu\nu} = \bg_{\mu\nu}+\delta g_{\mu\nu}\equiv \bg_{\mu\nu}+\delta g^{(1)}_{\mu\nu}+\delta g^{(2)}_{\mu\nu},
\label{ap11}
\end{equation}
with
\begin{eqnarray}
\delta g^{(1)}_{\mu\nu} &=& h_{\mu\nu}+ m \bg_{\mu\nu} h,\nonumber\\
\delta g^{(2)}_{\mu\nu} &=& \omega h_{\mu\rho}h^\rho_\nu + m h h_{\mu\nu}
 +m\left(\omega-\frac{1}{2}\right) \bg_{\mu\nu} h_{\alpha\beta}^2 + \frac{1}{2} m^2 \bg_{\mu\nu} h^2 ,
\label{ap111} 
\end{eqnarray}
with $m$ and $\omega$ free parameters and the upper index $i$ in $\delta g^{(i)}$ counts the number of $h$ fields. Also, we define $\delta g^{(i)}\equiv \bg^{\mu\nu}\delta g^{(i)}_{\mu\nu}$.

Expanding the Ricci scalar $R$ in powers of $h$, one obtains
\begin{equation}
R = \br + R^{(1)} + R^{(2)} + \ldots,
\label{Ricciexp}
\end{equation}
with
\begin{equation}
R^{(1)} = \bnabla_\mu \bnabla_\nu \delta g^{(1)\mu\nu} - \bnabla^2\delta g^{(1)} - \delta g^{(1)\mu\nu} \br_{\mu\nu},
\label{Ricci_1}	
\end{equation}
and
\begin{eqnarray}
R^{(2)} &=& \delta g^{(1)\mu\nu}\bnabla_\mu \bnabla_\nu\delta g^{(1)} - 2\delta g^{(1)\mu\nu} \bnabla_\nu \bnabla_\alpha \delta g^{(1)}_{\mu}{}^{\alpha} + \delta g^{(1)\mu\nu} \bnabla^2 \delta g^{(1)}_{\mu\nu}  + \frac{3}{4}\bnabla_\alpha \delta g^{(1)}_{\mu\nu} \bnabla^\alpha \delta g^{(1)\mu\nu}  \nonumber \\
&-&\frac{1}{4} \bnabla_\nu \delta g^{(1)} \bnabla^\nu\delta g^{(1)}- \bnabla_\mu \delta g^{(1)\mu\nu} \bnabla_\alpha \delta g^{(1)}_{\nu}{}^{\alpha} + \bnabla^\nu \delta g^{(1)} \bnabla^\alpha \delta g^{(1)}_{\nu\alpha} - \frac{1}{2} \bnabla_\nu \delta g^{(1)}_{\mu\alpha} \bnabla^\alpha \delta g^{(1)\mu\nu} \nonumber\\
&+& \bnabla_\mu \bnabla_\nu \delta g^{(2)\mu\nu} -  \bnabla^2 \delta g^{(2)} + \delta g^{(1)\mu\nu} \delta g^{(1)\alpha\beta} \br_{\mu\alpha\nu\beta} - \delta g^{(2)\mu\nu} \br_{\mu\nu} .
\label{Ricci_2}
\end{eqnarray}

\section{Hessians}\label{hessians&Cutoffs}

In this appendix we collect all the hessians that enter the flow equation \eqref{fe4}. Notice that the following expressions contain contributions coming from the gravitational action, gauge-fixing term \eqref{gf1}, Faddeev-Popov ghosts and auxiliary fields.  

\begin{equation}
\Gamma^{\mu\nu\alpha\beta}_{\mathrm{TT}} = -\frac{1}{2}\left[\left(-\bnabla^2+\frac{2}{d}\frac{\br}{d-1}\right)f^\prime_k (\bar{R})-(2\omega-1)(1+dm)\left(f_k (\br)-\frac{2}{d}\br f^{\prime}_k (\br)\right)\right]  \mathbb{1}^{\mu\nu,\alpha\beta} \,,
\label{hess1}
\end{equation}

\begin{equation}
\Gamma^{\mu\nu}_{\xi\xi} = \left(-\bnabla^2-\frac{\br}{d}\right)\left[(2\omega-1)(1+dm)\left(f_k (\br)-\frac{2}{d}\br f^{\prime}_k (\br)\right)+\frac{Z_N}{\alpha}\left(-\bnabla^2-\frac{\br}{d}\right)\right]\bg^{\mu\nu}\,,
\label{hess2}
\end{equation}

\begin{eqnarray}
\Gamma_{\sigma\sigma}&=&\frac{d-1}{2d}\left[\frac{2}{d}(d-1)f^{\prime\prime}_k (\br)\left(\bnabla^2+\frac{\br}{d-1}\right)\bnabla^2-\frac{d-2}{d}f^{\prime}_k (\br)\bnabla^2+(2\omega-1)(1+dm)\left(f_k (\br)-\frac{2}{d}\br f^{\prime}_k (\br)\right)\right.\nonumber\\
&-&\left.\frac{2 Z_N}{\alpha}\frac{d-1}{d}\left(\bnabla^2+\frac{\br}{d-1}\right)\right]\left(\bnabla^2 + \frac{\br}{d-1}\right)\bnabla^2\,,
\label{hess3}
\end{eqnarray}

\begin{eqnarray}
\Gamma_{hh}&=&\left(\frac{1+dm}{d}\right)^2\left[(d-1)^2 f^{\prime\prime}_k (\br)\left(-\bnabla^2-\frac{\br}{d-1}\right)^2+\frac{(d-1)(d-2)}{2}f^{\prime}_k (\br)\left(-\bnabla^2-\frac{\br}{d-1}\right)\right.\nonumber\\
&+&\left.\frac{d}{4}\left(\frac{2(2\omega-1)}{1+dm}+d\right)\left(f_k (\br)-\frac{2}{d}\br f^{\prime}_k (\br)\right)-\frac{\beta^2 Z_N}{\alpha}\bnabla^2\right]\,,
\label{hess4}
\end{eqnarray}

\begin{equation}
\Gamma_{\sigma h} = \Gamma_{h\sigma} = \frac{(d-1)(1+dm)}{d^2}\left[(d-1)f^{\prime\prime}_k (\br)\left(-\bnabla^2-\frac{\br}{d-1}\right)+\frac{1}{2}(d-2)f^{\prime}_k (\br)+\frac{\beta Z_N}{\alpha}\right]\left(\bnabla^2 + \frac{\br}{d-1}\right)\bnabla^2\,,
\label{hess5}
\end{equation}

\begin{equation}
\Gamma^{\mu\nu}_{C^\mathrm{T} C^\mathrm{T}} = \bg^{\mu\nu}\left(\bnabla^2+\frac{\br}{d}\right)\,,
\label{hess6}
\end{equation}

\begin{equation}
\Gamma_{C C} = 2\frac{1+\beta-d}{d}\left(\bnabla^2-\frac{\br}{1+\beta-d}\right)\bnabla^2\,,
\label{hess7}
\end{equation}

\begin{equation}
\Gamma^{\mu\nu}_{\chi\chi} = \bg^{\mu\nu}\left(\bnabla^2+\frac{\br}{d}\right)\,,
\label{hess8}
\end{equation}

\begin{equation}
\Gamma_{\theta\theta} = \left(\bnabla^2+\frac{\br}{d-1}\right)\bnabla^2\,,
\label{hess9}
\end{equation}

\begin{equation}
\Gamma_{\bar{\phi}\phi} = -\bnabla^2\,,
\label{hess10}
\end{equation}

\begin{equation}
\Gamma^{\mu\nu}_{\zeta\zeta} = \bg^{\mu\nu}\left(\bnabla^2+\frac{\br}{d}\right)\,,
\label{hess11}
\end{equation}

\begin{equation}
\Gamma_{\psi\psi}=-\left(\bnabla^2+\frac{\br}{d-1}\right)\bnabla^2\,.
\label{hess12}
\end{equation}
with $\mathbb{1}^{\mu\nu,\alpha\beta} = (1/2) (\bg^{\mu\alpha}\bg^{\nu\beta}+\bg^{\mu\beta}\bg^{\nu\alpha})$.
It should be emphasized here that in the expressions of the Hessians shown above, we have factorized the term $f_k (\br)-(2/d) \br f^{\prime}_k (\br)$ whose zero corresponds the equation of motion of \eqref{fr1} projected on the background. One sees that the parametrization dependence on $(\omega,m)$ always appears as an overall factor or proportional to the equation of motion.
Then, if one imposes the equation of motion, i.e., the on-shell condition, one easily sees that the parametrization dependence drops. For simplicity, we don't write the primes which take into account spurious modes for each operator written above.

\section{Heat Kernel} \label{heatkernel}

Along this paper we compute functional traces with the following structure
\begin{align}
\Tr_{(s)}\big[ W(-\CDbar^2) \mathbb{1}_{(s)} \big],
\end{align}
where $W(-\CDbar^2)$ denotes some function of the Bochner-Laplacian operator and $\mathbb{1}_{(s)}$ represents the identity operator acting on the space of scalars ($s=0$), transverse vectors ($s=1$) and transverse-traceless symmetric tensors ($s=2$). Usually, this computation can be performed in terms of the Heat Kernel expansion \cite{Buchbinder:1992rb,Avramidi:2000bm,Percacci:2017fkn}, namely
\begin{align}
\Tr_{(s)}\big[ W(-\CDbar^2) \mathbb{1}_{(s)} \big] = \frac{1}{(4\pi)^{d/2}}\sum_{n=0}^\infty \int d^dx\, \sqrt{\bar{g}} \,\, Q_{d/2-n}[W] \,\, \textmd{tr}\big[ b_{2n}(-\CDbar^2) \mathbb{1}_{(s)}  \big],
\end{align}
where $b_{2n}(-\CDbar^2)$ denote the (non-integrated) Heat Kernel coefficients associated with the operator $-\CDbar^2$. In addition, the so-called $Q$-functional can be written in terms of the following expression (for arbitrary real $n$)
\begin{align}
Q_n[W] = \frac{(-1)^k}{\Gamma(n+k)} \int_0^\infty dz \, z^{n+k-1} \,\frac{d^k W(z)}{dz^k} ,
\end{align}
where $k$ denotes some (arbitrary) positive integer satisfying the following restriction $n+k>0$. 

Furthermore, we cast the values of $\mathrm{tr}\big[b_\textbf{n}(-\CDbar^2)\mathbb{1}_{(\textbf{s})})\big]$, i.e., the heat kernel coefficients for each spin sector $s$ and for each derivative order $n$, evaluated over a sphere $S^4$ in the Table \ref{table:HKCOEFFS}.

\begin{table}[htb]
\begin{tabular}{|c||*{7}{c|}}
\hline
\backslashbox{$\textbf{s}$}{$\textbf{n}$}
&\makebox[3em]{$0$}&\makebox[3em]{$2$}&\makebox[3em]{$4$}
&\makebox[3em]{$6$}&\makebox[3em]{$8$}&\makebox[3em]{$10$}&\makebox[3em]{$12$}\\
\hline\hline
&&&&&&&\\
 $0$ & $1$ & $\frac{1}{6}\bar{R}$ & $\frac{29}{2160}\bar{R}^2$ & $\frac{37}{54432}\bar{R}^3$ & $\frac{149}{6531840}\bar{R}^4$ & $\frac{179}{431101440}\bar{R}^5$ & $-\frac{1387}{201755473920}\bar{R}^6$ \\
&&&&&&&\\
\hline
&&&&&&&\\
 $1$ & $3$ & $\frac{1}{4}\bar{R}$ & $-\frac{7}{1440}\bar{R}^2$ & $-\frac{541}{362880}\bar{R}^3$ & $-\frac{157}{2488320} \bar{R}^4$ & -- & -- \\
&&&&&&&\\
\hline
&&&&&&&\\
 $2$ &  5 & $-\frac{5}{6}\bar{R}$ & $-\frac{1}{432}\bar{R}^2$ & $\frac{311}{54432}\bar{R}^3$ & $\frac{109}{1306368}\bar{R}^4$ & -- & -- \\
&&&&&&&\\
\hline
\end{tabular}
\caption{Heat kernel coefficients on $S^4$.}
\label{table:HKCOEFFS}
\end{table}

Since we have performed traces over differential constrained fields associated with the York decomposition, we have to remove some spurious modes which do not contribute to the l.h.s. of the FRG equation. As usual this question was indicated with the inclusion of primes over the trace operation. As a matter of fact, the ``primed'' trace can be computed in terms of the usual trace by means of the following expression
\begin{align}
\Tr^{\prime \cdots \prime}_{(s)}\big[ W(-\CDbar^2) \mathbb{1}_{(s)} \big] = \Tr_{(s)}\big[ W(-\CDbar^2) \mathbb{1}_{(s)} \big] - \sum_{l \in M} D_l(d,s) W(\lambda_l(d,s)),
\end{align}
where $M=\{0,1,2,\cdots,m-1\}$ for scalar fields, $M=\{1,2,\cdots,m\}$ for the case of vectors and so on ($m$ denotes the number of spurious modes). In addition, $\lambda_l(d,s)$ represents the $l$-th eigenvalue of the $d$-dimensional Bochner-Laplacian acting in space of spin-$s$ fields and $D_l(d,s)$ denotes the degree of degeneracy associated with $\lambda_l(d,s)$. Below we present some expressions for $\lambda_l$ and $D_l$ (evaluated over the sphere $S^4$) which has been used along this work
\begin{align}
\lambda_l(4,s) = \frac{(l+3)\,l-s}{12} \,\bar{R},
\end{align}
\begin{subequations}
	\begin{align}
	D_l(4,0) = \frac{(2 l+3) \,(l+2)!}{6\, l!},
	\end{align}
    \begin{align}
    D_l(4,1) = \frac{1}{2} l (l+3) (2 l+3).
    \end{align}
\end{subequations}


\section{Linear and exponential parametrization: numerical results for fixed points and critical exponents}\label{numres}

\subsection{Collection of the numerical results taking $\beta\to 0$ according to the order of truncation $N$.}

\begin{table}[htb]\scriptsize
\begin{tabular}{|c|c|c|c|c|c|c|c|c|c|c|}\hline\hline
\makebox[3em]{$(\omega,m)$} & \makebox[3em]{N} & \makebox[3em]{$\tilde{\Lambda}^\ast$} & \makebox[3em]{$\tilde{G}^\ast$} & \makebox[3em]{$g^\ast_0$} & \makebox[3em]{$g^\ast_1$} & \makebox[3em]{$g^\ast_2$}
& \makebox[3em]{$g^\ast_3$} & \makebox[3em]{$g^\ast_4$} & \makebox[3em]{$g^\ast_5$} & \makebox[3em]{$g^\ast_6$}\\\hline
 & 1 & 0.1293 & 0.9842 & 5.2261 & -20.2143 & -- & -- & -- & -- & -- \\\cline{2-11}
 & 2 & 0.1204 & 1.4532 & 3.2978 & -13.6899 & 1.6139 & -- & -- & - & -- \\\cline{2-11}
 & 3 &0.1322 & 1.0155 & 5.1789 & -19.5905 & 0.7163 & -7.3666 & -- & -- & --\\\cline{2-11}
$(0,0)$ & 4 & 0.1250 & 0.9764 & 5.0953 & -20.3749 & 0.3310 & -7.9945 & -4.9547 & -- & --\\\cline{2-11}
 & 5 & 0.1253 & 0.9780 & 5.0994 & -20.3415 & 0.3466 & -7.7524 & -4.7746 & -0.4112 & -- \\\cline{2-11}
 & 6 & 0.1220 & 0.9580 & 5.0672 & -20.7662 & 0.0653 & -8.4164 & -6.9364 & -0.8994 & 3.2256 \\\hline\hline
 & 1 & $0.2130$ & $1.8611$ & $4.5543$ & $-10.6898$ & -- & -- & -- & -- & --\\\cline{2-11}
 & 2 & 0.2117 & 1.8502 & 4.5530 & -10.7528 & 0.0166 & -- & -- & -- & --\\\cline{2-11}
 & 3 & 0.7563 & 9.062 & 3.3199 & -2.1948 & 2.2341 & -0.7894 & -- & -- & --\\\cline{2-11}
$\left(\frac{1}{2},0\right)$ & 3 & 0.1871 & 1.5315 & 4.8604 & -12.9903 & 0.5985 & 0.1587 & -- & -- & -- \\\cline{2-11}
 & 4 & 0.2794 & 3.493 & 3.1828 & -5.6949 & 1.4525 & 0.48974 & 0.0855  & -- & --\\\cline{2-11}
& 4 & 0.1694 & 1.3755 & 4.9012 & -14.4638 & 0.9469 & 0.2670 & 0.0328  & -- & --\\\cline{2-11}
& 5 & 0.4335 & 5.109 & 3.3766 & -3.8943 & 2.5626 & -0.2955 & 0.1403 & 0.0378 & --\\\cline{2-11}

& 5 & 0.1491 & 1.3202 & 4.4943 & -15.0692 & 1.1932 & 0.3310 & 0.0585 & 8.8872$\times 10^{-3}$ & --\\\cline{2-11}
& 6 & 0.361 & 4.267 & 3.3647 & -4.6620 & 2.3788 & -0.00786 & -0.1489 & -0.0442 & -0.00662 \\\cline{2-11}
& 6 & 0.1082 & 1.3839 & 3.1108 & -14.3753 & 1.4132 & 0.3569 & 0.0723 & 0.0164 & 2.9223$\times 10^{-3}$ \\\hline\hline
\end{tabular}
\caption{Collection of the fixed points and the dimensionless parameters $\tilde{G}^\ast$ and $\tilde{\Lambda}^\ast$ for $\beta\to 0$ related to truncations from $R$ to $R^6$ in the linear and exponential parametrization. All dimensionless couplings $g^\ast$ have been multiplied by $10^3$.}
\end{table}

\begin{table}[htb]\scriptsize
\begin{tabular}{|c|c|c|c|c|c|c|c|c|}\hline\hline
\makebox[3em]{$(\omega,m)$} & \makebox[3em]{N} & \makebox[3em]{$\theta_1$} & \makebox[3em]{$\theta_2$} & \makebox[3em]{$\theta_3$}
& \makebox[3em]{$\theta_4$} & \makebox[3em]{$\theta_5$} & \makebox[3em]{$\theta_6$} & \makebox[3em]{$\theta_7$}\\\hline
 & 1 & 2.3824-2.1682i & 2.3824+2.1682i & -- & -- & -- & -- & \\\cline{2-9}
 & 2 & 1.7460-1.9667i & 1.7460+1.9667i & 25.3713 & -- & -- & -- &  \\\cline{2-9}
 & 3 & 2.8804-2.3186i & 2.8804+2.3186i & 2.0680 & -3.5466 & -- & -- & -- \\\cline{2-9}
$(0,0)$ & 4 & 3.0312-2.5737i & 3.0312+2.5737i & 1.6344 & -3.0192 & -5.2476 & -- & -- \\\cline{2-9}
 & 5 & 2.5393-2.7234i & 2.5393+2.7234i & 1.8536 & -3.9829-5.5518i & -3.9829+5.5518i & -4.2034 & -- \\\cline{2-9}
 & 6 & 2.5911-2.5754i & 2.5911+2.5754i & 1.3801 & -4.2009-5.1360i & -4.2009+5.1360i & -3.9152 & -8.4705 \\\hline\hline
  & 1 & 1.9651-2.4602i & 1.9651+2.4602i & -- & -- & -- & -- & --\\\cline{2-9}
 & 2 & 2.3541-1.3334i & 2.3541+1.3334i & -14.0103 & -- & -- & -- & --\\\cline{2-9}
 & 3 & 4.3792 & 0.9969 & -2.1823 & -7.5179 & -- & -- & --\\\cline{2-9}
$\left(\frac{1}{2},0\right)$ & 3 & 1.9970-1.8763i & 1.9970+1.8763i & -23.6101 & -42.9915 & -- & -- & -- \\\cline{2-9}
& 4 & 3.8619+1.1964i & 3.8619-1.1964i& 2.3719 & -2.3054 & -7.6558 & -- & -- \\\cline{2-9}
& 4 & 1.4877-2.1637i & 1.4877+2.1637i & -22.2537 & -1.6897-52.4913i  & -1.6897+52.4913i & -- & -- \\\cline{2-9}
 & 5 & 4.5842 & 2.5658 & -2.6039+1.5312i & -2.6039-1.5312i & -5.9929 & -16.325 & -- \\\cline{2-9}

& 5 & 0.8547-2.0726i & 0.8547+2.0726i & 30.6260-23.1427i & 30.6260+23.1427i & -21.3017 & -105.9410 & -- \\\cline{2-9}
& 6 & 5.508+2.849i & 5.508-2.849i & -2.612+2.644i & -2.612-2.644i & -7.3045 & - 15.923 & -34.678 \\\cline{2-9}
& 6 & 0.6977-1.5086i & 0.6977+1.5086i & 21.3403 & 37.6352 & -15.6291-76.8264i & -15.6291+76.8264i & -17.9181
\\\hline\hline
\end{tabular}
\caption{Collection of the critical exponents' for $\beta\to 0$ related to linear and exponential parametrization corresponding to truncations from $R$ to $R^6$.}
\end{table}
\newpage

\subsection{Collection of numerical results taking $\beta\to-\infty$ according to the order of truncation $N$.}

\begin{table}[htb]\scriptsize
\begin{tabular}{|c|c|c|c|c|c|c|c|c|c|c|}\hline\hline
\makebox[3em]{$(\omega,m)$} & \makebox[3em]{N} & \makebox[3em]{$\tilde{G}^\ast$} & \makebox[3em]{$\tilde{\Lambda}^\ast$} & \makebox[3em]{$g^\ast_0$} & \makebox[3em]{$g^\ast_1$} & \makebox[3em]{$g^\ast_2$}
& \makebox[3em]{$g^\ast_3$} & \makebox[3em]{$g^\ast_4$} & \makebox[3em]{$g^\ast_5$} & \makebox[3em]{$g^\ast_6$}\\\hline
 & 1 & 0.1341 & 1.0368 & 5.1447 & -19.1892 & -- & -- & -- & -- & --  \\\cline{2-11}
 & 2 & 0.0961 & 1.1387 & 3.3570, & -17.4716 & 2.8955 & -- & -- & -- & -- \\\cline{2-11}
$(0,0)$ & 3 &  0.1180 & 0.9464 & 4.9600 & -21.0219 & -1.4932 & -13.6691 & -- & -- & -- \\\cline{2-11}
 & 4 & 0.0763 & 0.6679 & 4.5450 & -29.7845 & -21.4321 & -70.7808 & -70.4563 & -- & --  \\\cline{2-11}
 & 5 & 0.0751 & 0.6597 & 4.5268 & -30.1556 & -21.5084 & -77.6290 & -79.6259 & 27.4329 & --\\\cline{2-11}
 & 6 & 0.1300 & 1.0149 & 5.0954 & -19.6022 & -0.2955 & -5.8584 & -1.5900 & -3.0728 & 1.6702 \\\hline\hline
 & 1 & 0.2330 & 2.7451 & 3.3774 & -7.2473 & -- & -- & -- & -- & --\\\cline{2-11}
 & 2 & 0.2026 & 2.2498 & 3.5838 & -8.8427 & -2.7598 & -- & -- & -- & --\\\cline{2-11}
$\left(\frac{1}{2},m\right)$ & 3 & 0.1583 & 1.7374 & 3.6243 & -11.4505 & -6.7583 & -3.6280 & -- & -- & -- \\\cline{2-11}
 & 4 & 0.2426 & 2.7923 & 3.4566 & -7.1248 & -0.3962 & 0.3601 & 0.1253 & -- & --\\\cline{2-11}
& 5 & 0.2141 & 2.3942 & 3.5579 & -8.3093 & -1.8365 & 0.0811 & 0.1950 & 0.0663 & --\\\cline{2-11}
 & 6 & 0.2037 & 2.6879 & 3.0160 & -7.4014 & 0.6575 & 0.3404 & 0.2990 & -0.0476 & 0.0201\\\hline\hline
\end{tabular}
\caption{Collection of the fixed points and the dimensionless parameters $\tilde{G}^\ast$ and $\tilde{\Lambda}^\ast$ for $\beta\to -\infty$ related to truncations from $R$ to $R^6$ in the linear and exponential parametrization. All dimensionless couplings $g^\ast$ have been multiplied by $10^3$.}
\end{table}

\begin{table}[htb]\scriptsize
\begin{tabular}{|c|c|c|c|c|c|c|c|c|}\hline\hline
\makebox[3em]{$(\omega,m)$} & \makebox[3em]{N} & \makebox[3em]{$\theta_1$} & \makebox[3em]{$\theta_2$} & \makebox[3em]{$\theta_3$}
& \makebox[3em]{$\theta_4$} & \makebox[3em]{$\theta_5$} & \makebox[3em]{$\theta_6$} & \makebox[3em]{$\theta_7$}\\\hline
 & 1 & 2.4545 -2.5767i & 2.4545 +2.5767i & -- & -- & -- & -- & -- \\\cline{2-9}
 & 2 & 2.5148-1.8917i & 2.5148+1.8917i & 7.8779 & -- & -- & -- & --  \\\cline{2-9}
$(0,0)$ & 3 & 2.6394-2.1159i & 2.6394+2.1159i & 4.0655 & -1.8134 & -- & -- & -- \\\cline{2-9}
 & 4 & 4.3836-2.8471i & 4.3836+2.8471i & 3.0636 & -0.2609 & -5.7246 &  -- & -- \\\cline{2-9}
 & 5 & 2.5904-7.1776i & 2.5904+7.1776i & 3.7292 & -0.3354 & -3.2286 & -8.1730 & -- \\\cline{2-9}
 & 6 & 3.0894-3.4346i & 3.0894+3.4346i & 5.6203 & -2.4728 & -4.5203 & -12.2586-2.8747i & -12.2586+2.8747i \\\hline\hline
 & 1 & 4.0 & 2.4415 & -- & -- & -- & -- & -- \\\cline{2-9}
 & 2 & 4.0 & 1.7496 & -19.1745 & -- & -- & -- & -- \\\cline{2-9}
$\left(\frac{1}{2},m\right)$ & 3 & 4.0 & 0.4374 & 20.7814 & -11.4433 & -- & -- & -- \\\cline{2-9}
 & 4 & 4.0 & 1.9279 & -4.4716 & -7.1001 & -10.0966 & -- & -- \\\cline{2-9}
 & 5 & 4.0 & 0.8724 & 25.3874 & -5.4564 & -5.9866 & -7.8517 & -- \\\cline{2-9}
 & 6 & 4.0 & 2.1819 & 4.9551 & -4.1972-1.1378i & -4.1972+1.1378i & -4.7917-0.3080i & -4.7917+0.3080i \\\hline\hline
\end{tabular}
\caption{Collection of the critical exponents for $\beta\to -\infty$ related to linear and exponential parametrization corresponding to truncations from $R$ to $R^6$.}
\end{table}

\bibliography{refs}

\end{document}